\documentclass[apj,numberedappendix,appendixfloats]{emulateapj}
\usepackage{xspace,graphics,graphicx,longtable}
\usepackage{subfigure}
\usepackage{appendix}
\usepackage{lscape}
\usepackage{natbib}
\newcommand{\mkms}{{\rm \; km\;s^{-1}}}
\newcommand{\kms}{km~s$^{-1}$ }

\slugcomment{Submitted to ApJ} 

\shorttitle{Ubiquitous Collimated Outflows at $z \sim 0.5$ }
\shortauthors{Rubin et al.}

\begin{document}

\title{Evidence for Ubiquitous Collimated Galactic-Scale Outflows along the Star-Forming Sequence at $z \sim 0.5$ }
\author{Kate H. R. Rubin\altaffilmark{1}, J. Xavier Prochaska\altaffilmark{1,2}, David C. Koo\altaffilmark{2}, Andrew C. Phillips\altaffilmark{2}, Crystal L. Martin\altaffilmark{3}, \& Lucas O. Winstrom\altaffilmark{4}}
\altaffiltext{1}{Max-Planck-Institut f\"ur Astronomie, K\"onigstuhl 17, 69117 Heidelberg, Germany; rubin@mpia.de}
\altaffiltext{2}{Department of Astronomy and Astrophysics, UCO/Lick Observatory, University of California, 1156 High Street, Santa Cruz, CA 95064}
\altaffiltext{3}{Department of Physics, University of California, Santa Barbara, CA 93106}
\altaffiltext{4}{Newman Laboratory of Elementary Particle Physics, Cornell University, Ithaca, NY 14853}

\begin{abstract}
We present an analysis of the \ion{Mg}{2} $\lambda \lambda 2796, 2803$ and \ion{Fe}{2} $\lambda \lambda 2586, 2600$ absorption line profiles in 
individual spectra of 105 galaxies at $0.3 < z < 1.4$.  The galaxies, drawn from redshift surveys of the GOODS fields and the Extended Groth Strip, 
fully sample the range in star formation rates (SFRs) occupied by the star-forming sequence with stellar masses $\log M_*/M_{\odot} \gtrsim 9.5$ at $0.3 < z < 0.7$.  Using the Doppler shifts of the \ion{Mg}{2} and \ion{Fe}{2} absorption lines as tracers of cool gas kinematics, we detect large-scale winds in $66\pm5\%$ of the galaxies.  High-resolution \emph{Hubble Space Telescope}/Advanced Camera for Surveys imaging and our spectral analysis indicate that the outflow detection rate depends primarily on galaxy orientation: winds are detected in $\sim89\%$ of galaxies having inclinations ($i$) $< 30^{\circ}$ (face-on), while the wind detection rate is only $\sim45\%$ in objects having $i > 50^{\circ}$ (edge-on).  Combined with the comparatively weak dependence of the wind detection rate on intrinsic galaxy properties (including SFR surface density), this suggests that biconical outflows are ubiquitous in normal, star-forming galaxies at $z \sim 0.5$, with over half of the sample having full wind cone opening angles of $\sim100^{\circ}$.  We find that the wind velocity is correlated with host galaxy $M_*$ at $3.4\sigma$ significance, while the equivalent width (EW) of the flow is correlated with host galaxy SFR at $3.5\sigma$ significance, suggesting that hosts with higher SFR may launch more material into outflows and/or generate a larger velocity spread for the absorbing clouds.  
The large ($> 1$ \AA) \ion{Mg}{2} outflow EWs typical of this sample are rare in the context of \ion{Mg}{2} absorption studies along QSO sightlines probing the extended halos of foreground galaxies, implying that this wind material is not often detected at impact parameters $> 10$ kpc.
Assuming that the gas is launched into dark matter halos with simple, isothermal density profiles, the wind velocities measured for the bulk of the cool material ($\sim 200-400\mkms$) are sufficient to enable escape from the halo potentials only for the lowest-$M_*$ systems in the sample.  However, the highest-velocity gas in the outflows 
typically carries sufficient energy to reach distances of $\gtrsim50$ kpc, and may therefore be a viable source of cool material for the massive circumgalactic medium observed around bright galaxies at $z\sim0$.  
\end{abstract}

\section{Introduction}

Over the past decade, large-scale redshift surveys have significantly advanced our 
understanding of the buildup of luminous structures over cosmic time.  
The Sloan Digital Sky Survey (SDSS; \citealt{York2000}) has provided an exceptionally detailed picture of the properties of galaxies in the local universe, establishing their bimodal distribution in the color-magnitude diagram \citep[e.g.,][]{Blanton2003b} and the galaxy stellar mass function to high precision \citep{Bell2003}.  Surveys reaching into the more distant universe (e.g., COMBO-17, DEEP2, the NEWFIRM Medium-Band Survey; \citealt{Bell2004,Davis2003,Whitaker2011}) have revealed a
gradual decline in the characteristic galaxy mass scale hosting strong star formation activity \citep[`downsizing';][]{Cowie1996,Bell2005,Bundy2006}, with a concomitant, 
factor of $\sim10$ increase in the mass density of red galaxies from $z\sim2.2$ to today \citep{Bell2004,Faber2007,Brammer2011}.  

Theoretical efforts to understand these galaxy distributions, as well as the observed metal enrichment of the intergalactic medium \citep[IGM; e.g.,][]{SongailaCowie1996,Simcoe2004,Adelberger2005}, have invariably invoked feedback mechanisms in order to reconcile the discrepancies between these observations and predictions based on the assumption of efficient galaxy growth within the framework of hierarchical structure formation \citep[e.g.,][]{WhiteFrenk1991, SomervillePrimack1999,SpringelHernquist2003,OppenheimerDave2006}.  For instance, studies of hydrodynamical simulations which follow the gas accretion and buildup of stellar mass in galaxy halos formed in a cosmological context \citep[e.g.,][]{Keres2005,Keres2009,Oppenheimer2010} compare the predicted local galaxy stellar mass function to that observed \citep{Bell2003}, finding that the simulations overproduce the number of galaxies at every mass scale.  
By implementing an `ejective'  stellar feedback mechanism, 
\citet{Oppenheimer2010} were able to suppress stellar mass buildup sufficiently to bring the present-day stellar mass function into accord with observations.  Similar results have been achieved in semi-analytic models invoking stellar feedback \citep[e.g.][]{Guo2011}.  

The overall success of these efforts in predicting the distribution of luminous matter at $z\sim0$ motivates further testing of the predicted co-evolution of galaxies and the kinematics and spatial distribution of non-luminous baryons in their surroundings (i.e., the circumgalactic medium, or CGM).  The large-scale galactic outflow phenomenon that inspired the implementation of `ejective feedback' is now understood to be a common feature of vigorously star-forming galaxies in the local universe and out to $z\sim6$ \citep{Heckman2000,Ajiki2002,Shapley2003,Martin2005,Rupke2005b,Weiner2009}.
However, the ubiquity of feedback among \emph{all} star-forming galaxies is a presupposition of these models that is not constrained by the vast majority of studies of star-formation driven outflows in the local universe, which have traditionally focused on observations of winds around extreme starburst or merger systems.  
Further, this ubiquity may run contrary to the conventional notion that the surface density of star formation activity must reach a critical threshold in order to launch a large-scale wind \citep{McKeeOstriker1977,Heckman1990,Heckman2002,Kornei2012}.  
In their survey of winds in more typical, massive ($\log M_*/M_{\odot} > 10.4$) star-forming galaxies at $z\sim0$, \citet{ChenNaI2010} have recently demonstrated that outflows are indeed pervasive among their sample; however, observational evidence for ubiquitous outflows from more distant star-forming galaxies has remained tentative \citep[e.g.,][]{Weiner2009,RubinTKRS2009}.  In particular, because these studies rely on the absorption signature of cool gas in spectroscopy of faint galaxy continua to detect winds, much of their analysis is limited to measurement of gas kinematics in higher signal-to-noise (S/N), coadded spectra of many tens or hundreds of galaxies, and thus they cannot explore the dispersion in outflow properties among their galaxy samples.  
In one of the first studies to surpass these limitations, \citet{Martin2012} detected winds in nearly half of their deep, individual spectra of $\sim200$ $z\sim1$ galaxies having $\log M_*/M_{\odot} \gtrsim 9.4$.  
They found no evidence that this detection rate depends on the intrinsic properties of the host galaxies (e.g., $M_*$ or SFR), positing that 
 while winds  are a typical feature of star-forming galaxies at this epoch, their detection depends strongly on viewing angle.  
 

The implementation of a variety of feedback `recipes' in galaxy formation models provides another point of comparison with observations, where a `recipe' is a 
scaling law relating outflow velocities and gas mass to intrinsic host galaxy properties.  
In particular, \citet{Oppenheimer2010} favor a recipe in which wind velocity scales with galaxy velocity dispersion ($\sigma_\mathrm{gal}$) and wind mass is inversely proportional to $\sigma_\mathrm{gal}$.  This scaling is motivated by measurements of outflow kinematics in nearby extreme starbursting systems with absolute SFRs covering nearly 4 orders of magnitude \citep[$0.1~M_{\odot}~\rm yr^{-1} < \mathrm{SFR} < 1000~M_{\odot}~\rm yr^{-1}$;][]{Martin2005}.  However, studies of scaling laws in larger samples of more typical star-forming galaxies having a comparatively narrow range in SFR 
have remained inconclusive.  The weak rise in outflow velocity measured from analysis of coadded spectra of galaxies of increasing SFR and $M_*$ at $z\sim1.4$ by \citet{Weiner2009} is suggestive of a positive correlation; however, other work has concluded
 that outflow velocity is independent of host galaxy SFR and $M_*$ over 2.5 dex in SFR \citep[i.e., $1~M_{\odot}~\rm yr^{-1} < \mathrm{SFR} < 500~M_{\odot}~\rm yr^{-1}$;][]{Rupke2005b,ChenNaI2010,Martin2012}.  Consistent, detailed analysis of outflow kinematics in high S/N spectroscopy of a galaxy sample spanning a large dynamic range in intrinsic host galaxy properties is an important step toward establishing empirical models that may guide future theoretical studies.


A final and fundamental prediction of cosmological galaxy formation models addresses the ultimate fate of gas expelled from star-forming galaxies in large-scale winds.  The modest velocities and high gas masses of winds implemented in \citet{OppenheimerDave2006} and \citet{Oppenheimer2010} keep much of the outflowing material within the gravitational potential well of the host halos such that it supplies a massive gas reservoir in the galaxy environs (the CGM).  
This material is subsequently re-accreted on timescales of 1-2 Gyr, providing the dominant source of fuel for star formation at $z< 1$.  
Observations of warm absorbing material along QSO sightlines surrounding low-redshift, $L^*$ galaxies have already confirmed the ubiquitous presence of such extended gas reservoirs, and find that the mass of material is comparable to that contained within the host galaxy interstellar medium \citep[ISM;][]{Tumlinson2011,Prochaska2011b,Werk2013}.  The kinematics and gas masses of winds at the time of launch are thus expected to be modest in comparison to the escape velocity of the host galaxy halos, and yet sufficient to maintain this massive gas reservoir in the CGM.

In this work, we use rest-frame near-UV spectroscopy of a sample of 105 galaxies at $0.3 < z < 1.4$ to examine the kinematics of cool ($T \lesssim10^4~K$) gas traced by \ion{Mg}{2} $\lambda \lambda 2796,2803$ and \ion{Fe}{2} $\lambda \lambda 2586, 2600$ absorption.  This sample, drawn from redshift surveys of the GOODS and EGS fields \citep[e.g.,][]{Wirth2004,Davis2003,LeFevre2005,Szokoly2004}, fully covers the SFR-$M_*$ parameter space occupied by star-forming galaxies with $\log M_*/M_{\odot} \gtrsim 9.5$ at $z\sim0.5$, permitting exploration of outflow properties over the entire breadth of the star-forming sequence at $z > 0.3$ for the first time.  The \ion{Mg}{2} and \ion{Fe}{2} absorption detected in the foreground of the galaxy stellar continuum is assumed to trace outflowing gas if the line profiles are blueshifted with respect to the galaxy systemic velocity.  The high S/N of our spectra ($3~\rm pixel^{-1} \lesssim S/N \lesssim 30~pixel^{-1}$, with pixel widths $\sim90\mkms$) permits measurement of outflow velocity and equivalent width (EW) and constrains the outflow column density, covering fraction and velocity width for each individual galaxy in our sample.  Furthermore, the extensive ancillary multiwavelength data available in the GOODS and EGS fields facilitates detailed constraints on host galaxy morphology, SFR, $M_*$, and SFR surface density ($\Sigma_\mathrm{SFR}$).  
We use our sample to explore the dependence of the incidence of outflows and wind velocity and EW on these intrinsic galaxy properties as well as on the viewing angle of the host galaxy disks.  This exploration, together with the constraints on outflow energetics provided by our line profile analysis, are leveraged to address the fate of the detected wind material and its contribution to the CGM.  

We describe our sample selection, observations, and supplementary data in \S\ref{sec.obs}.  We briefly describe our measurements of galaxy systemic velocities in \S\ref{sec.redshifts}, giving further details in Appendix A.  Measurements of host galaxy properties are described in \S\ref{sec.galprop}.  The details of our line profile analysis and the classification of our sample based on absorption kinematics are described in \S\ref{sec.modeling}.  Section \ref{sec.fitresults} contains a summary of our spectral fitting results and analysis of the sensitivity of our line profile modeling to the presence of winds.  Readers interested primarily in our results on the relationship between outflow detection rates, velocities and EWs and host galaxy properties may wish to start with \S\ref{sec.bigresults}.  We discuss the implications of these results in the context of the frequency and morphology of outflows and their physical impact on their host galaxies and the surrounding CGM in \S\ref{sec.discussion}, and conclude in \S\ref{sec.conclusions}.
We adopt a $\rm \Lambda CDM$ cosmology with $h_{70} = H_0 / (70~\rm km~s^{-1}~Mpc^{-1}) $, $\rm \Omega_M = 0.3$, and $\rm \Omega_{\Lambda} = 0.7$.  Where it is not explicitly written, we assume $h_{70} = 1$.  Magnitudes quoted are in the AB system.

\section{Sample Selection, Observations, and Supplementary Data}\label{sec.obs}

Because one of the primary goals of this work is to compare cool gas kinematics with galaxy orientation, morphology, and the spatial distribution of star formation, 
our galaxy sample is drawn from pre-existing photometric and spectroscopic redshift surveys in fields with deep imaging taken with the \emph{HST} Advanced Camera for Surveys.  
Specifically, we targeted galaxies in the GOODS fields \citep{Giavalisco2004} and the AEGIS survey field \citep[the Extended Groth Strip or EGS;][]{Davis2007}.  
High-quality galaxy redshifts 
from the Team Keck Treasury Redshift Survey \citep[TKRS;][]{Wirth2004} in the GOODS-N field 
and the DEEP2 survey \citep{Davis2003} of the EGS, both of which 
are optimized for the range $0.2 \lesssim z \lesssim 1.4$, 
aided in the sample selection.  
 We used several different surveys of the GOODS-S field to select objects, 
including DEEP2, the VIRMOS VLT DEEP survey 
\citep{LeFevre2005}, 
and FORS spectroscopy from \citet{Szokoly2004} and ESO VLT program 170.A-0788(B) (PI:
Cesarsky).  \citet{Croom2001}, \citet{Colless2001}, and \citet{Mignoli2005} provided some additional 
spectroscopic redshifts.  We also used photometric redshifts from COMBO-17 \citep{Wolf2004} and 
\citet{Zheng2004}.  
To ensure coverage of at least the \ion{Mg}{2} doublet transition at $\lambda_\mathrm{rest} \sim 2800$ \AA\
with our chosen spectroscopic setup (see \S\ref{sec.specsetup}), we targeted galaxies having $z > 0.3$.
Finally, to obtain the S/N level required to detect absorption lines against the stellar continua in 
reasonable exposure times, we selected galaxies having $B_\mathrm{AB} < 23$, where $B$-band photometry was obtained from \citet[][GOODS-N]{Capak2004}, 
\citet[][GOODS-S]{Giavalisco2004}, and CFHT imaging of the EGS \citep{Davis2003,Coil2004}.

\subsection{Keck/LRIS Spectroscopy}\label{sec.specsetup}
Spectroscopic observations were carried out using the Low Resolution Imaging Spectrometer (LRIS) with the Atmospheric Dispersion Corrector on Keck 1 \citep{Cohen1994} in multislit
mode on 2008 May 30-31 UT, 2008 October 2-3 UT, 2008 November 27-28 UT, and 2009 April 3 UT.  Seeing conditions varied over the course of the program (FWHM $\sim 0.6\arcsec - 1.6\arcsec$), and a $0.9\arcsec$ slit width was used for all slitmasks.  We used the $\rm 600~ l~mm^{-1}$ grism blazed at 4000~\AA~on the blue side and the $\rm 600~ l~mm^{-1}$ grating blazed at 7500~\AA~on the red side with the D560 dichroic.  This setup affords nearly contiguous wavelength coverage between $\sim 3200$~\AA~ and $\sim 8200$~\AA. 
The FWHM resolution ranges between $400~\rm km~s^{-1}$ at $\sim3200$~\AA\ and $180~\rm km~s^{-1}$ at $\sim8200$ \AA, and is determined from the measured width of arc lamp lines over this wavelength interval.  In practice, the median FWHM resolution for our sample spectra is $274\mkms$ at $\lambda_\mathrm{rest} \sim 2800$ \AA\ and $286\mkms$ at $\lambda_\mathrm{rest} \sim 2600$ \AA.
 We obtained  between four and eight $\sim1800$ sec (2-4 hour)
exposures for each slitmask.  See Table~\ref{tab.obs_summary} for details on our chosen fields and exposure times.  

The data were reduced using the XIDL LowRedux\footnote{http://www.ucolick.org/$\sim$xavier/LowRedux/} data reduction pipeline.  The pipeline includes bias subtraction and flat-fielding, slit finding, 
wavelength calibration, object identification, sky subtraction, cosmic ray rejection, and relative flux calibration.  
Typical rms errors in the wavelength solution are 0.17 \AA\ on the blue side and 0.04 \AA\ on the red side.  
Wavelength calibrations were adjusted for flexure by applying an offset calculated from the cross correlation of the observed sky spectrum with a sky spectral template.  
Vacuum and heliocentric corrections were then applied.

\tabletypesize{\scriptsize}
\begin{deluxetable*}{lcccccc}
\tablecolumns{7}
\tablecaption{Summary of Observed Fields \label{tab.obs_summary}}
\tablewidth{0pt}
\tablehead{\colhead{Field} & \colhead{R. A.} & \colhead{Declination} & \multicolumn{2}{c}{Exposure Time} & \colhead{Number of Spectra} & \colhead{Date}\\ 
 & \colhead{J2000} & \colhead{J2000} & Blue & Red & &}
\startdata
GOODS-N & 12:36:24.21 & +62:11:46.0 & $5 \times 1800 + 1560$ sec & $2 \times 1800 + 4 \times 1680$ sec & 12 & 2008 May 30\\ 
GOODS-N & 12:37:03.82 & +62:16:23.1 & $4 \times 1800$ sec & $4 \times 1750$ sec & 15 & 2008 May 31\\ 
EGS            & 14:17:16.75 & +52:29:03.6 & $6 \times 1800$ sec & $3 \times 1750 + 3 \times 1450$ sec & 13 & 2008 May 30\\  
EGS            & 14:20:47.03 & +53:08:18.0 & $3 \times 1800 + 2 \times 1500$ sec &  $3 \times 1750 + 2 \times 1450$ sec & 12 & 2008 May 31\\ 
EGS 	           & 14:19:29.13 & +52:50:00.8 & $5 \times 1840 + 1789$ sec & $5 \times 1800 + 1792$ sec & 13 & 2009 Apr 03\\
GOODS-S  & 03:32:32.67 & -27:45:24.5 &  $6 \times 1800$ sec & $6 \times 1800$ sec & 16 & 2008 Oct 02\\   
GOODS-S  & 03:32:29.97 & -27:43:54.5 &  $5 \times 1800 + 2 \times 1250$ sec  & $5 \times 1800 + 2 \times 1200$ sec & 5 & 2008 Oct 03 \\ 
GOODS-S  & 03:32:31.25 & -27:49:58.2 &  $4 \times 1800 + 4 \times 1500$ sec  & $4 \times 1800 + 4 \times 1500$ sec & 9 & 2008 Nov 27\\ 
GOODS-S & 03:32:33.54  & -27:53:14.3 &  $8 \times 1800$ sec &  $8 \times 1800$ sec & 10 & 2008 Nov 28 \\ 
\enddata
\end{deluxetable*}

\subsection{Supplementary Data}
We use the high quality \emph{HST}/ACS imaging available in both the GOODS fields and the EGS \citep{Giavalisco2004,Davis2007}.  The GOODS imaging covers a $10\arcmin \times 16\arcmin$ area in each of the fields with the ACS F435W, F606W, F775W and F850LP bands ($B_{435}$, $V_{606}$, $i_{775}$ and $z_{850}$).  The limiting surface brightness at $1\sigma$ in a $\rm 1~sq. \arcsec$ aperture in the F850LP band is $\mu_{AB} = 27.3~\rm mag~arcsec^{-2}$ \citep[][version 1 release]{Giavalisco2004}.  We use the mosaic data in each band with a pixel scale $\rm 0.03\arcsec ~ pix^{-1}$.  
The EGS has been imaged in the $V_{606}$ and F814W ($i_{814}$) bands to a limiting magnitude of 28.10 in $i_{814}$ over a $10\arcmin \times 67\arcmin$ area with a pixel scale $\rm 0.03\arcsec ~ pix^{-1}$ \citep{Davis2007}.  

In addition, we take advantage of the extensive ground-based optical and near-IR broad-band photometry available in these fields, as well as photometry from deep Spitzer/IRAC and MIPS imaging.  In GOODS-N, we use photometry provided by the MOIRCS Deep Survey \citep{Kajisawa2011}, which includes MOIRCS near-IR photometry, $U$-band photometry from \citet{Capak2004}, and  \emph{HST}/ACS optical photometry from \citet{Giavalisco2004}.  In GOODS-S, we use the FIREWORKS catalog \citep{Wuyts2008}, which includes broad-band photometry from the ESO/MPG 2.2m WFI $U_{38}$ and optical bands and the VLT/ISAAC $JHK_s$ bands.  Both the MOIRCS Deep Survey and FIREWORKS also include Spitzer/IRAC and MIPS photometry from the \emph{Spitzer Space Telescope} Legacy Program \citep{Dickinson2003}.
In the EGS, we use the catalogs published in \citet{Barro2011}, which incorporate $u$-band and optical photometry from the CFHTLS\footnote{www.cfht.hawaii.edu/Science/CFHTLS-DATA/}, near-IR photometry from \citet{Bundy2006}, 4 bands of IRAC imaging from \citet{Barmby2008}, and MIPS $24\mu$m imaging from MIPS GTO and FIDEL surveys.  In addition, we use NUV fluxes from the source catalog of the $GALEX$ public data release GR6 in each of our fields.  Further details on the photometry we use throughout this paper are given in Appendix~\ref{sec.appen_photometry}.

\section{Redshifts}\label{sec.redshifts} 
One of the main goals of our analysis is to determine the speed of cool gas relative to 
its host galaxy's disk or star-forming regions; therefore, accurate redshift measurements are of primary importance.  A full description of our method of redshift determination is given in Appendix~\ref{sec.appen_redshifts}, but here we summarize the most pertinent details.  Redshift values are derived using an IDL code adapted from the publicly available programs developed for the SDSS.  This code calculates the best-fit lag between observed spectra and a linear combination of SDSS galaxy eigenspectra.  We prefer redshift measurements based on stellar absorption, as stellar continuum emission better traces the systemic velocity of the associated ensemble of dark matter and stars than nebular emission from \ion{H}{2} regions \citep[e.g.,][]{Rodrigues2012}.  Therefore, where possible (i.e., where the stellar continuum S/N is sufficient), we mask nebular emission lines in the data prior to redshift fitting.  

\begin{figure}
\includegraphics[angle=90,width=\columnwidth]{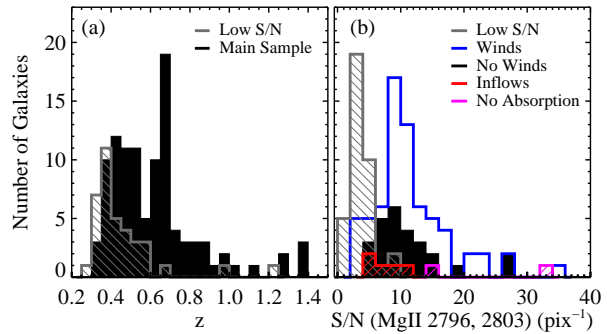}
\caption[]{\emph{(a)} Redshift distribution of the portion of the sample having spectra with sufficient $\rm S/N$ to constrain cool gas kinematics (black).  The redshift distribution of the portion of the sample with insufficient $\rm S/N$ for cool gas kinematical constraints is shown in gray.  The latter galaxies are systematically at lower redshifts than those with sufficient S/N, such that the \ion{Mg}{2} and \ion{Fe}{2} transitions are shifted to the blue extreme of our spectral coverage.
\emph{(b)}   Distribution of mean spectral $\rm S/N$ measured at rest wavelengths $\rm 2770~\AA < \lambda_{rest} < 2780~\AA$ and $\rm 2810~\AA < \lambda_{rest} < 2820~\AA$ for galaxies with no detected inflow or winds (black), winds (blue), inflows (red), no detected absorption (magenta), and for galaxies whose spectra have insufficient $\rm S/N$ to characterize cool gas kinematics (gray).  See Section~\ref{sec.mcmc} for further discussion of these categorizations.
     \label{fig.zstonhist}}
\end{figure}

From a comparison between our redshifts and those measured by the TKRS and AEGIS surveys (see Figure~\ref{fig.redshifts}c), we estimate that our measurements have an rms uncertainty of $28\mkms$.  This is consistent with the redshift offsets found for galaxies which were observed more than once during our LRIS survey (which have a mean offset of 19 \kms and a maximum offset of $32\mkms$).  The redshift distribution of the portion of the sample for which cool gas kinematic measurements are possible is shown in black in Figure~\ref{fig.zstonhist}a.  
The median redshift of this sample is 0.619, and the minimum and maximum redshifts are 0.310 and 1.384.  The redshift distribution of the portion of the sample whose spectra have insufficient $\rm S/N$ for constraints on cool gas kinematics is shown in gray; this distribution is shifted to lower redshifts because of 
increasing atmospheric extinction, declining instrument sensitivity, and declining galaxy continuum emission blueward of 3500 \AA.

\section{Host Galaxy Properties}\label{sec.galprop}

\subsection{Inclination and Morphology}\label{sec.inclination}

We estimate the inclination ($i$) of our galaxies by simply assuming that $i$ depends on the ratio of a galaxy's semi-minor ($b$) and semi-major ($a$) axes as
$\cos i = b/a$.  
Such an assumption holds exactly for circular, infinitely thin disks, and generally yields an inclination within $\sim10^{\circ}$ of the `true' inclination for more realistic galaxies with smooth, triaxial structures (A.\ van der Wel, in preparation).  
We measure this axis ratio in the reddest available \emph{HST}/ACS passband ($i_{814}$ for galaxies in the EGS and $z_{850}$ for the remainder of the sample) in order to trace the spatial distribution of the entire stellar population as closely as possible.  To measure galaxy major and minor axes, we first obtain a segmentation map for each object created by SExtractor \citep{BertinArnouts1996}.  For galaxies in the EGS, we use segmentation maps generated from the sum of the $V_{606}$ and $i_{814}$ images provided by J. Lotz and described in \citet{Lotz2006}.  For galaxies in GOODS, we create SExtractor segmentation maps from the $z_{850}$-band images with the detection threshold set to 0.6$\sigma$ and a minimum detection area of 16 pixels.  We visually inspect the segmentation map for each object, 
adjusting it by hand for a small subset to improve object deblending.  
We then compute the light-weighted center and second-order moments of each galaxy image.  Arithmetic combinations of these measurements yield the maximum and minimum spatial rms of the galaxy light profile (i.e., the major and minor axes) as described in the SExtractor documentation.\footnote{http://www.astromatic.net/software/sextractor} 

Our assumption that the ratio of these axis measurements can be used to estimate a galaxy's orientation 
breaks down completely for galaxies with morphological irregularities.  
For galaxies undergoing mergers, for instance, an inclination is not easily defined.  We therefore perform visual morphological classification for the purpose of distinguishing between disk-like systems and objects that are morphologically disturbed.   Following loosely the classification scheme of \citet{Abraham1996}, we divide galaxies into 6 categories after inspecting their $i_{814}$ or $z_{850}$-band images: compact, E/SO, Sab, S, Ir, and merger.  Both this classification and the inclination measured for each galaxy are listed in Table~\ref{tab.gals}.
 
\subsection{Rest-Frame Magnitudes and Colors}


We derive rest-frame $M_B$ and $U-B$ colors for our sample from \emph{HST}/ACS and ground-based optical photometry and near-IR photometry using the code KCORRECT \citep{Blanton2007}.  Specifically, for galaxies in the GOODS-N field, we use ground-based $U$-band photometry, $b_{435}v_{606}i_{775}z_{850}$, and $JHK_s$ measurements provided by the MOIRCS Deep Survey \citep{Kajisawa2011}.  For galaxies in the EGS, we use
ground-based $ugriz$ and $J$, $K$, and $K_s$ photometry from \citet{Barro2011}, and 
for galaxies in the GOODS-S field we use $U_{38}BVRI$ photometry, $z_{850}$ measurements, and VLT/ISAAC $JK_s$ photometry from the FIREWORKS survey \citep{Wuyts2008}.  As the GOODS-N and -S photometric catalogs report aperture photometry, we apply an aperture correction calculated from the ratio between the total $K_s$-band flux and the flux measured in the appropriate aperture.  
In addition, we apply a correction for Galactic reddening from the maps of \citet{SFD1998} to the photometry in all three fields, and then fit each galaxy spectral energy distribution (SED) using KCORRECT.  

To test our method, we compare our $M_B$ and $U-B$ values for galaxies in GOODS-N to those derived in \citet{Weiner2006}, who used photometry from independent source catalogs (\citealt{Giavalisco2004} and \citealt{Capak2004}) and a different $K$-correction procedure \citep{Weiner2005,Willmer2006}.  We find a median offset of $-0.13$ mag in $M_B$ values with a dispersion of $0.15$ mag, and a median offset of $0.065$ mag in $U-B$ with a dispersion of $0.092$.  We also compare our rest-frame photometry for galaxies in the EGS to that described in \citet{Weiner2007}, who used observed $BRI$ photometry and the same $K$-correction procedure as that of \citet{Weiner2006}.  For this field we find a median offset of $-0.02$ mag in $M_B$ values with a dispersion of $0.30$ mag, and a median offset of $0.025$ mag in $U-B$ with a dispersion of $0.120$ mag.  From this consistency between our rest-frame photometry and that of \citet{Weiner2006,Weiner2007}, we conclude that our $K$-correction method is indeed robust.

\subsection{Total $M_*$ and SFR}\label{sec.mstarsfr}

To calculate $M_*$ and total SFR for our sample, we fit SEDs to broadband photometry over observed wavelengths 2400 \AA\ to 24$\mu$m.  The photometry available differs slightly from field to field; however, in all of the fields we include measurements from the {\it GALEX} NUV band, optical photometry in several bands centered at $\sim 3800$ \AA\ through 8500 \AA, near-IR $J$ and $K$ or $K_{s}$ photometry, measurements in the four IRAC bands at 3.6, 4.5, 5.8 and 8.0 $\mu$m, and MIPS 24 $\mu$m fluxes.  

We use the SED fitting code MAGPHYS, described in \citet{daCunha2008, daCunha2011}.  In brief, MAGPHYS combines the stellar population synthesis models of \citet{BC2003} with dust emission from both molecular clouds and an ambient, diffuse ISM \citep[e.g.,][]{CharlotFall2000} to simultaneously fit the observed photometry.  
Several dust `components' are implemented (e.g., polycyclic aromatic hydrocarbons, and hot, warm ($30-60$ K), and cold ($15-25$ K) dust).  For each galaxy, the code builds a library of UV-to-IR SEDs assuming an exponential star formation history with variable galaxy age and star formation timescale and with bursts of star formation superimposed at randomly-selected epochs.  
The resulting SEDs span a wide range of plausible physical parameters (SFR, $M_*$, dust luminosity, temperature of the dust in the ambient ISM and birth clouds, etc.).  
The code then computes the $\chi^2$ value for each model SED with respect to the observed photometry and builds marginalized likelihood distributions for each physical parameter.   
A demonstration of this approach is given in Appendix~\ref{sec.appen_photometry}.  All quantities are computed assuming a Chabrier IMF \citep{Chabrier2003}.  

As SFR and $M_*$ estimates are of particular importance for our analysis, we compare our measurements to those of \citet[][hereafter B11]{Barro2011}, who used a more traditional approach to estimate these quantities from the same observed photometry adopted here for the EGS.  Briefly, they fitted SEDs to this photometry and interpolated the resulting best fit to find a luminosity for each object at 2800 \AA \ in the rest-frame.  They converted this luminosity to a $\rm SFR_{UV}$, uncorrected for extinction, using the relation given in \citet{Kennicutt1998}.  They also fitted dust emission templates to the 24$\mu$m flux measured for each galaxy to calculate its IR luminosity and converted this to a $\rm SFR_{IR}$ again using \citet{Kennicutt1998}.  They recommend summing these values, such that the total SFR is given by $\rm SFR = SFR_{IR} + SFR_{UV}$.  In the following, we compare our SFR measurements against this sum from B11.
 
After adjusting their results to a Chabrier IMF, 
we find good agreement between MAGPHYS and B11 $M_*$ values, with a mean offset in $\log M_*/M_{\odot}$ of -0.038 dex and a dispersion of 0.31 dex.  Less consistent are measurements of SFR, for which MAGPHYS yields values that are systematically 0.305 dex lower than B11, with a dispersion of 0.33 dex.  We discuss possible explanations for this offset in detail in Appendix~\ref{sec.appen_photometry}, concluding that it is most likely due to differences in the assumed sources of dust heating between the two methods.  While the \citet{Kennicutt1998} calibration (and B11) assumes that a galaxy's IR emission arises entirely from reprocessed UV photons emitted by young stars, MAGPHYS accounts for contributions to dust heating from the full stellar population, attributing a fraction of the total IR emission to heating by older stars.  This results in lower MAGPHYS-based SFRs, particularly for galaxies with low specific SFR (sSFR; i.e., with a larger contribution to dust heating made by older stellar populations).  
Because of the low dispersion in this offset, 
and because MAGPHYS explicitly forces consistency between $M_*$, dust luminosity, and SFR,
we consider our MAGPHYS results sufficiently robust for comparisons of star formation activity within our galaxy sample, but caution the reader that comparisons to samples for which SFR has been measured via a different method must be made with care.

\begin{figure}
\includegraphics[angle=0,width=\columnwidth]{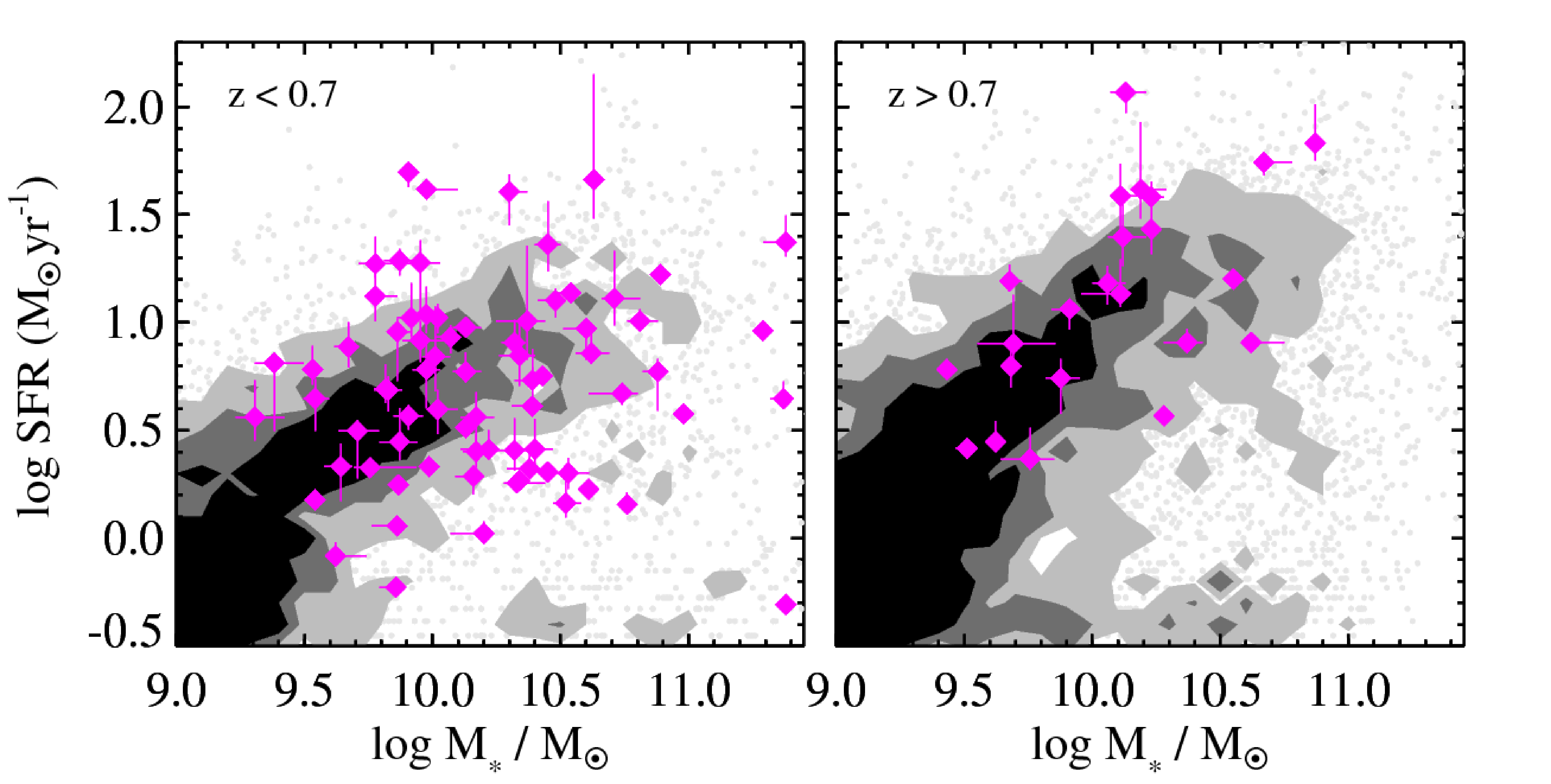}
\caption[]{Location of our galaxy sample on the star-forming sequence at $z<0.7$ (left) and $z>0.7$ (right).  The SFR-$M_*$ distribution of the B11 sample at $0.4 < z < 0.7$ (left) and $0.7 < z < 1.0$ (right) and converted to a Chabrier IMF is shown with gray contours and dots for comparison. 
Our sample fully covers the star-forming sequence down to $\log M_*/M_{\odot} \sim 9.5$ at $z<0.7$, and more sparsely covers the higher-SFR edge of the galaxy distribution at higher redshifts.
     \label{fig.sfsequence}}
\end{figure}

The SFR-$M_*$ distribution of our sample is shown in Figure~\ref{fig.sfsequence} in magenta.  
The full B11 EGS sample is shown for comparison with gray contours.  Several of our sample galaxies sit below the main B11 star-forming sequence; this is likely due to differences in the techniques used to measure SFRs for the two samples as discussed.  Specifically, galaxies with sSFR $< -9.5~\rm yr^{-1}$ (see Appendix Figure~\ref{fig.checksfrs}) tend to have MAGPHYS-based SFRs which are lower by $\sim0.5$ dex than SFRs calculated by B11.  Thus, galaxies in our sample having $\log \mathrm{SFR}~ [M_{\odot}~\rm yr^{-1}] < 0.5$ and $\log M_*/M_{\odot} \sim 10.0$, for example, would be predicted to lie directly on the star-forming sequence by B11.  
We reiterate, however, that our favored MAGPHYS-based SFRs allow for consistent comparison of SFRs among our full galaxy sample, 
and discuss absorption kinematic measurements for individual galaxies with particularly low SFRs in \S\ref{sec.lowsfr}. 
Finally, we note that our sample fully covers the star-forming sequence down to $\log M_*/M_{\odot} \sim 9.5$ at $z<0.7$.  At $z>0.7$, we tend to sample the higher-SFR edge of the sequence.   

\subsection{SFR Surface Density ($\Sigma_\mathrm{SFR}$)}\label{sec.sfrsd}

High spatial concentrations of star formation activity have long been suspected of 
playing an integral role in the driving of galactic outflows.  
Beginning with \citet{McKeeOstriker1977}, and most recently in \citet{Murray2011}, theoretical studies have predicted that a ``threshold" 
$\Sigma_\mathrm{SFR}$ must be met in order for a galaxy to launch a large-scale wind.  The existence of such a threshold is proposed 
for a variety of assumed wind physics (e.g., in studies of both momentum- and energy-driven winds), and yet observational constraints on 
its value have remained weak \citep{LehnertHeckman1996,Martin1999,Heckman2002,Dahlem2006}.  \citet{Heckman2002} noted that both local and $z\sim3$ starburst galaxies which are known to exhibit outflows satisfy the criterion $\Sigma_\mathrm{SFR} > 0.1~ M_{\odot}~\rm yr^{-1}~kpc^{-2}$, suggesting that the proposed threshold $\Sigma_\mathrm{SFR}$ must be close to or below this value.  
\citet{Kornei2012} searched for winds traced by \ion{Fe}{2} absorption in a sample of $\sim70$ star-forming galaxies at $0.7 < z < 1.3$, finding that outflows tend to reach higher velocities as $\Sigma_\mathrm{SFR}$ increases.  This result suggests that $\Sigma_\mathrm{SFR}$ may have physical relevance for the driving of winds; however, because much of this sample has $\Sigma_\mathrm{SFR}$ values higher than $0.1~ M_{\odot}~\rm yr^{-1}~kpc^{-2}$, 
this threshold remains untested.
\citet{Law2012} identified a similar trend among their sample of $z\sim2-3$ star-forming systems, but found that it is driven primarily by an even stronger anti-correlation between galaxy size and wind velocity.
Here, we characterize the spatial distribution of star formation 
in our sample with the aim to constrain the value of the proposed $\Sigma_\mathrm{SFR}$ threshold for driving winds, as well as to search for correlations between outflow properties and $\Sigma_\mathrm{SFR}$ 
for comparison to the results discussed above.

To calculate $\Sigma_\mathrm{SFR}$ for our sample, we combine our measurements of total SFR (\S\ref{sec.mstarsfr}) with an analysis of the galaxies' flux distributions in the bluest available \emph{HST}/ACS passband.  For the galaxies in GOODS, $b_{435}$ imaging ($\lambda_\mathrm{obs} \sim 4318$ \AA) traces emission at $\lambda_\mathrm{rest} \sim 2400 - 3322$ \AA\ for the bulk of the sample at $0.3 < z < 0.8$, providing maps of the distribution of young stellar populations with minimal contamination from older stars \citep{Kennicutt1998}.  For EGS galaxies, the $V_{606}$ imaging at 
$\lambda_\mathrm{obs} \sim 5919$ \AA\  covers $\lambda_\mathrm{rest} \sim 3290 - 4550$ \AA\ at $0.3 < z < 0.8$, and therefore includes emission both from very young stellar populations and from stars older than $\sim1$ Gyr.  We nevertheless treat the $V_{606}$ imaging as a direct tracer of the spatial distribution of star formation activity for EGS galaxies, 
and note that size scales computed from these images likely overestimate the true spatial extent of stars with ages $<1$ Gyr.  

For GOODS galaxies, we first generate SExtractor segmentation maps from the $b_{435}$-band images as described in \S\ref{sec.inclination}.  For EGS galaxies, we again adopt segmentation maps produced by J. Lotz \citep{Lotz2006}.  We then renormalize the total flux in all pixels assigned to a given galaxy by its total SFR, which results in a pixel map of the galaxy's star-formation intensity.   
We calculate semi-major and semi-minor axes ($a$ and $b$) from each galaxy's SFR map as described in \S\ref{sec.inclination}, and calculate a ``global" $\Sigma_\mathrm{SFR}$ using the equation $\Sigma_\mathrm{SFR, global} = \mathrm{SFR}/(\pi a^2)$.  Here we assume that the true physical radius of each galaxy is best characterized by $a$, as was done for our calculations of galaxy inclination (i.e, $\Sigma_\mathrm{SFR, global}$ has been deprojected).\footnote{This assumption results in $\Sigma_\mathrm{SFR, global}$ values up to 0.4 dex lower than those which adopt the mean of $a$ and $b$ as the true physical radius, with a median offset of 0.16 dex.}
While most of our discussion of wind properties will focus on $\Sigma_\mathrm{SFR, global}$, we also compute the average $\Sigma_\mathrm{SFR}$ of all pixels in each galaxy, 
 as well as the maximum $\Sigma_\mathrm{SFR}$ pixel value in each galaxy ($\Sigma_\mathrm{SFR, max}$).
 Errors on the various $\Sigma_\mathrm{SFR}$ measurements are calculated assuming that the uncertainty is dominated by the error in the total SFR.

\section{Absorption Line Profile Analysis}\label{sec.modeling}


\subsection{Equivalent Width Measurements}
We measure equivalent widths (EWs) of absorption features after normalizing the spectra to the continuum level.  This level is determined via a linear fit to the continuum around each absorption doublet.  We use the flux in the rest wavelength ranges 2551 - 2570 \AA \ and 2647 - 2665 \AA \ to fit the continuum level for the spectral region around the \ion{Fe}{2} $\lambda\lambda2586, 2600$ lines, and use the rest wavelength ranges 2770 - 2780 \AA \ and 2810 - 2820 \AA \ to fit the continuum level for the region around the \ion{Mg}{2} $\lambda\lambda2796, 2803$ doublet.  We then use a feature-finding code described in \citet{Cooksey2008} to identify absorption lines and measure their boxcar EW.   
Briefly, this code first convolves each spectrum with a Gaussian having a FWHM $ = 100\mkms$.  The resultant pixels are then grouped into features having a significance $> 3\sigma$; i.e., with $\mathrm{EW}/\sigma_\mathrm{EW} \ge 3$.  
We modified the code slightly to separate blended \ion{Mg}{2} doublet lines at the wavelength corresponding to the maximum flux level between their systemic wavelengths.  EWs of each transition and the S/N in selected continuum regions are listed in Tables~\ref{tab.mglines} and \ref{tab.felines}.



\subsection{Absorption Line Modeling}\label{sec.mcmc}

\subsubsection{Line Profile Description}
To quantify the kinematics and absorption strength of \ion{Mg}{2} and \ion{Fe}{2} line profiles in our spectra, 
we construct two distinct models to describe our data.  Both of these models assume that the line profile shape is 
due entirely to the absorption of continuum emission by foreground \ion{Mg}{2} or \ion{Fe}{2} ions.  However, 
continuum photons will be absorbed by gas both in front of and surrounding the galaxy, and must be re-emitted in turn, such that the excited ions decay directly back to the ground state or
to lower-energy fine-structure levels.
This scattering process has been shown to give rise to P Cygni-like 
line profiles for \ion{Mg}{2} and emission in fine-structure transitions in the case of \ion{Fe}{2} \citep{Rubin2011,Prochaska2011}.  
P Cygni emission may in principle also affect both of the \ion{Fe}{2} transitions studied here, although such emission has not been observed 
and is expected to be much weaker than that observed in \ion{Mg}{2} due to the presence of accessible \ion{Fe}{2} fine-structure levels 
\citep{Prochaska2011}.  The overall EW and morphology of such emission is highly sensitive to the gas geometry, density profile and dust content, 
and we therefore do not attempt to model it here.  Instead, 
for those spectra in our sample which exhibit significant \ion{Mg}{2} P Cygni emission (e.g., see panels showing EGS12008589, EGS12027896, EGS13050565, or EGS13058718 in Figure~\ref{fig.allspecs}), we mask affected pixels (marked in red) and reset their value to the continuum level before performing our model fits.  Even in profiles which do not exhibit significant emission, this scattering may fill in the absorption trough near the systemic velocity, shifting the 
deepest part of the trough blueward and decreasing the apparent gas covering fraction ($C_f$; \citealt{Prochaska2011}).  We use a `two-component' model, described in detail below, to mitigate these effects; however, radiative transfer analyses and higher-resolution spectra are required to improve our understanding of 
such processes in the context of galactic winds \citep[e.g.,][]{Prochaska2011}.

Here we note that the presence of \ion{Mn}{2} $\lambda\lambda 2576.877,2594.499,2606.462$ absorption in very close proximity to the \ion{Fe}{2} transitions of interest may also affect our analysis.  These \ion{Mn}{2} transitions are interstellar in origin, and likely arise in gas of a similar temperature and density to that traced by \ion{Fe}{2}.  However, as Mn is a factor of $\sim100$ less abundant than Fe \citep{Asplund2009}, these lines are rare, and appear only in our highest-S/N spectra.  Because absorption in the \ion{Mn}{2} 2594 transition may shift the fitted centroid of the \ion{Fe}{2} 2600 transition blueward, we mask pixels which exhibit absorption at the wavelengths of these three \ion{Mn}{2} transitions and replace their values with the continuum level.  Figure~\ref{fig.allspecs} shows the locations of these masks in red.  

The first model, or `one-component' model, parameterizes 
 the normalized flux as a function of wavelength as implemented by \citet{Rupke2005a} and \citet{Sato2009}.  This model 
describes the line intensity as $I(\lambda) = 1 - C_f (\lambda) + C_f (\lambda) e^{-\tau(\lambda)}$, where $C_f (\lambda)$ is 
the gas covering fraction as a function of $\lambda$.  We assume the optical depth can be written as a Gaussian, 
$\tau (\lambda) = \tau_0 e^{-(\lambda - \lambda_0)^2 / (\lambda_0 b_D / c)^2}$, where $\tau_0$ and $\lambda_0$ are the central optical depth and 
central wavelength of the line, and $b_D$ is the Doppler parameter or velocity width of the absorption.  
Although $C_f (\lambda)$ may certainly vary as 
a function of wavelength, as the observed absorption likely arises from multiple gas clouds with different projected sizes and velocities \citep[see, e.g.,][]{MartinBouche2009},  we do not attempt to constrain this variation at our relatively low spectral resolution, 
and assume instead that $C_f$ is constant.  

When more than one transition of the same species is available (with known oscillator strengths, $f_0$), simultaneous fitting of these transitions may break the degeneracy between the $C_f$ and $\tau_0$ parameters, as the relative depths of the lines are independent of $C_f$.  We therefore model the two lines in each of the \ion{Mg}{2} $\lambda \lambda 2796, 2803$ and \ion{Fe}{2} $\lambda \lambda 2586, 2600$ transitions simultaneously, writing the line profile intensity $I(\lambda) = 1 - C_f(\lambda) + C_f(\lambda) e^{-\tau_\mathrm{blue}(\lambda) - \tau_\mathrm{red}(\lambda)}$.   Here, $\tau_\mathrm{blue}$ and $\tau_\mathrm{red}$ are the optical depths of the two lines for a given species.  In the case of \ion{Mg}{2}, $\tau_\mathrm{0, blue} / 2 = \tau_\mathrm{0, red}$, while for the \ion{Fe}{2} lines 
$\tau_\mathrm{0, blue}  = 0.2877 \times \tau_\mathrm{0, red}$ \citep{Morton2003}.  Because we may write the column density, $N$, as a function of $\tau_0$ and $b_D$, 
\begin{eqnarray}
N \mathrm{[cm^{-2}]}= \frac{\tau_0 b_D [\mkms]}{1.497 \times 10^{-15} \lambda_0 [\mathrm{\AA}] f_0}, 
\end{eqnarray}
we describe our normalized line profile intensity as a function of four parameters: $\lambda_{0,1}$, $b_{D,1}$, $C_{f,1}$, and $N_1$, where the subscript `1' denotes one-component model parameters.  

Our second model assumes that the same line profiles arise from two velocity `components', rather than one (as assumed above).  This is motivated by the multiple means by which a galaxy may produce \ion{Mg}{2} and \ion{Fe}{2} absorption.  Not only do these resonance transitions trace the kinematics of cool, photoionized gas flows, but they also likely trace absorption in the ISM of the galaxies, the large-scale kinematics of which are governed by the dynamics of the galaxies' stellar and dark components \citep[e.g.,][]{Weiner2009,RubinTKRS2009}.  Damped Ly$\alpha$ absorption systems, for instance, which may trace gas embedded in galactic disks \citep[e.g.,][]{Krogager2012}, are known to simultaneously give rise to strong \ion{Mg}{2} absorption \citep{Ellison2006,Zwaan2008}.  Furthermore, stellar atmospheres may also absorb in the \ion{Mg}{2} and \ion{Fe}{2} transitions studied here \citep{RubinTKRS2009,Coil2011}, and the resulting line profiles will likewise exhibit the velocity centroid and spread of the galaxies' stellar component.  

To model this stellar and interstellar absorption, we introduce a second model absorption component with a central wavelength ($\lambda_0$) fixed at the galaxy systemic velocity and having $\log N_\mathrm{MgII} \ge 14.4$ or $\log N_\mathrm{FeII} \ge 15.0$, with a constant $C_f = 1$.  By adopting the latter two constraints, we assume that the stellar continuum emission is fully covered by a uniform screen of saturated ISM absorption.  Studies of the ISM in nearby star-forming galaxies \citep[e.g., the THINGS survey;][]{Leroy2008,Bagetakos2011} show that interstellar \ion{H}{1} gas and the galaxies' stellar components have comparable scale heights and are typically cospatial, although the \ion{H}{1} often extends to much larger radii.  The observed \ion{H}{1} distribution is by no means uniform; i.e., the line-of-sight column densities vary across the galactic disks.  However, because our data cannot constrain variations in $C_f$ as a function of either spatial location or velocity, and because the gas likely surrounds the stellar emission, we consider our simplifying assumption of full gas coverage well-justified.   

In addition to this `fixed-velocity' absorption component ($I_\mathrm{sys} (\lambda)$), we include a `flow' component ($I_\mathrm{flow} (\lambda)$) with a floating $\lambda_0$, $b_D$, $C_f$, and $N$, as in the `one-component' model described above.  The normalized flux of the line profile is then 
given by $I(\lambda) = I_\mathrm{sys} (\lambda) I_\mathrm{flow} (\lambda)$, which assumes that the two components overlap spatially at a given wavelength.  
This two-component model therefore describes the data with six free parameters: $N_\mathrm{sys}$, $b_{D, \rm sys}$, $\lambda_\mathrm{0, flow}$, 
$b_{D, \rm flow}$, $C_{f, \rm flow}$, and $N_\mathrm{flow}$.

Finally, before comparing either the one- or two-component models to the data, we convolve the model line profiles with a Gaussian having a FWHM equal to 
the velocity resolution of each spectrum near either \ion{Fe}{2} or \ion{Mg}{2}.  Because the instrumental resolution is comparable to the expected flow velocities and widths (with FWHMs ranging between $\sim 150$ and $445\mkms$), this convolution has a significant qualitative effect on the shape of our model line profiles, and we consider this step crucial to estimating robust model parameter constraints.

\subsubsection{Bayesian Parameter Constraints}\label{sec.bayes_constraints}

We assume that the logarithm of the likelihood function is given by the distribution of $\chi^2 / 2$ for each model.
We developed a code which samples the posterior probability density function (PPDF) for each model using the Multiple-Try Metropolis Markov Chain Monte Carlo technique \citep{Liu2000}.  The code is written in ROOT/RooFit, an object-oriented framework written in C++ \citep{BrunRademakers1997}, and calls the Metropolis sampler contained in the RooStats package, a publicly-available set of statistical tools built on top of RooFit.  

We adopt uniform probability densities over the allowed parameter intervals as priors, adjusting these intervals slightly for each model and transition.  In fitting the one-component models for both \ion{Mg}{2} and \ion{Fe}{2}, we implement the following parameter ranges:
$20 \mkms \le b_{D,1} \le 450 \mkms$, $0 \le C_{f,1} \le 1$, and $9 \le \log N_1 \le 22$.  The latter constraints on $N_1$ are quite liberal, and allow for both optically thin and fully saturated absorption.    
For the \ion{Mg}{2} one-component absorption models, we allow the central wavelength of the 2803 \AA\ doublet line ($\lambda_{0,1}$) to vary within $\pm 700\mkms$ of its rest wavelength (2803.53 \AA), such that $2796.99~\mathrm{\AA} \le \lambda_{0,1} \le 2810.08~\mathrm{\AA}$.  For the \ion{Fe}{2} one-component models, we reduce this velocity range slightly to exclude regions which may be contaminated by \ion{Mn}{2} absorption (i.e., $2595.0~ \mathrm{\AA} \le \lambda_{0,1} \le 2604.0~ \mathrm{\AA}$,  where  $\lambda_{0,1}$ now refers to the central wavelength of the 2600.17 \AA\ transition).

For our two-component models, we retain the same limits on $\lambda_{0,\rm flow}$, $b_{D,\rm flow}$, and $C_{f, \rm flow}$ as listed above.  
However, because the likelihood space is much larger for these models, we adopt more stringent constraints on $N_\mathrm{flow}$ to prevent likelihood sampling for $N_\mathrm{flow}$ values which are unphysically large (i.e., which demand flow-component hydrogen column densities of $N_\mathrm{H, flow} > 10^{23}~\rm cm^{-2}$).  Adopting conservative dust depletion factors consistent with those measured in the local Galactic ISM \citep[+0.5 dex for Mg and +1.0 dex for Fe;][]{Jenkins2009}, and assuming solar abundance ratios and no ionization correction, this limit implies $N_\mathrm{flow}$(\ion{Fe}{2})$< 10^{17.5}~\rm cm^{-2}$ and $N_\mathrm{flow}$(\ion{Mg}{2})$< 10^{18.0}~\rm cm^{-2}$.  Our limits on flow-component column densities are therefore $9 \le \log N_\mathrm{flow}$(\ion{Fe}{2})$\le 17.5$ and 
$9 \le \log N_\mathrm{flow}$(\ion{Mg}{2})$\le 18.0$.

For the systemic components of our two-component models, we adopt the same limits on the Doppler parameter as described above ($20 \mkms \le b_{D,\rm sys} \le 450 \mkms$).  We further impose the same upper limits on $N_\mathrm{sys}$ as those adopted for $N_\mathrm{flow}$, again to avoid sampling unphysical regions of likelihood space.  Finally, we force these components to be optically thick, as they are meant to account for both galactic disk ISM and absorption from stellar atmospheres.  The optically thick limit implies $15.0 \le \log N_\mathrm{sys}$(\ion{Fe}{2}) $ \le 17.5$ and $14.4 \le \log N_\mathrm{sys}$(\ion{Mg}{2}) $\le 18.0$.

\begin{figure}
\includegraphics[angle=0,width=\columnwidth]{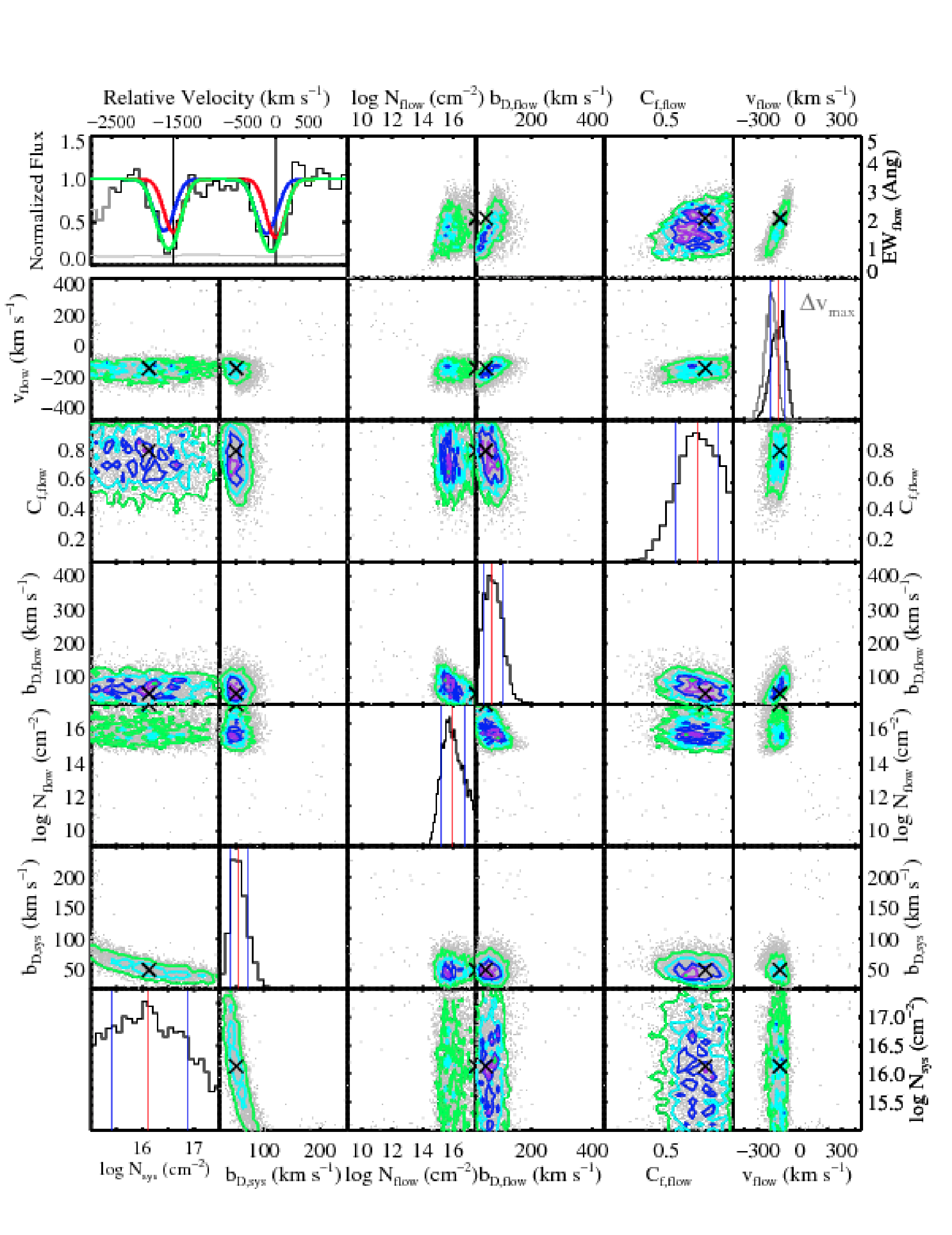}
\caption[]{Example of fitting code output.  The \ion{Fe}{2} absorption profile for EGS13003705 is shown in the upper left panel in black, 
with the error in each pixel plotted in gray at a normalized flux level $\sim0.1$.  A portion of the spectrum that is likely contaminated by 
\ion{Mn}{2} 2576 absorption has been masked prior to fitting (marked in gray).  The vertical black lines mark the systemic velocity of each transition.  The red line shows the maximum-likelihood $I_\mathrm{sys}(\lambda)$, the blue line shows 
the maximum-likelihood $I_\mathrm{flow}(\lambda)$, and the green line shows the combined maximum-likelihood intensity profile, $I(\lambda)$.  Note that we do not make use of the maximum-likelihood parameter values in our analysis, and instead adopt the median of the marginalized PPDF as the `best' value of each parameter; here the maximum-likelihood fit is shown solely for illustrative purposes.  The remaining panels show marginalized PPDFs.  In panels for which the x- and y-axes are the same, one-dimensional PPDFs are shown.  The median and $\pm34$th-percentile values of each parameter are indicated with red and blue vertical lines.   Two-dimensional PPDFs are shown in the off-diagonal panels.  
The maximum-likelihood value of each parameter is shown with a black cross for reference.       
In the panel showing the marginalized PPDF for $v_\mathrm{flow}$ (black), we also show the PPDF of $\Delta v_\mathrm{max}$ in gray.  
   \label{fig.fitdemo}}
\end{figure}

Our code produces marginalized PPDFs for each of the parameters listed above.  It also generates the marginalized PPDF as a function of the EW of each absorption component by computing $\mathrm{EW_X} = \int (1 - I_\mathrm{blue}(\lambda)) d\lambda$, where $I_\mathrm{blue}(\lambda) = 1 - C_f + C_f e^{\tau_\mathrm{blue}(\lambda)}$ is the line intensity profile for either the one-component model or the flow or systemic components of the two-component model (and $\rm EW_X$ is either $\rm EW_1$, $\rm EW_{flow}$, or $\rm EW_{sys}$).
Finally, it generates a marginalized PPDF for the quantity $\lambda_\mathrm{0, max} = \lambda_\mathrm{0,flow} \times (1 - \frac{b_{D,\rm flow}}{\sqrt{2} c})$, which we use to assess the maximum velocity extent of the `flow' absorption component blueward of the systemic velocity (see Figure~\ref{fig.fitdemo}).  

We visually inspect the PPDFs for each fit to evaluate the success of this procedure (e.g., Figure~\ref{fig.fitdemo}).  We require that each PPDF be populated with at least 500
points (in general the shortest acceptable Markov chain has 500 steps, with only two exceptions having $> 350$ steps) in order to permit robust estimates of the $\pm34$th- and $\pm47.5$th-percentile parameter values.  In cases for which the algorithm did not sufficiently sample the PPDF, we inflated the error on each pixel slightly (by $\lesssim20\%$) and repeated the procedure.  For a handful of line profiles, the PPDFs were dominated by combinations of parameters which 
tend to correct for a slightly overestimated continuum level -- i.e., very large $b_{D,\rm flow}$ values in combination with small $C_{f, \rm flow}$ or $N_\mathrm{flow}$.  To prevent this part of parameter space from being sampled, we tightened our priors on either $b_{D, \rm flow}$ (typically adopting the range $20 \mkms \le b_{D, \rm flow} \le 200 \mkms$), $N_\mathrm{flow}$ (such that $\log N_\mathrm{flow} > 13.5$ or $14~\rm [cm^{-2}]$), or 
$C_{f,\rm flow}$ (such that $C_{f, \rm flow} > 0.35$)
in $\sim 5$ cases for each transition and model. 

The code output for the two-component model fit of the \ion{Fe}{2} line profile for EGS13003705 is shown in Figure~\ref{fig.fitdemo}.  The upper left-most panel shows the observed \ion{Fe}{2} absorption lines (black) and the maximum-likelihood model (green).  The blue and red lines show the `flow' and `systemic' components of this maximum-likelihood model.  The remaining panels show one- and two-dimensional marginalized PPDFs for all model parameters.  Note that we do not make use of the maximum-likelihood model in any of our analysis, as we prefer to adopt the median of the marginalized PPDF as the `best' value of each parameter (shown with red vertical lines).  

This Figure illustrates several important aspects of our fitting procedure.  First, our assumption that the absorption at systemic velocity ($I_\mathrm{sys}$) is saturated and fully covers the continuum emission tends to maximize the portion of the absorption profile that is attributed to this component.  In other words, if we were to allow $C_{f,\rm sys}$ to vary, allowing the absorption depth of $I_\mathrm{sys}$ to decrease, this would both
shift the PPDF for $\rm EW_{flow}$ to higher values and increase the range of the PPDF for $v_\mathrm{flow} = c(\lambda_\mathrm{0,flow} - \lambda^{2600}_\mathrm{rest})/\lambda^{2600}_\mathrm{rest}$ to include lower wind velocities (where $\lambda^{2600}_\mathrm{rest}$ refers to the rest wavelength of the \ion{Fe}{2} 2600 transition).  Our adopted fitting procedure therefore tends to yield the largest values of $v_\mathrm{flow}$ that are consistent with the data.  Furthermore, these assumptions force the $I_\mathrm{sys}$ profile shape to be tightly constrained by the data redward of systemic velocity.  
While changes in the value of $N_\mathrm{sys}$ do not change the shape of $I_\mathrm{sys}$ significantly (all allowed values of $N_\mathrm{sys}$ yield saturated lines), panels showing the distribution of $b_{D, \rm sys}$ values in  
Figure~\ref{fig.fitdemo} indicate that this parameter is constrained to lie within a quite narrow range ($41\mkms < b_{D, \rm sys} < 71 \mkms$) and that it is not highly covariant with other model parameters, including $v_\mathrm{flow}$.

Second, we note that the PPDF for the $C_{f, \rm flow}$ parameter is broad, with $\pm34$th-percentile values in the range $0.57 < C_{f,\rm flow} < 0.90$.  Further,  our constraint on $N_\mathrm{flow}$ must be interpreted as a lower limit, as its marginalized PPDF indicates high likelihoods at the upper limit of the allowed parameter range.  
$\rm EW_{flow}$ has a weak dependence on $N_\mathrm{flow}$ and a stronger dependence on $C_{f,\rm flow}$ and $b_{D,\rm flow}$.  These latter parameters exhibit a weak covariance (see the $C_{f,\rm flow}$ vs.\ $b_{D,\rm flow}$ panel) which, along with the large allowed ranges in $N_\mathrm{flow}$ and $C_{f,\rm flow}$, leads us to instead rely primarily on $\rm EW_{flow}$ (top row) for quantitative comparison of the flow absorption strength among different line profiles.

Third, the panel showing the one-dimensional PPDF for $v_\mathrm{flow}$ (black) also shows the PPDF for the quantity 
$\Delta v_\mathrm{max} = v_\mathrm{flow} - b_{D, \rm flow} / \sqrt 2 = c(\lambda_\mathrm{0,max} - \lambda^{2600}_\mathrm{rest})/\lambda^{2600}_\mathrm{rest}$. 
We have introduced this quantity in order to characterize the `maximum' velocity achieved by the wind, as well as to 
ameliorate the effect of the apparent covariance between $v_\mathrm{flow}$ and $b_{D, \rm flow}$ evident in some of our fitting results (that is, a lower wind velocity is permitted as the flow component broadens).  The PPDF of $\Delta v_\mathrm{max}$ in this case is correspondingly more narrow than that of $v_\mathrm{flow}$ (by $\sim20 \mkms$).  In general, the $\pm34$th-percentile uncertainty interval determined from the PPDF of $\Delta v_\mathrm{max}$ is 
within $25\mkms$ of the width of the corresponding uncertainty interval determined for $v_\mathrm{flow}$ in 80\% of our two-component fits to the \ion{Mg}{2} line profile, and is narrower in most of the remaining cases.
We therefore prefer $\Delta v_\mathrm{max}$ as a robust indicator of the maximum velocity extent of the flow component, and invoke it in much of our analysis below.

Finally, we note that several of the \ion{Mg}{2} line profiles in our sample have negligible absorption 
at and redward of $v = 0\mkms$ (e.g., EGS12012586, EGS12012905, EGS12027896), presumably due to the resonant scattering of \ion{Mg}{2} photons discussed above.  In these cases, our two-component model (which always includes absorption at $v = 0\mkms$) fits  our data poorly, 
and we adopt our one-component fitting results to quantify outflow properties (and assume $\rm EW_{flow} = EW_1$, $v_\mathrm{flow} = v_{0,1}$, and $\Delta v_\mathrm{max} = v_{0,1} - b_{D, 1} / \sqrt 2 $).  
These line profiles may indeed trace interstellar absorption as well as outflows, even if they also exhibit significant resonant emission, and we cannot assume that 
the one-component $\Delta v_\mathrm{max}$ is an `equivalent' quantity to the two-component $\Delta v_\mathrm{max}$ used in instances of significant systemic absorption.  
Moreover, resonant emission is expected to affect \emph{every} \ion{Mg}{2} line profile, including cases in which the profile does not rise above the continuum level, 
likely suppressing the EW of the systemic component and possibly enhancing $\rm EW_{flow}$ as a result (see \citealt{Prochaska2011}, \citealt{Martin2012}, and \citealt{Erb2012} for further discussion).  We do not attempt to disentangle these complex effects here, and simply
assume that resonant emission has sufficiently reduced the effect of ISM absorption such that the one-component model results provide the best constraints on the maximum outflow velocities and absorption strengths for spectra without significant absorption at $v = 0\mkms$.  These spectra are referenced as having `No Systemic Absorption' (NSA) in the relevant Figures and text below.


\subsubsection{Classification of Absorption Kinematics}\label{sec.classes}

Based on our fitting results, we define several distinct classes of absorption kinematics for our sample:

\begin{itemize}
	
	\item{\bf Wind (in \ion{Mg}{2} and/or \ion{Fe}{2}):}  Our primary criterion for the detection of a wind is that $> 95\%$ of the marginalized PPDF for 
		the one-component model must lie at velocities $< 0 \mkms$ ($P_\mathrm{out,1} > 0.95$).  For many of our galaxies, the one-component model fits 
		to both the \ion{Fe}{2} and \ion{Mg}{2} transitions
		yield $P_\mathrm{out,1} > 0.95$; however, a substantial fraction of the sample satisfies this criterion in one transition and not the other.  
		We therefore maintain a distinction between, e.g.,  those galaxies  with detected winds in \ion{Mg}{2} but without detected winds in \ion{Fe}{2}, 
		and those galaxies with winds detected in both transitions.
		The former objects are labeled `winds (Mg)' in Tables~\ref{tab.mglines} and \ref{tab.felines} and Figure~\ref{fig.allspecs}, and the latter objects 
		are labeled `winds (Mg, Fe)'.  Note that objects may fall into the `winds (Mg)' class \emph{either} because the \ion{Fe}{2} one-component fit does 
		not satisfy our $P_\mathrm{out,1} > 0.95$ criterion, or because we lack spectral coverage of the \ion{Fe}{2} transition.
		  
		Two galaxies in our sample, TKRS5379 and J033231.36-274725.0, do not satisfy the criterion described above, and yet 
		exhibit substantial `flow' components in the \ion{Mg}{2} transition (with $\rm EW_{flow} > 1$ \AA).  This is due to a strong blue `wing' 
		of absorption with a shallow line depth relative to the absorption at systemic velocity in both cases.  As this wing is likely due to outflow, 
		we make an exception to our $P_\mathrm{out,1}$-based criterion for these two objects, and classify them as `wind' galaxies.
	
	\item {\bf Inflow:} Galaxies are classified as hosts of cool gas inflow if $> 95\%$ of the marginalized PPDF for 
		the one-component model lies at velocities $> 0\mkms$ ($P_\mathrm{in,1} > 0.95$)
		for \emph{both} 
		\ion{Fe}{2} and \ion{Mg}{2} wherever coverage of both ions is available.  
		If coverage of one ion is missing but the $P_\mathrm{in,1} > 0.95$ criterion is met by the other line profile, the galaxy is placed in this class.
		These criteria are more stringent than those used to classify `wind' galaxies, as we wish to be conservative in claiming 
		detection of this rare phenomenon.  Three of our galaxies exhibit $P_\mathrm{in,1} > 0.95$ as a result of our fit to the \ion{Mg}{2} profile but 
		have an \ion{Fe}{2} profile which does not satisfy this criterion, while six of our galaxies exhibit $P_\mathrm{in,1} > 0.95$ in our fit to the 
		\ion{Fe}{2} profile but not in our fit to \ion{Mg}{2}.  We place these nine objects in the `systemic' absorption category, described below.
		The six objects which fully satisfy our inflow criteria have been discussed in \citet{Rubin2012}.
	
	\item {\bf Systemic:}  Galaxy spectra which fail to satisfy either our `wind' criteria or our `inflow' criteria, but which have significant metal-line
		absorption near systemic velocity such that $\rm EW_1 > 0.5$ \AA\ for \emph{either} \ion{Fe}{2} or \ion{Mg}{2}, are placed in this class.
		The marginalized PPDFs for the one-component model fits to these spectra have $\pm34$th-percentile velocity widths of $< 125\mkms$.
	
	\item {\bf No Absorption:} Two of our spectra have high S/N ($> 10~\rm pix^{-1}$), and yet do not exhibit either \ion{Fe}{2} or \ion{Mg}{2} absorption 
		(EGS12012471 and TKRS6709).  The data place a $3\sigma$ upper limit on the strength of undetected absorption at $\rm EW_{2796} < 0.2$ \AA.
		These 
		spectra instead have broad ($> 3000\mkms$) \ion{Mg}{2} emission, indicative of the presence of broad-line AGN.  We exclude these two objects from the remainder of our analysis.
	
	\item {\bf Low S/N:}  Because our target selection included objects with red continua, several of our spectra do not have sufficient S/N and/or sufficiently strong absorption to constrain our 
		models.  We find that in these cases, the PPDFs for the one-component model central velocities 
		are either quite broad $> 200\mkms$ (for 35 objects), or are dominated by large continuum 	
		fluctuations (for 4 additional objects).  Figure~\ref{fig.zstonhist} shows the redshift and S/N distribution of these spectra.  We note that the few spectra which fall into this category 
		despite having $\rm S/N > 7~pix^{-1}$ have very weak absorption lines (with $\rm EW_{2796} < 1$ \AA).
		We exclude all of these objects from the remainder of our analysis. 	

\end{itemize}

\subsubsection{Consistency of Fitting Results}

We observed six of our targets in two separate mask pointings and on different nights, obtaining two independent spectra for each.  The data are of sufficient quality to constrain absorption kinematics in both spectra for four of these objects (J033225.26-274524.0, J033231.36-274725.0, J033234.18-274554.1, and J033237.96-274652.0).  We find that our kinematic classifications and constraints on most model parameters are consistent among each of these pairs, though 
the constraints on a few model parameters from different spectra of the same object can be significantly different.   For example, the two spectra of J033221.36-274725.0 yield $C_{f, \rm flow}$ values for \ion{Mg}{2} of $0.32^{+0.07}_{-0.05}$ and $0.19\pm0.03$.  We note also that the $P_\mathrm{1, out}$ values (0.989 and 0.937) differ significantly and straddle our primary criterion for the `wind' class (although we made an exception to that criterion in this case; see \S\ref{sec.classes}).  Even more striking are the differences in our two spectra of J033237.96-274652.0, which exhibit \ion{Mg}{2} emission lines that differ in EW by $\sim0.5 -1$ \AA.  This may be due to a change in the portion of the wind covered by slits at two distinct positions and orientations \citep{Prochaska2011}.  In this case, 
the $\rm EW_{flow}$ values measured for both \ion{Mg}{2} and \ion{Fe}{2} are significantly offset (i.e., $<0.46$ \AA\ vs. $0.86\pm0.17$ \AA\ for \ion{Mg}{2}, and $<0.06$ \AA\ vs. $0.47^{+0.24}_{-0.21}$ \AA\ for \ion{Fe}{2}).  These measurements illustrate the potential of \ion{Mg}{2} emission observations for constraining the spatial location of the outflowing gas around distant galaxies \citep[e.g.,][]{Rubin2011,Martin2013}.  We discuss the \ion{Mg}{2} and \ion{Fe}{2} emission properties of our sample in Rubin et al.\ 2013 (in preparation).  



\section{Tracing Cool Gas Flows with \ion{Mg}{2} and \ion{Fe}{2} Line Profiles in Galaxy Spectroscopy}\label{sec.fitresults}

Based on the criteria described in \S\ref{sec.classes}, of a total sample of 105 unique objects (with 109 high-quality spectra), we detect winds in either \ion{Mg}{2} or \ion{Fe}{2} in $66 \pm 5\%$ of the galaxies, inflow in $6\pm2\%$ of the galaxies, and absorption only at systemic velocity (no winds/inflow) in $27\pm4\%$ of the galaxies.  Among the 61 galaxies with winds whose spectra cover both \ion{Mg}{2} and \ion{Fe}{2}, we detect winds in both transitions in $61\%$ of these objects, while winds are detected in only \ion{Mg}{2} in 22 ($36\%$) galaxies, and in only \ion{Fe}{2} in 2 ($3\%$) galaxies.  In this section, 
we explore our sensitivity to detecting winds and compare wind velocities and absorption strengths obtained from different model fitting methods and transitions.  


\subsection{Sensitivity}\label{sec.sensitivity}

Our ability to constrain absorption-line kinematics with our spectra is strongly dependent on the spectral S/N and resolution.  That is, spectra which fall into the `wind', `inflow', or `systemic' categories must generally yield marginalized PPDFs which are narrow in velocity space, and which are not dominated by large continuum uncertainties.  In addition, our ability to differentiate between galaxies with and without winds or inflow depends on the total amount of absorption in the metal-lines: for spectra exhibiting stronger total metal-line absorption, there is a higher probability that  a portion of that EW will be attributed to outflow.

\begin{figure}
\includegraphics[angle=90,width=\columnwidth]{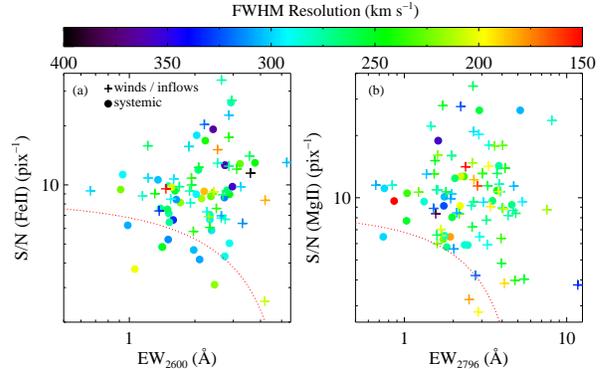}
\caption[]{\emph{(a)} S/N in the continuum surrounding the \ion{Fe}{2} transition vs.\ $\rm EW_{2600}$.  Objects having winds or inflows traced by \ion{Fe}{2} are shown with crosses, and objects without detected winds/inflows are shown with filled circles.  Points are color-coded according to spectral resolution, with the darkest points indicating the lowest resolution.  Our sensitivity to detecting winds/inflows in \ion{Fe}{2} depends on both spectral S/N and the overall strength of the absorption lines.  The red dotted line indicates our approximate detection limit, and was estimated by eye.   \emph{(b)}  S/N in the continuum surrounding the \ion{Mg}{2} transition vs.\ $\rm EW_{2796}$.  
The parameter space occupied by filled circles and crosses overlaps nearly completely in this panel, suggesting that at the S/N level of our spectral sample, our sensitivity to \ion{Mg}{2} flows is not strongly dependent on $\rm EW_{2796}$.
     \label{fig.stonew}}
\end{figure}

The left-hand panel in Figure~\ref{fig.stonew} compares the spectral S/N near the \ion{Fe}{2} transition with the total $\rm EW_{2600}$ for spectra exhibiting either winds or inflows in \ion{Fe}{2} (crosses) and spectra with \ion{Fe}{2} absorption classified as `systemic' (filled circles).  Points are color-coded by the FWHM velocity resolution near the \ion{Fe}{2} transitions.  The parameter space occupied by spectra in the `systemic' group extends to lower S/N and $\rm EW_{2600}$, suggesting that we are unable to detect winds/inflows in the S/N-$\rm EW_{2600}$ regime marked by the red dotted line, and additionally that our sensitivity to winds/inflows decreases as a function of both of these factors.  
On the other hand, the corresponding distributions for the \ion{Mg}{2} transition (panel \emph{(b)}) do not exhibit such an offset.  Indeed, there are very few spectra which occupy the space below the red dotted line (repeated from panel \emph{(a)}).  We conclude that analysis of \ion{Mg}{2} at our S/N permits a relatively complete census of winds for this sample, although we also note that the probability of detecting \ion{Mg}{2} winds/inflows increases marginally with S/N.  Specifically, in the range $\rm 5~pix^{-1} < S/N < 10~pix^{-1}$, our wind detection rate is $64\pm7$\%, which is lower than but consistent within the $1\sigma$ uncertainties with our wind detection rate in the range $\rm 10~pix^{-1} < S/N < 15~pix^{-1}$ ($76\pm7$\%).  
Because spectral S/N could be covariant with galaxy properties such as SFR and inclination ($i$), which in turn may be physically linked to the presence of outflows, we must exercise care in interpretating trends in outflow detection rates with SFR, $i$, etc.  This issue will be discussed further in \S\ref{sec.detrate}.  

Turning to the issue of spectral resolution, we note that in panel \emph{(a)}, several of the spectra in the `systemic' class have relatively low resolution (and are marked with dark blue/purple points).   Such low resolution could in principle tend to decrease our sensitivity to winds, and indeed we find that our \ion{Fe}{2} wind detection rate decreases from $53\pm6$\% at resolutions $200\mkms < \mathrm{FWHM} < 300\mkms$ to $35\pm11$\% at resolutions $\mathrm{FWHM} > 300\mkms$.  However, our \ion{Mg}{2} wind detection rate (panel \emph{(b)}) does not depend on resolution (i.e., we detect winds in $73\pm5$\% of spectra having $200\mkms < \mathrm{FWHM} < 300\mkms$ and in $75\pm11$\% of spectra having $\mathrm{FWHM} > 300\mkms$).  We conclude that variable spectral resolution does not significantly affect our ability to detect winds or inflows; however, it may nevertheless affect our constraints on $\Delta v_\mathrm{max}$ and $\rm EW_{flow}$.

\begin{figure}
\includegraphics[angle=90,width=\columnwidth]{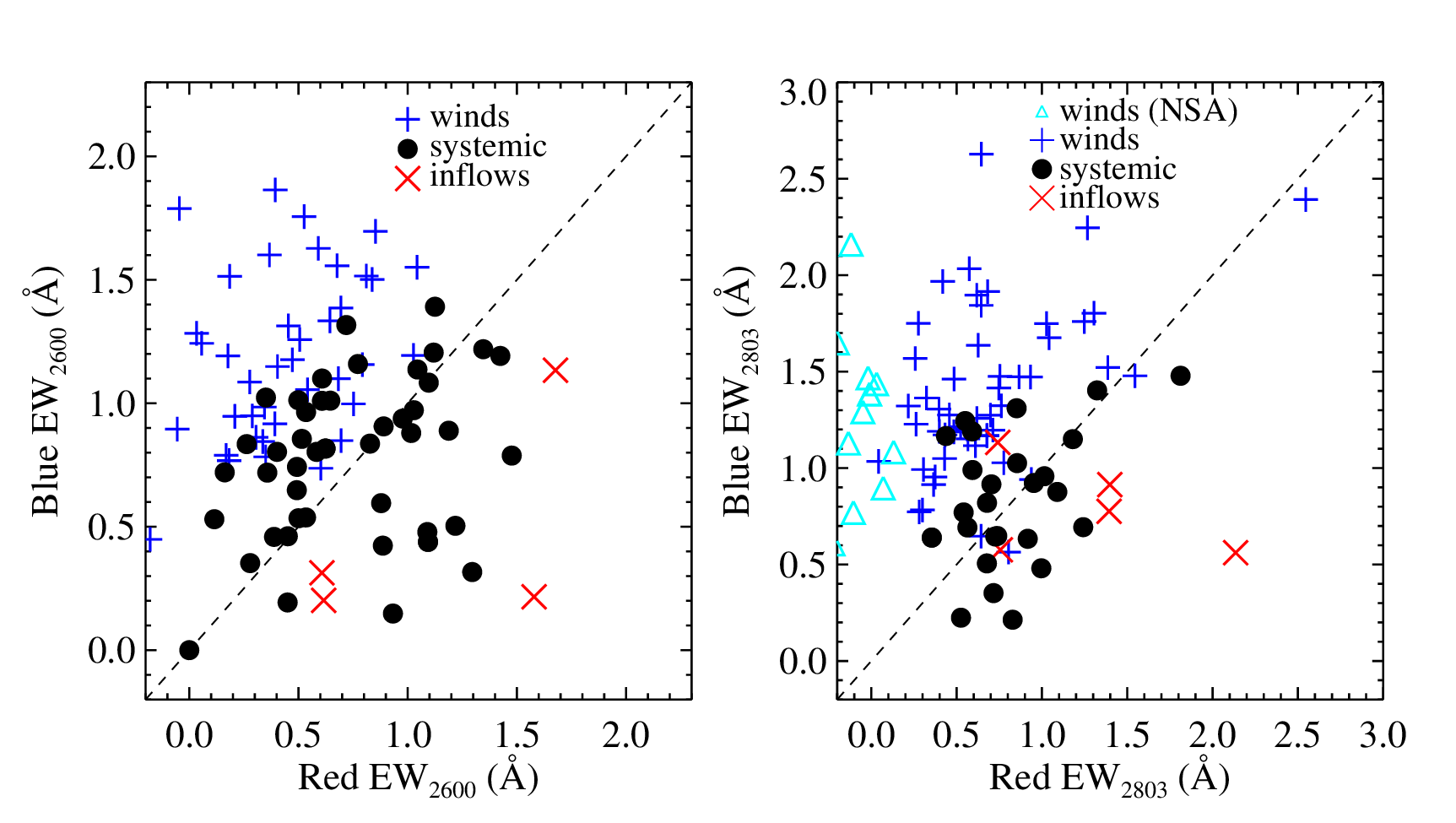}
\caption[]{\emph{Left:}  $\rm EW_{2600}$ at $v < 0\mkms$ (`blue') vs.\ $\rm EW_{2600}$ at $v > 0\mkms$ (`red') for galaxies with winds detected in \ion{Fe}{2} (blue crosses), no winds/inflow (black filled circles), and inflows (red crosses).  The dashed line shows a 1:1 relation.  Several galaxies without detected winds have large blue $\rm EW_{2600}$ ($> 0.8$ \AA), comparable to blue $\rm EW_{2600}$  values measured for galaxies with detected winds.   \emph{Right:}  Same as left-hand panel, but for the \ion{Mg}{2} 2803 transition.  Spectra which exhibit negligible absorption at $v > 0 \mkms$ are shown with cyan triangles (indicated in the legend as `NSA' systems, as they have No Systemic Absorption).  
     \label{fig.redblue}}
\end{figure}

Our sensitivity to winds/inflows is also necessarily dependent on the relative absorption strength of gas at all velocities within each galaxy halo along the line of sight.  For instance, strong and broad absorption redward of systemic velocity is almost always attributed to a `systemic' absorption component in our analysis, and will decrease the EW that is attributed to outflow (as compared with a spectrum that has much weaker `systemic' absorption).  Furthermore, the presence of strong absorption associated with \emph{inflowing} gas may mask the presence of outflow, as our wind detection criterion leverages the central velocity of the full line profile.  The left panel of Figure~\ref{fig.redblue} shows the distribution of 2600 \AA\ absorption line EWs measured at $v < 0\mkms$ (`blue') and $v> 0\mkms$ (`red') for objects with winds, inflows, and `systemic' absorption.  Points showing galaxies with detected winds are, by construction, located in the upper left part of the plot, as their absorption profiles are dominated by gas at velocities blueward of systemic.  However, nearly half of the objects without winds detected in \ion{Fe}{2} (26 of 47) have blue $\rm EW_{2600}$ values $> 0.8$ \AA, or as large as the minimum blue $\rm EW_{2600}$ values exhibited by galaxies with winds.  A similar fraction of objects without winds detected in \ion{Mg}{2} (14 of 27) have large blue $\rm EW_{2803}$ values consistent with those measured for wind galaxies (see right-hand panel).  
Therefore, winds of similar strength to those detected in bonafide `wind' galaxies could also exist in these systems but be masked by systemic absorption.
We address this point again in \S\ref{sec.detrate}, identifying the galaxies with large blue $\rm EW_{2600}$ or $\rm EW_{2803}$ values and exploring whether their inclusion in our `winds' subsample affects our conclusions.  

All of these issues are further explored in Appendix~\ref{sec.synthetic}, in which we generate synthetic spectra of the \ion{Mg}{2} profile at different velocity resolutions tracing winds with varying maximum velocities and interstellar absorption strengths.  We then apply our one- and two-component model fitting procedures.  In brief, 
we find that larger ISM velocity widths ($b_{D, \rm ISM}$) 
may reduce the probability that a line profile will satisfy our criterion for the `wind' class, and will
weaken our constraints on $\Delta v_\mathrm{max}$ and $\rm EW_{flow}$.  
For instance, nearly all synthetic spectra generated with input wind velocities of $250\mkms$ or greater and having $\rm S/N = 9~pix^{-1}$ meet our `wind' criterion 
if $b_{D, \rm ISM} = 20-60\mkms$.  However, this criterion is met with decreasing frequency if $b_{D, \rm ISM} = 100-120\mkms$, particularly at lower input wind velocities.
Synthetic spectra generated with input wind velocities up to $300\mkms$ and $b_{D, \rm ISM}$ between $20\mkms$ and $120\mkms$ all yield $\rm EW_{flow}$ values of $2.28 - 2.88$ \AA, but with error bars of $\pm 0.48$ \AA\ for $b_{D, \rm ISM} = 20\mkms$ and $\pm 1.15$ \AA\ for $b_{D, \rm ISM} = 120\mkms$.  For lower wind velocities ($< 300\mkms$), it is increasingly likely that 
a significant fraction of the PPDF of $\rm EW_{flow}$ will overlap with 0 \AA\ as $b_{D, \rm ISM}$ increases, limiting our ability to constrain the strength of the wind in such galaxies.  Furthermore, when a significant outflow component is recovered, the resulting values of $\Delta v_\mathrm{max}$ at each input wind velocity are consistent within their error bars over the full range of $b_{D, \rm ISM}$ values tested.  However, the uncertainties on $\Delta v_\mathrm{max}$ may be a factor of 5-10 larger for $b_{D, \rm ISM} = 120\mkms$ than for $b_{D, \rm ISM} = 20\mkms$.  

We find that degrading the spectral resolution does not significantly affect the results of our one-component model fits (and hence our ability to detect winds). However, it does have a similar effect to that described above on our two-component model fits.
We are less likely to recover a significant `flow' component at lower spectral resolutions, particularly with larger input $b_{D, \rm ISM}$ values and lower input wind velocities.  However, when this `flow' component is successfully recovered, the resulting $\rm EW_{flow}$ and $\Delta v_\mathrm{max}$ values are consistent within their  uncertainties for all of the resolutions tested.  The uncertainties on $\Delta v_\mathrm{max}$ are a factor of $\sim2$ larger for spectral resolutions FWHM $ = 360\mkms$ than for FWHM $ = 190\mkms$, while the uncertainties on $\rm EW_{flow}$ tend to be $0.2-0.5$ \AA\ larger for synthetic spectra with FWHM $ = 360\mkms$.  Because spectral resolution is entirely independent of host galaxy properties, this variable sensitivity will simply introduce an additional source of scatter into our wind measurements.  However, this analysis suggests that wind studies at higher spectral resolution will allow increased leverage in assessing variation in outflow EWs and velocities with changing galaxy properties.  



\subsection{Wind Velocity and Absorption Strength Indicators}\label{sec.indicators}

\begin{figure}
\includegraphics[angle=90,width=\columnwidth]{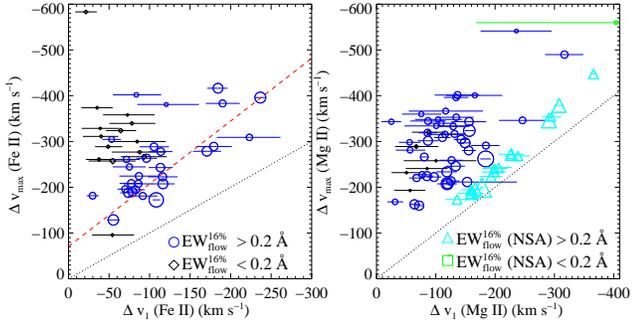}
\caption[]{\emph{Left:}  Comparison of $\Delta v_\mathrm{max}$ (obtained from our two-component model fits) and $\Delta v_1$ measured for the \ion{Fe}{2} transition.  Systems for which $> 84\%$ of 
the PPDF for $\rm EW_{flow}$ ($\rm EW_{flow}^{16\%}$) is $> 0.2$ \AA\ are shown in blue, and the remaining systems are shown in black.  The point sizes scale with the inverse of the uncertainty on $\Delta v_\mathrm{max}$.  The black dotted line marks a 1:1 relation.  $\Delta v_\mathrm{max}$ is correlated with $\Delta v_1$ at $3.0\sigma$ significance, and is 
generally $\sim100\mkms$ blueward.  The red dashed line shows a linear fit to the blue points, with a slope of 1.37 and intercept of $-71\mkms$.   \emph{Right:}  Same as left-hand panel, for \ion{Mg}{2}.  Cyan points indicate velocities measured for spectra with no absorption at $v > 0\mkms$ (indicated as `NSA' systems, with No Systemic Absorption), and thus lie very close to the 1:1 relation (black dotted line).  The green point marks a system with no absorption at $v > 0\mkms$, and which has $\rm EW_{flow}^{16\%} < 0.2~\AA$.  Here, $\Delta v_\mathrm{max}$ exhibits significantly more scatter at a given $\Delta v_1$ than in the left-hand panel; however, $\Delta v_\mathrm{max}$ measurements for both transitions are generally $100-300\mkms$ higher than $\Delta v_1$.
     \label{fig.ncomp1ncomp2}}
\end{figure}

In Figure~\ref{fig.ncomp1ncomp2} we compare measurements of wind velocity obtained from our one-component ($\Delta v_1 = c(\lambda_{0,1} - \lambda_\mathrm{rest})/\lambda_{0,1}$) and two-component ($\Delta v_\mathrm{max}$) model fits.  The left-hand panel shows results for galaxies with winds detected in \ion{Fe}{2}.  First, we note that several of these spectra 
yield small values of $\rm EW_{flow}$; i.e., a significant portion of the $\rm EW_{flow}$ PPDF lies at $< 0.01$ \AA,  
resulting in weak constraints on the velocity of this flow component.  
We therefore differentiate between spectra yielding $\rm EW_{flow}$ values for which $> 84\%$ of the marginalized PPDF is $> 0.2$ \AA\ ($\rm EW_{flow}^{16\%} > 0.2$ \AA; blue), and spectra yielding smaller $\rm EW_{flow}^{16\%}$ values (black; $\sim 1/3$ of the systems shown).  
This division places nearly all fits having $\pm34$th-percentile probability intervals for $\Delta v_\mathrm{max}$ broader than $200\mkms$  into the latter category.  However, 
as there are only 4 systems with $\rm EW_{flow}^{16\%}$ in the interval $\rm 0.01~\AA < EW_{flow}^{16\%} < 0.2~\AA$, our choice to separate fits at $\rm EW_{flow}^{16\%} = 0.2$ \AA\ 
does not significantly affect our results, and is done simply to select systems with statistically-significant flow components at our velocity resolution.
The point size in Figure~\ref{fig.ncomp1ncomp2} is scaled by the inverse of the uncertainty on $\Delta v_\mathrm{max}$.  Calculation of the Spearman's rank correlation coefficient for  the $\Delta v_\mathrm{max}$ and $\Delta v_1$ values of the blue points 
indicates that they 
are correlated with  $3.0\sigma$ significance, with $\Delta v_\mathrm{max}$ generally $\sim 100\mkms$ blueward of $\Delta v_1$, although the velocities may be discrepant by as much as $300\mkms$.  A linear fit to these data (shown with the red dashed line) yields a slope of 1.37 and an intercept of $-71\mkms$.  

The same measurements are shown for the \ion{Mg}{2} transition in the right-hand panel.  Here, 45 out of 51 systems have 
$\rm EW_{flow}^{16\%} > 0.2$ \AA\ and are plotted in blue.  In contrast to \ion{Fe}{2}, these measurements are not significantly correlated, 
yielding a sum-squared difference of ranks only 
$1.9\sigma$ from the null-hypothesis expected value, and with 
the range in $\Delta v_\mathrm{max}$ values at a given $\Delta v_1$ being quite broad.  This may be due to, e.g., a broader range in \ion{Mg}{2} absorption strength near systemic velocity as compared with \ion{Fe}{2}, or simply to a broader range in the maximum wind velocities probed.  Spectra which exhibit little or no absorption at $v > 0 \mkms$ are shown in cyan and green; by definition these points lie close to the 1:1 relation, and at slightly lower $\Delta v_\mathrm{max}$ on average than the blue points at a given $\Delta v_1$.  Because the latter measurements have a distribution which is qualitatively different from those derived from our two-component analysis (and appear to underestimate the maximum wind velocity), we flag them throughout the remaining analysis.  Generally speaking, this Figure suggests that our \ion{Mg}{2} absorption profiles are more variable in morphology than our \ion{Fe}{2} profiles, but that $\Delta v_\mathrm{max}$ measurements for both transitions are $100-300\mkms$ higher than $\Delta v_1$.  
 
 \begin{figure}
\includegraphics[angle=90,width=\columnwidth]{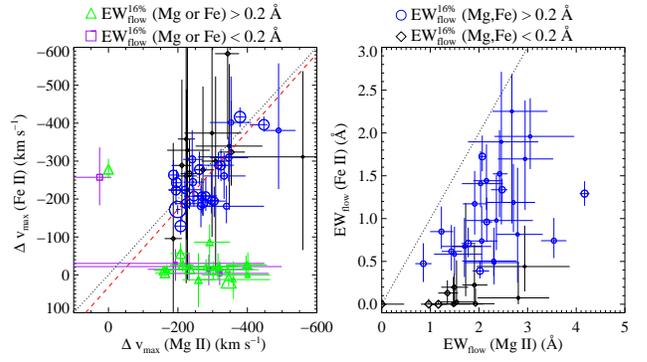}
\caption[]{\emph{Left:}  Comparison of $\Delta v_\mathrm{max}$ measured from \ion{Mg}{2} and \ion{Fe}{2}.  Objects with winds detected in both transitions are indicated with open circles and diamonds.  Blue circles mark cases for which $\rm EW_{flow}^{16\%} > 0.2$ \AA\ for \emph{both} transitions, and black diamonds are used for the remaining `wind' galaxies.  Green triangles indicate $\Delta v_\mathrm{max}$ values for objects with winds detected in only one transition and having $\rm EW_{flow}^{16\%} > 0.2$ \AA, while purple squares show  $\Delta v_\mathrm{max}$ values for objects with winds detected in only one transition and having $\rm EW_{flow}^{16\%} < 0.2$ \AA.    The point size is scaled by the inverse of the combined uncertainties on $\Delta v_\mathrm{max}$ values.  The black dotted line shows a 1:1 relation, and the red dashed line shows a linear fit to the blue points.  \emph{Right:}  Comparison of $\rm EW_{flow}$ values measured in the \ion{Fe}{2} and \ion{Mg}{2} transitions for objects with winds detected in both ions.  Blue circles and black diamonds show objects with $\rm EW_{flow}^{16\%} > 0.2$ \AA\ for both \ion{Mg}{2} and \ion{Fe}{2} and objects with $\rm EW_{flow}^{16\%} < 0.2$ \AA\ in at least one transition, respectively.  The point size is scaled by the inverse of the combined uncertainties on $\rm EW_{flow}$.  The black dotted line shows a 1:1 relation.  
For spectra with significant $\rm EW_{flow}$ components, maximum wind velocities measured from both transitions are tightly correlated, with \ion{Fe}{2} absorption extending to velocities only $\sim30 \mkms$ lower than \ion{Mg}{2} absorption.  However, $\rm EW_{flow}$(\ion{Fe}{2}) values exhibit a large scatter at a given $\rm EW_{flow}$(\ion{Mg}{2}), and are typically $\gtrsim1$ \AA\ lower.
    \label{fig.compareMgFe}}
\end{figure}

\begin{figure*}
\includegraphics[angle=90,width=\textwidth]{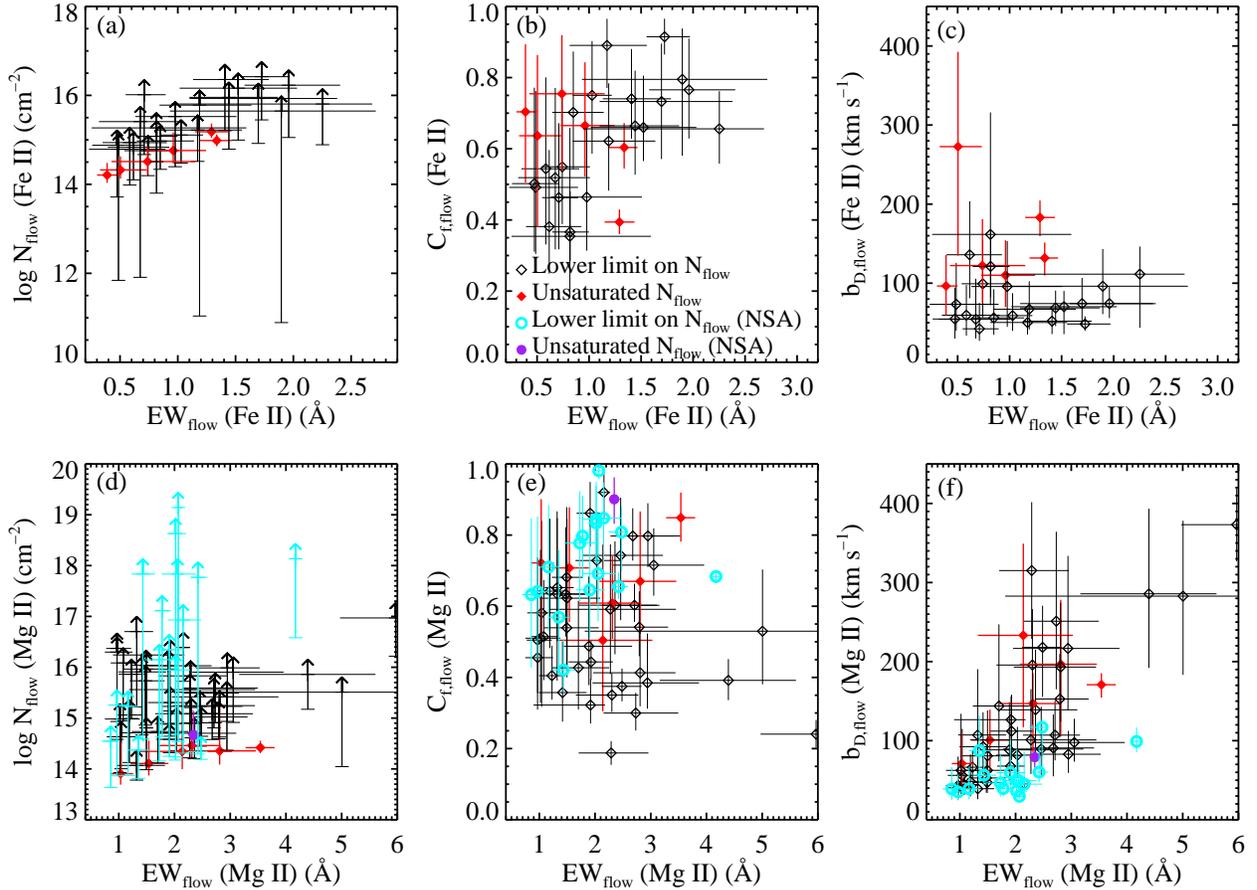}
\caption[]{\emph{Top Row:} Constraints on $\log N_\mathrm{flow}$, $C_{f, \rm flow}$, and $b_{D, \rm flow}$ as a function of $\rm EW_{flow}$ for the \ion{Fe}{2} transition.  Systems for which the line profiles are inconsistent with completely saturated `flow' component absorption are shown in red, and the remaining systems are marked in black.  \emph{Bottom Row:}  Same as the top row, for the \ion{Mg}{2} transition.  Line profiles which exhibit no significant absorption at $v > 0\mkms$ and which are saturated are marked in cyan; unsaturated profiles with no absorption at $v>0\mkms$are marked with purple circles.
     \label{fig.ewNCfbD}}
\end{figure*}

Figure~\ref{fig.compareMgFe} compares wind properties measured from the \ion{Mg}{2} and \ion{Fe}{2} transitions.  We find that for spectra which yield $\rm EW_{flow}^{16\%} > 0.2$ \AA\ (blue open circles), the measured $\Delta v_\mathrm{max}$ for both ions are quite similar.  A linear fit to these points (red dashed line) yields a slope of $1.0 \pm 0.1$ and intercept of $32 \pm 34 \mkms$, indicating that when it traces an outflow, \ion{Fe}{2} absorption extends to velocities only slightly lower ($\sim30\mkms$) than  \ion{Mg}{2}  absorption.  However, as shown in the right-hand panel, 
although $\rm EW_{flow}$ values measured for \ion{Fe}{2} are significantly correlated (at the $2.7\sigma$ level) with $\rm EW_{flow}$ values measured for \ion{Mg}{2}, 
\ion{Fe}{2} $\rm EW_{flow}$ values are significantly lower 
(by $\gtrsim 1$ \AA) than those measured for \ion{Mg}{2}.  This is indicative of either lower \ion{Fe}{2} wind absorption covering fractions, column densities, or velocity widths, or a combination of these factors.    
We note that $\Delta v_\mathrm{max}$ and $\rm EW_{flow}$ measured for \ion{Fe}{2} are not significantly correlated, whereas  $\Delta v_\mathrm{max}$ and $\rm EW_{flow}$ for \ion{Mg}{2} are strongly correlated (at $4.4\sigma$ significance).  This is likely due to the larger range in values of $\rm EW_{flow}$ obtained from analysis of \ion{Mg}{2}: values of $\rm EW_{flow} > 2.5$ \AA\ all occur at $\Delta v_\mathrm{max} < -250\mkms$.  We return to this issue in \S\ref{sec.discussion}.

Finally, we investigate the physical significance of $\rm EW_{flow}$.  Figure~\ref{fig.ewNCfbD} compares $\rm EW_{flow}$ with our model constraints on parameters which together determine $\rm EW_{flow}$: $N_\mathrm{flow}$, $C_{f, \rm flow}$, and $b_{D, \rm flow}$.  We highlight a few line profiles for which the 84th-percentile PPDF value of $N_\mathrm{flow}$ is $< 10^{15.5}~\rm cm^{-2}$;  these profiles have shapes which are inconsistent with higher `flow' component column densities (shown in red and purple).  The remaining profiles yield lower limits on $N_\mathrm{flow}$, and the range of the upper 95\% of the $N_\mathrm{flow}$ PPDF for each object is indicated.   
We find that there is considerable scatter in $N_\mathrm{flow}$ and $C_{f,\rm flow}$ values at a given $\rm EW_{flow}$ measured for the \ion{Mg}{2} transition, with no appreciable trend in either parameter.  The \ion{Mg}{2} $\rm EW_{flow}$  seems instead to depend most strongly on $b_{D, \rm flow}$, with the highest $b_{D, \rm flow}$ values occurring at the largest $\rm EW_{flow}$.  For \ion{Fe}{2}, however, $b_{D, \rm flow}$ exhibits no dependence on $\rm EW_{flow}$ (panel \emph{(c)}), and most of the line profiles are well-fit with $b_{D, \rm flow} \lesssim 100\mkms$.  Instead, $\rm EW_{flow}$ 
increases with larger $C_{f, \rm flow}$ and increasing lower limits on $N_\mathrm{flow}$ (or, in the cases of unsaturated profiles, increasing $N_\mathrm{flow}$ values).  We again attribute the differences between the trends in fitted parameters with $\rm EW_{flow}$ for \ion{Mg}{2} and \ion{Fe}{2} to the wider variety of profile morphology exhibited by \ion{Mg}{2}: the latter extends to higher velocities, typically with weaker absorption at systemic velocity; whereas our fitting technique tends to yield similar \ion{Fe}{2} model parameter constraints from galaxy to galaxy.  Regardless of these differences, however, 
these results suggest that for \ion{Fe}{2}, larger values of $\rm EW_{flow}$ can be attributed to a larger amount of outflowing material and/or a more widespread distribution of this material.  We return to this point in \S\ref{sec.discussion} in our discussion of the mass carried by the detected winds.  

\subsection{Summary}
In \S\ref{sec.sensitivity}, we found that we are approximately equally sensitive to winds traced by \ion{Mg}{2} over the full range of spectral S/N of our sample.  While our sensitivity to outflows traced by \ion{Fe}{2} is reduced for our lowest-S/N spectra, our use of both transitions in detecting winds mitigates this effect, such that we are not significantly biased against wind detection at lower S/N.  In \S\ref{sec.indicators}, we showed that our measurement of the maximum wind velocity ($\Delta v_\mathrm{max}$) is tightly correlated with the velocity centroid of one-component model fits ($\Delta v_1$) for the \ion{Fe}{2} transition, but exhibits substantial scatter for the \ion{Mg}{2} transition, and is generally $100-300\mkms$ higher than $\Delta v_1$ for both transitions.  Our comparison of $\Delta v_\mathrm{max}$ measured for \ion{Mg}{2} and \ion{Fe}{2}  showed that these quantities have nearly a 1:1 relation.  Our measurement of the EW of absorbing gas in outflows ($\rm EW_{flow}$) is typically $\gtrsim1$ \AA\ higher for the \ion{Mg}{2} transition than for \ion{Fe}{2}.  Finally, increasingly large $\rm EW_{flow}$ values measured for \ion{Fe}{2} are associated with larger outflow column densities and more substantial covering fractions for the wind material.

\section{Outflows and their Host Galaxies}\label{sec.bigresults}

In this section, we examine the relationship between wind detection rates, velocities, and absorption strengths and the galaxy properties measured in \S\ref{sec.galprop}.

\subsection{Trends in Wind Detection Rates with Intrinsic Galaxy Properties}\label{sec.detrate}

\begin{figure}
\includegraphics[angle=90,width=\columnwidth]{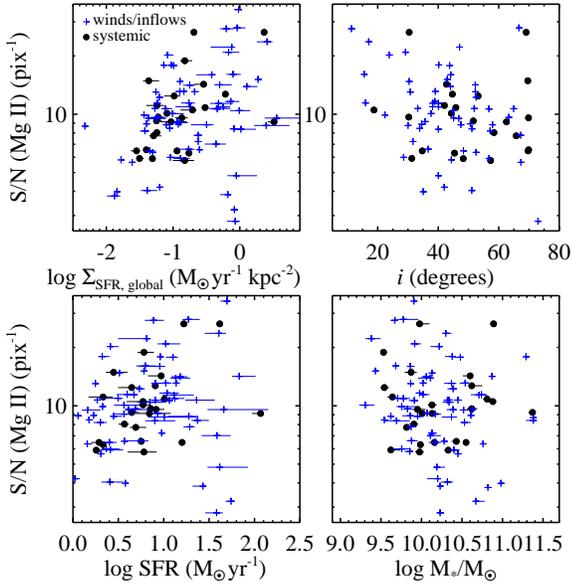}
\caption[]{Comparison of $\rm S/N$ around the \ion{Mg}{2} transition with the galaxy properties explored in \S\ref{sec.bigresults}: 
$\Sigma_\mathrm{SFR,global}$ (top left), inclination (top right), SFR (bottom left), and $\log M_*/M_{\odot}$ (bottom right).  Galaxies with either detected winds or 
inflows are indicated with blue crosses, and galaxies with systemic absorption are marked with black filled circles.  
Spearman's rank correlation coefficients obtained for $\rm S/N$ and each of the properties shown, along with the significance with which the sum-squared 
difference of ranks deviates from its null hypothesis value,
are $r_\mathrm{S} = 0.35$ and $3.4\sigma$ ($\Sigma_\mathrm{SFR,global}$), $r_\mathrm{S} = -0.25$ and $2.2\sigma$ (inclination), $r_\mathrm{S} = 0.30$ and $2.9\sigma$ (SFR), and $r_\mathrm{S} = -0.15$ and $1.5\sigma$ ($\log M_*/M_{\odot}$).  These values and the above panels indicate that spectral S/N is only weakly correlated with the galaxy properties of interest, with the strongest correlation ($r_\mathrm{S} = 0.35$) exhibited between S/N and $\Sigma_\mathrm{SFR,global}$.
     \label{fig.galprop_ston}}
\end{figure}

Before exploring the dependence of the wind detection rate on galaxy $M_*$, SFR, $\Sigma_\mathrm{SFR}$, and morphology, we first examine whether our sensitivity to winds is dependent on these same galaxy properties.  Figure~\ref{fig.galprop_ston} compares S/N near the \ion{Mg}{2} transition with galaxy properties measured as described in \S\ref{sec.galprop}.  Symbols mark both galaxies with detected winds or inflows (blue crosses) and those with `systemic' absorption (black filled circles).  The Figure, as well as Spearman's rank correlation coefficients, 
indicate that S/N is only weakly correlated with each of these properties, with the largest correlation coefficient $r_\mathrm{S} = 0.35$ for $\Sigma_\mathrm{SFR,global}$.  Our ability to detect winds even at the lowest S/N (and hence at the lowest $\Sigma_\mathrm{SFR,global}$, SFR, etc.) and the weak dependence of outflow detection rate on S/N (discussed in \S\ref{sec.sensitivity}), combined with the weakness of the correlations presented here, suggest that our sensitivity to winds does not depend significantly on host galaxy properties.  Hence, our measured wind detection rates reflect the underlying wind physics, rather than our sensitivity limit.

\begin{figure*}
\includegraphics[angle=90,width=\textwidth]{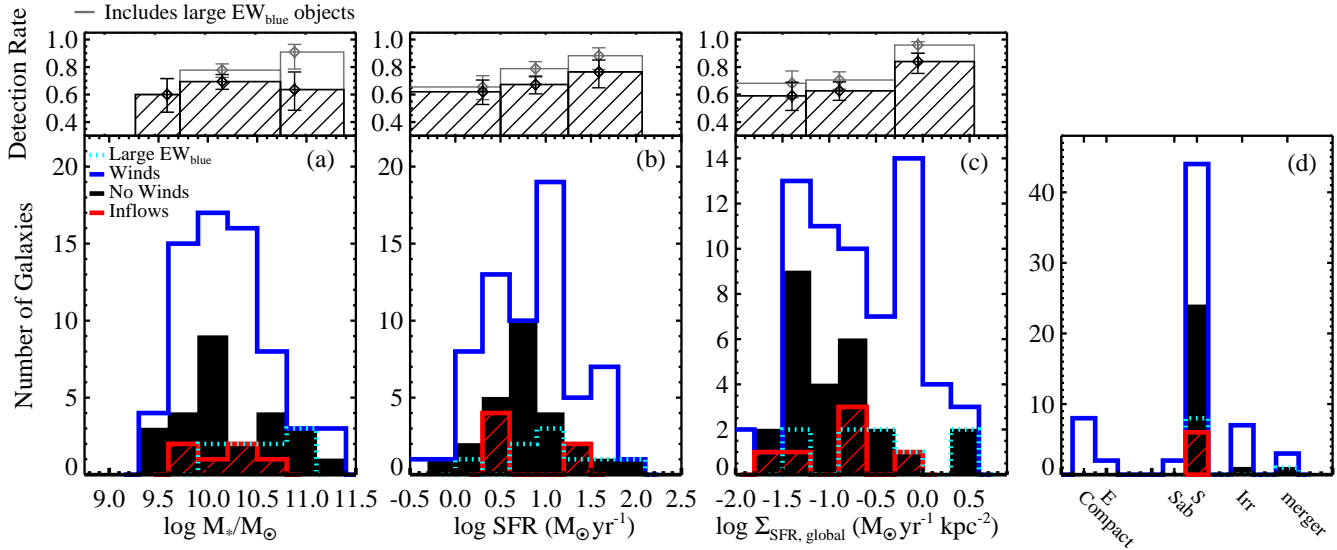}
\caption[]{\emph{Bottom Row:} Distribution of $\log M_*/M_{\odot}$ \emph{(a)}, $\log \rm SFR$ \emph{(b)}, $\log \Sigma_\mathrm{SFR,global}$ \emph{(c)}, and morphology \emph{(d)} for galaxies with 
detected winds (blue solid line), systemic absorption (black filled histogram), inflows (red histogram), and for `systemic' galaxies having large ($> 1.1$ \AA) $\rm EW_{2600}$ or $\rm EW_{2803}$ at $v < 0\mkms$ ($\rm EW_{blue}$; cyan dotted line).  `Compact' indicates compact morphologies, and `E' indicates early-type (E/S0) morphologies.  \emph{Top Row:} Wind detection rate as a function of $\log M_*/M_{\odot}$ (left), $\log \rm SFR$ (middle), and $\log \Sigma_\mathrm{SFR,global}$ (right).  The gray histograms show the detection rate assuming galaxies with large $\rm EW_{blue}$ are driving winds, while the black histograms show the detection rate that results when these galaxies are assumed to not drive winds.  Error bars indicate the standard binomial Wilson score 68\% confidence intervals.
     \label{fig.galprop_hist}}
\end{figure*}

Figure~\ref{fig.galprop_hist} shows the distribution of galaxy $M_*$, SFR, and $\Sigma_\mathrm{SFR, global}$ values for objects with detected winds (blue), systemic absorption (black), and inflows (red).  We additionally define a subsample of objects which have been placed in the `systemic' class, but which have large ($> 1.1$ \AA) $\rm EW_{2600}$ or $\rm EW_{2803}$ at velocities blueward of systemic ($v < 0\mkms$).  As discussed in \S\ref{sec.sensitivity}, these galaxies have EWs at $v < 0\mkms$ ($\rm EW_{blue}$) which are larger than those exhibited by many of our `wind' galaxies (Figure~\ref{fig.redblue}), suggesting that these objects may in fact be driving winds while also having strong	absorption due to inflowing gas which shifts the line profile toward the systemic velocity.  
We caution that such large $\rm EW_{blue}$ values in galaxies in the `systemic' class need not arise from winds, and may instead be due to strong, broad \ion{Fe}{2} or \ion{Mg}{2} stellar absorption features which are more likely to occur in high-$M_*$ galaxies with older stellar populations \citep{RubinTKRS2009,Coil2011}; however, we do not attempt to predict the strength of this absorption via stellar population synthesis modeling here.
The $M_*$, SFR, and $\Sigma_\mathrm{SFR}$ distributions of this new, `large $\rm EW_{blue}$' subsample are shown with the dotted cyan lines in Figure~\ref{fig.galprop_hist}.  The top panels show the measured wind detection rate in three bins for each galaxy property.  We include galaxies in both the `systemic' and `inflow' classes as non-detections.  The black histogram assumes that objects with large $\rm EW_{blue}$ (and classed as `systemic') do not drive winds, while the gray histogram shows the detection rates that result from assuming these galaxies do indeed drive winds.  

Galaxies with and without winds (the blue and black/red histograms, respectively) exhibit similar distributions in $M_*$ and SFR, and a Kolmogorov-Smirnov (K-S) test indicates they are likely drawn from the same parent population.  However, the $\Sigma_\mathrm{SFR}$ distributions of wind and no-wind galaxies have only a 3\% probability of being drawn from the same parent population, with wind galaxies generally having higher $\Sigma_\mathrm{SFR}$ than galaxies without winds.  These differences are also reflected in the wind detection rates shown in the top panels: the rates for the three selected bins in $M_*$ and SFR are consistent within their $1\sigma$ uncertainties, while the highest $\Sigma_\mathrm{SFR}$ bin has a wind detection rate over $1\sigma$ above that of the middle $\Sigma_\mathrm{SFR}$ bin ($0.84\pm0.07$ vs.\ $0.63\pm0.07$).  
We remind the reader that these $\Sigma_\mathrm{SFR}$ values have been deprojected to account for the orientation of each system; if `projected' $\Sigma_\mathrm{SFR}$ values are used (assuming that the galaxy radius is the average of its semi-major and semi-minor axes), the wind detection rates in the three $\Sigma_\mathrm{SFR}$ bins shown in Figure~\ref{fig.galprop_hist} are $0.86\pm0.13$, $0.53\pm0.07$, and $0.87\pm0.06$, such that the difference in wind detection rates in the latter two bins remains statistically significant.

Even given these trends, however, we find that winds are detected in galaxies over the full ranges of $M_*$, SFR, and $\Sigma_\mathrm{SFR}$ exhibited by this sample.  In particular, we find no evidence for a `threshold' $\Sigma_\mathrm{SFR}$ below which winds are not driven.  On the contrary, we detect winds in galaxies whose \emph{maximum} $\Sigma_\mathrm{SFR}$ ($\Sigma_\mathrm{SFR, max}$; see \S\ref{sec.sfrsd}) is as low as $\sim0.04~M_{\odot}~\rm yr^{-1}~kpc^{-2}$.

Our conclusions change slightly if we move the `large $\rm EW_{blue}$' objects from the `systemic' class to the `wind' class.  These galaxies tend to have high $M_*$ (see cyan histogram in Figure~\ref{fig.galprop_hist}a), and their inclusion in the `wind' subsample therefore increases the wind detection rate in the highest-$M_*$ bin to a value $\sim1\sigma$ higher than in the middle-$M_*$ bin ($0.91\pm 0.08$ vs.\ $0.78\pm0.05$).  
Many of the `large $\rm EW_{blue}$' galaxies also have large SFR, such that if we assume that they host winds, the wind detection rate increases at higher SFRs (although the rates remain 
consistent within their uncertainties).
Finally, a few of these galaxies also have quite large $\Sigma_\mathrm{SFR}$, and thus increase the significance of the rise in detection rate 
between the middle- and high-$\Sigma_\mathrm{SFR}$ bins to
$\sim3\sigma$ (with rates of $0.71\pm0.06$ and $0.96\pm0.04$).  
As noted above, we cannot ultimately 
determine whether the strong systemic absorption in these objects is due to stellar atmospheres, interstellar material, or halo gas;
however, these results suggest
that galaxies with higher $M_*$, SFR, and $\Sigma_\mathrm{SFR}$ may be more likely to drive outflows which are camouflaged by 
strong absorption at velocities $ \ge 0\mkms$.


\begin{figure*}
\includegraphics[angle=90,width=\textwidth]{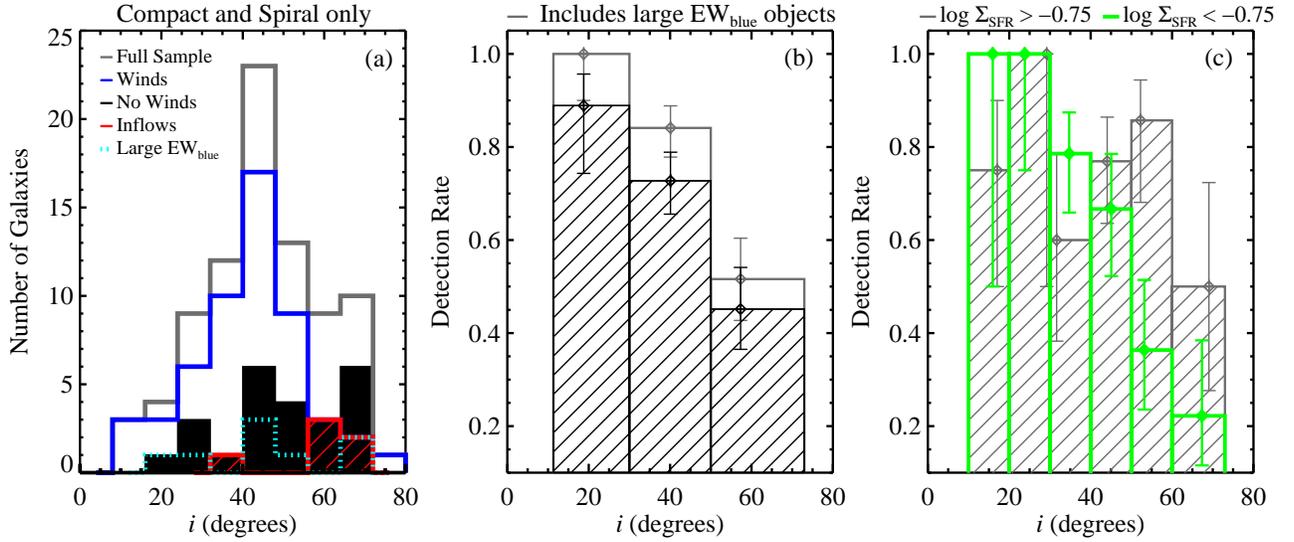}
\caption[]{
\emph{(a)}  Distribution of inclinations for compact and spiral galaxies.  Colors show absorption kinematic classes as listed in Figure~\ref{fig.galprop_hist}, with the gray histogram showing the distribution of inclinations for the full sample.  \emph{(b)} Wind detection rate as a function of inclination.  The gray and black histograms are calculated using the same assumptions described in the Figure~\ref{fig.galprop_hist} caption (top row).  Error bars indicate the standard binomial Wilson score 68\% confidence intervals.
\emph{(c)} Wind detection rate as a function of inclination for galaxies having $\log \Sigma_\mathrm{SFR, global}~[M_{\odot}~\rm yr^{-1}~kpc^{-2}] > -0.75$ (gray) and $\log \Sigma_\mathrm{SFR, global}~[M_{\odot}~\rm yr^{-1}~kpc^{-2}] < -0.75$ (green), and excluding galaxies with large $\rm EW_{blue}$ values.     Error bars are calculated as in panel \emph{(b)}.
     \label{fig.morph_hist}}
\end{figure*}

Figure~\ref{fig.galprop_hist}d shows the distribution of morphologies for galaxies with and without detected winds.  The vast majority of our sample has disk-like morphologies (Sab or S); however, several of our `wind' galaxies have disturbed morphologies suggestive of recent merger activity, or are very compact.  We detect winds at a rate $0.83\pm0.11$ in galaxies with irregular/disturbed morphologies, and at a rate $0.65\pm0.05$ for disk-like or compact galaxies, indicating that while there is a high probability of wind detection over the full range of morphologies exhibited by our sample, winds are somewhat more likely to be detected in disturbed systems.  

\subsection{Trends in Wind Detection Rates with Galaxy Orientation}\label{sec.detrate_orientation}

Figure~\ref{fig.morph_hist}a shows the distribution of inclinations for galaxies with and without winds, including only compact and disk-like systems.  We exclude galaxies with disturbed morphologies, as their inclinations cannot be determined from a simple axis ratio analysis.  The combined `No Winds' and `Inflows' distribution (black/red) is strongly skewed toward high inclinations (i.e., edge-on orientations), while the wind galaxies (blue histogram) tend to have lower inclinations.  These distributions (i.e., the `Winds' and the combined `No Winds' and `Inflows' distribution) have only a 0.1\% probability of being drawn from the same parent population.  Figure~\ref{fig.morph_hist}b  shows that the wind detection rate increases significantly (by $3.4\sigma$) as galaxy inclination decreases, from $0.45\pm0.09$ in edge-on galaxies (having $i > 50^{\circ}$), to $0.73\pm0.07$ in galaxies with $30^{\circ} < i < 50^{\circ}$, to $0.89\pm0.10$ in face-on ($i < 30^{\circ}$) galaxies.  {\it We conclude that orientation is the single most important factor in determining whether we will detect a wind from a given galaxy}, which suggests that many of the more edge-on galaxies in our sample with `systemic' absorption are driving winds which are  not oriented along our line of sight.  Furthermore, the near-ubiquity of winds in galaxies having $i < 30^{\circ}$ suggests that, under the assumption of a biconical wind morphology, the full cone opening angle of these outflows is almost always greater than $\sim 60^{\circ}$; i.e., close to or larger than  the opening angle measured for the nearby starburst M82 \citep[also $\sim60^{\circ}$;][]{Heckman1990,Walter2002}.  

As shown in \S\ref{sec.detrate}, we are marginally more likely to detect winds from galaxies with high $\Sigma_\mathrm{SFR, global}$.  The concomitant, strong dependence of detection rate on galaxy inclination motivates an attempt to disentangle the effects of both of these quantities on the 
overall likelihood of wind detection.  We start with the assumption that our measure of 
(deprojected) $\Sigma_\mathrm{SFR, global}$ is representative of the density of star formation activity in each galaxy as it is viewed face-on, such that we may test for changes in detection rates with varying inclination at a fixed $\Sigma_\mathrm{SFR, global}$. 
We divide our sample into two subsamples having $\log \Sigma_\mathrm{SFR, global}~[M_{\odot}~\rm yr^{-1}~kpc^{-2}] < -0.75$ and $> -0.75$, and show the wind detection rate as a function of galaxy inclination for these subsamples in Figure~\ref{fig.morph_hist}c.  The rates are consistently high ($\gtrsim60$\%) in both low- and high-$\Sigma_\mathrm{SFR}$ galaxies having $i < 50^{\circ}$, indicating that we are equally likely to detect winds regardless of $\Sigma_\mathrm{SFR}$, provided that the galaxies are close to face-on.  That is, at relatively low viewing angles ($i < 50^{\circ}$), $\Sigma_\mathrm{SFR}$ does not appear to 
be an important parameter in determining whether a detectable wind will be launched.

However, for galaxies having $50^{\circ} < i < 60^{\circ}$, the detection rate for low-$\Sigma_\mathrm{SFR}$ objects drops to $0.36 \pm 0.15$ while remaining high for the high-$\Sigma_\mathrm{SFR}$ subsample ($0.86\pm 0.13$).  
This $2.5\sigma$-significant difference in detection rates is striking given the good agreement in the detection rates at all other inclinations, and is not likely driven by differences in the S/N of the high- and low-$\Sigma_\mathrm{SFR}$ galaxy spectra, as the median S/N of the two subsamples in this inclination bin are similar ($10.6~\rm pix^{-1}$ and $8.2~\rm pix^{-1}$).
The difference suggests that the wind opening angles may be larger in galaxies with higher $\Sigma_\mathrm{SFR}$.  Specifically, the majority of galaxies with $\log \Sigma_\mathrm{SFR} ~[M_{\odot}~\rm yr^{-1}~kpc^{-2}] > -0.75$ have full cone opening angles of $\sim120^{\circ}$, while the majority of galaxies with lower $\Sigma_\mathrm{SFR}$ have full cone opening angles of only $\sim100^{\circ}$.  Because these subdivisions push the limits of the statistical power of our sample, it is important to test these constraints with a larger number of galaxies.  However, these results strengthen our previous conclusion that the detection of cool outflows is most fundamentally dependent on galaxy orientation above all other intrinsic galaxy properties.


\begin{figure*}
\includegraphics[angle=90,width=\textwidth]{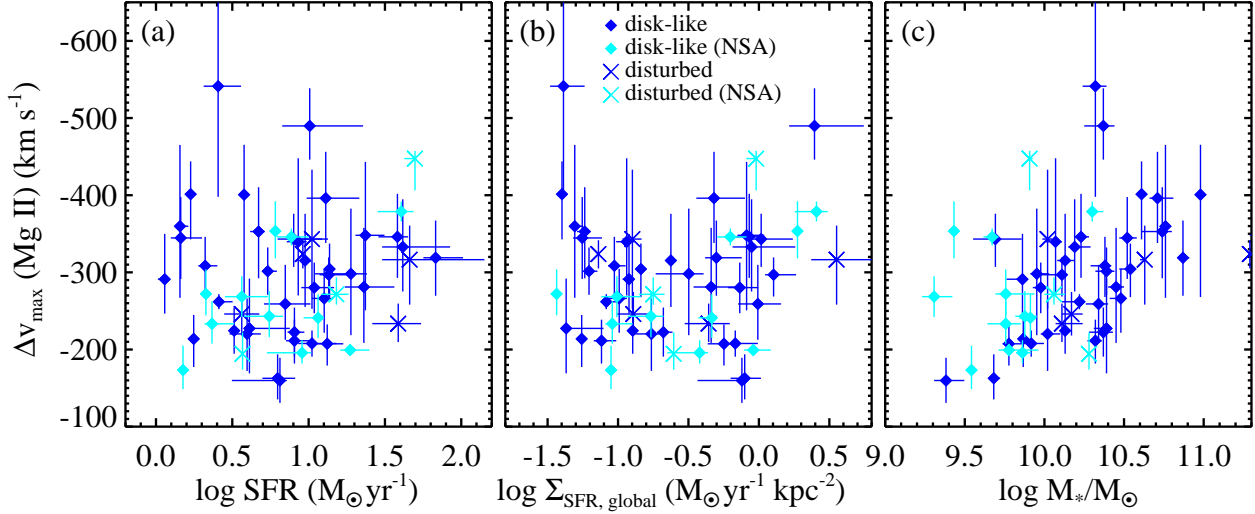}
\caption[]{$\Delta v_\mathrm{max}$ measured in the \ion{Mg}{2} transition vs.\ $\log \rm SFR$ \emph{(a)}, $\log \Sigma_\mathrm{SFR,global}$ \emph{(b)}, and $\log M_*/M_{\odot}$ \emph{(c)}.  Disk-like (spiral) and compact galaxies are marked with blue diamonds (for line profiles with significant absorption at systemic velocity) 
and cyan diamonds (for line profiles without systemic absorption, denoted `NSA').  Galaxies with disturbed morphologies are marked with blue crosses (with systemic absorption) 
and cyan crosses (without systemic absorption). 
     \label{fig.galprop_vel}}
\end{figure*}

\subsection{Trends in Wind Velocities and EWs with Intrinsic Galaxy Properties}\label{sec.velew}

We now examine the relationship between wind velocities and absorption strength and intrinsic galaxy properties.  Figure~\ref{fig.galprop_vel} compares $\Delta v_\mathrm{max}$ measured from the \ion{Mg}{2} transition with SFR, $\Sigma_\mathrm{SFR, global}$, and $M_*$.  We show regular, disk-like or compact galaxies with blue diamonds (for systems having \ion{Mg}{2} absorption at $v\sim0\mkms$) and cyan diamonds (for systems without \ion{Mg}{2} absorption at $v > 0\mkms$), and mark disturbed galaxies with blue and cyan crosses.  For all of these subsamples (taken together), we find no evidence for a correlation between $\Delta v_\mathrm{max}$ 
and either SFR or $\Sigma_\mathrm{SFR}$, indicating that 
these properties do not significantly influence the maximum velocities reached by cool outflows.  Figure~\ref{fig.galprop_vel}c demonstrates, however, a strong correlation (having $3.4\sigma$ significance and $r_\mathrm{S} = -0.46$) between $M_*$ and $\Delta v_\mathrm{max}$\footnote{All correlation coefficients reported in this section are computed after excluding two galaxies having $\mathrm{SFR} < 1~M_{\odot}~\rm yr^{-1}$, and excluding TKRS5379, which has $\Delta v_\mathrm{max}$ close to $-800\mkms$.  The latter object was placed in the `wind' class due to a significant blue wing in its \ion{Mg}{2} absorption profile; however, it did not meet our original selection criterion, and is an extreme outlier (see \S\ref{sec.classes}).}, indicating that star formation history and/or galaxy dynamics have a more direct physical link to maximum wind velocities than current star formation activity.  
Figure~\ref{fig.galprop_vel}c additionally shows that galaxies at lower $M_*$ ($\log M_*/M_{\odot} \lesssim 10$) are more likely to lack systemic \ion{Mg}{2} absorption (see cyan diamonds) than objects at higher $M_*$.  This may be due to a number of factors; e.g., stronger/broader interstellar absorption or enhanced dust content (which may both suppress resonant scattering) at higher $M_*$ \citep{Prochaska2011,Martin2012,Erb2012}.\footnote{Note that we detect no significant correlation between $M_*$ and $v_\mathrm{flow}$ ($r_\mathrm{S} = -0.12$ at 0.9$\sigma$ significance), and likewise find no significant correlation between $v_\mathrm{flow}$ and either SFR or $\Sigma_\mathrm{SFR}$.}  

\begin{figure*}
\includegraphics[angle=90,width=\textwidth]{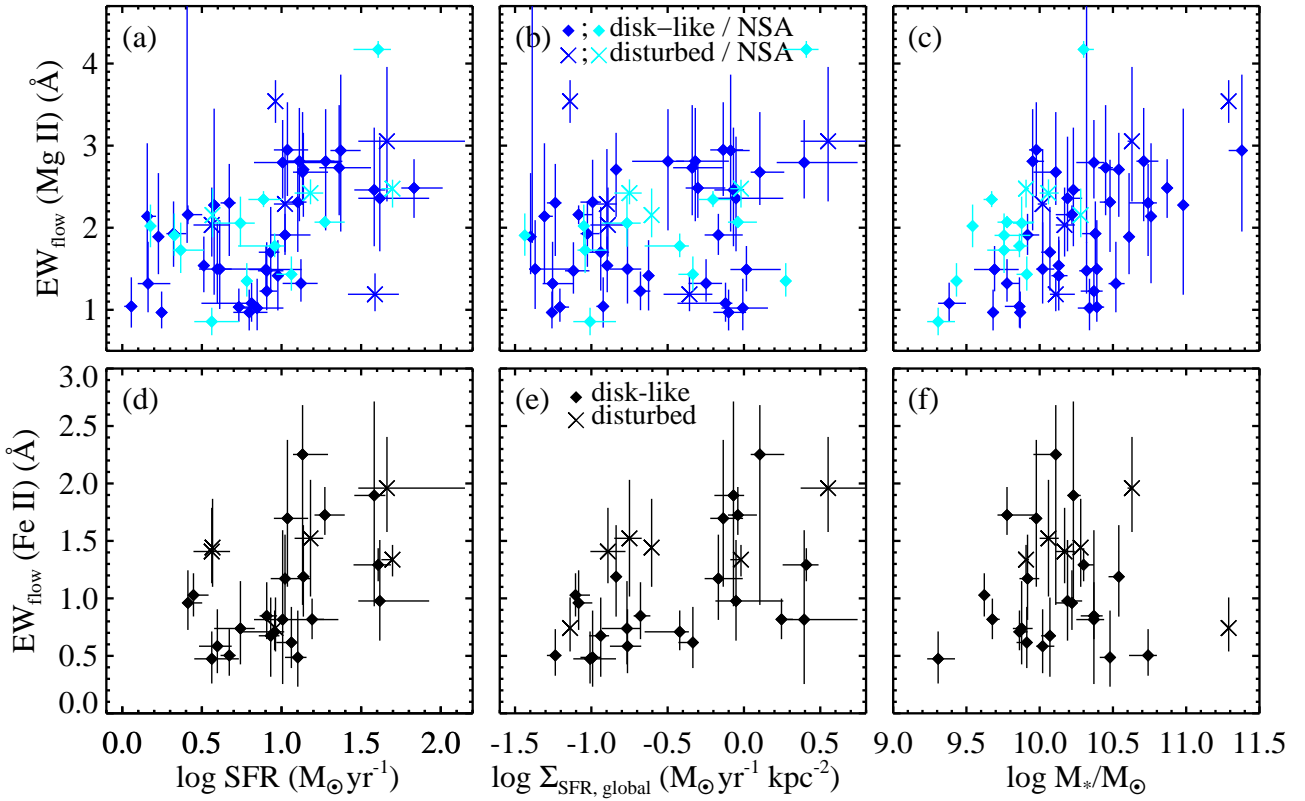}
\caption[]{\emph{Top Row:} $\rm EW_{flow}$ measured in the \ion{Mg}{2} transition vs.\  $\log \rm SFR$ (panel \emph{(a)}; $r_\mathrm{S} = 0.48$; $3.5\sigma$), $\log \Sigma_\mathrm{SFR,global}$ (panel \emph{(b)}; $r_\mathrm{S} = 0.21$; $1.5\sigma$), and $\log M_*/M_{\odot}$ (panel \emph{(c)}; $r_\mathrm{S} = 0.44$; $3.2\sigma$).  \ion{Mg}{2} profiles with and without systemic absorption are indicated with blue and cyan symbols as in Figure~\ref{fig.galprop_vel}.  \emph{Bottom Row:}  $\rm EW_{flow}$ measured in the \ion{Fe}{2} transition vs.\ $\log \rm SFR$ (panel \emph{(d)}; $r_\mathrm{S} = 0.46$; $2.4\sigma$), $\log \Sigma_\mathrm{SFR,global}$ (panel \emph{(e)}; $r_\mathrm{S} = 0.54$; $2.8\sigma$), and $\log M_*/M_{\odot}$ (panel \emph{(f)}; $r_\mathrm{S} = 0.07$; $0.4\sigma$).  Disk-like (spiral) and compact galaxies are marked with diamonds, and galaxies with disturbed morphologies are marked with crosses.  
     \label{fig.galprop_ew}}
\end{figure*}

\begin{figure*}
\includegraphics[angle=90,width=\textwidth]{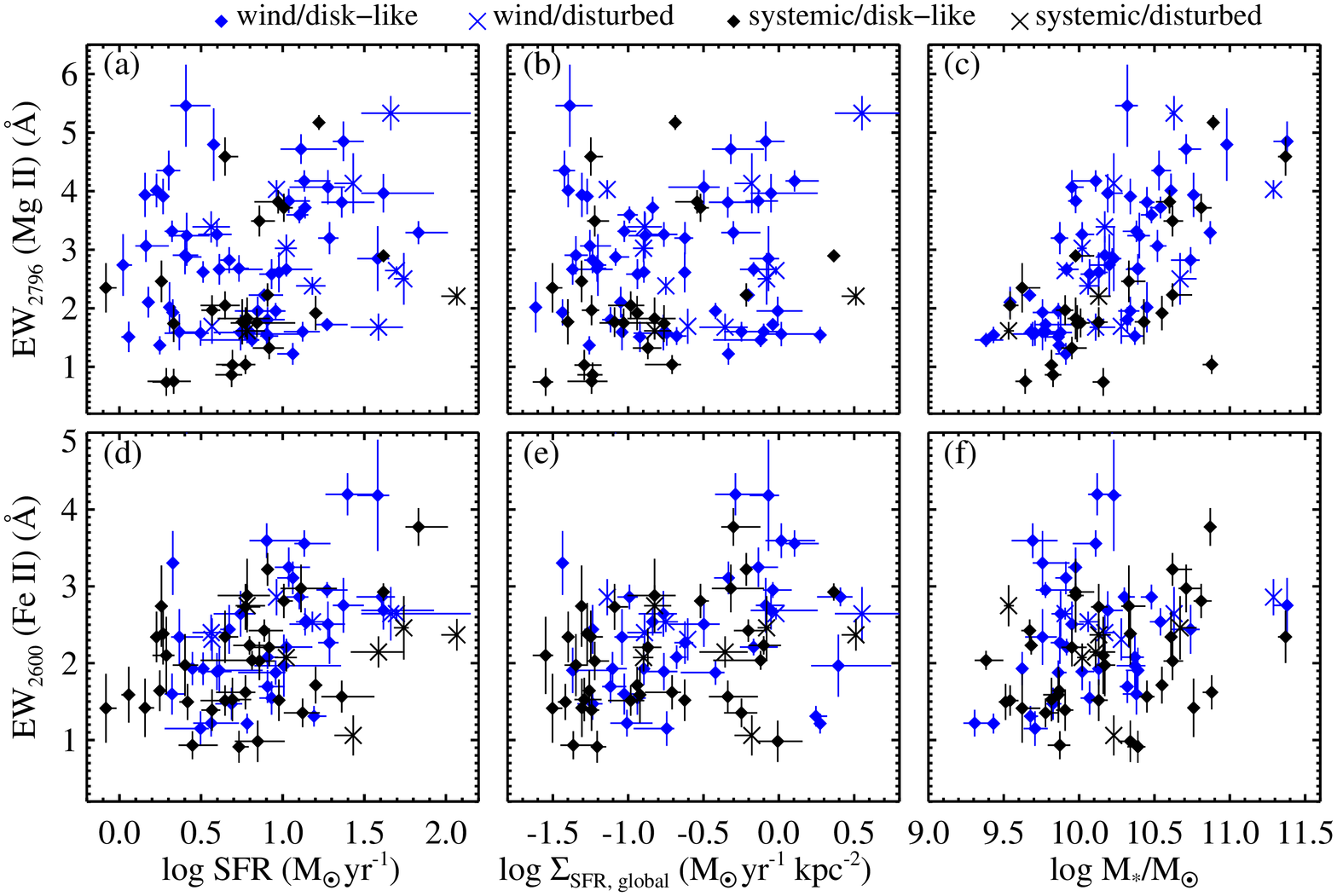}
\caption[]{$\rm EW_{2796}$ (top row) and $\rm EW_{2600}$ (bottom row) vs.\ $\log \rm SFR$ \emph{(a, d)}, $\log \Sigma_\mathrm{SFR,global}$ \emph{(b, e)}, and $\log M_*/M_{\odot}$ \emph{(c, f)}.  
Galaxies with detected winds or inflows are indicated in blue, and galaxies without detected winds or inflows are shown in black.
Disk-like (spiral) and compact galaxies (both with and without winds/inflows) are marked with diamonds, and galaxies with disturbed morphologies are marked with crosses.  
     \label{fig.galprop_ewtot}}
\end{figure*}

Figure~\ref{fig.galprop_ew} compares $\rm EW_\mathrm{flow}$ measured from \ion{Mg}{2} (top row) and \ion{Fe}{2} (bottom row) with host galaxy SFR, $\Sigma_\mathrm{SFR, global}$, and $M_*$.  Spearman's rank correlation coefficients and the corresponding statistical significance 
 for the quantities shown in each panel are given in the Figure caption.
While $\rm EW_{flow}$(\ion{Mg}{2}) and $\Sigma_\mathrm{SFR, global}$ do not exhibit a statistically significant correlation, we find that $\rm EW_{flow}$(\ion{Mg}{2}) and $M_*$ are correlated ($r_\mathrm{S} = 0.44$) with a significance of $3.2\sigma$, and that  $\rm EW_{flow}$(\ion{Mg}{2}) and SFR are correlated ($r_\mathrm{S} = 0.48$) with a significance of $3.5\sigma$.  
Similarly, $\rm EW_{flow}$ measured from \ion{Fe}{2} exhibits a $2.4\sigma$-significant correlation with SFR ($r_\mathrm{S} = 0.46$), and is additionally correlated at $2.8\sigma$ significance with $\Sigma_\mathrm{SFR}$ ($r_\mathrm{S} = 0.54$).  Unlike $\rm EW_{flow}$(\ion{Mg}{2}), however, $\rm EW_{flow}$(\ion{Fe}{2}) exhibits no significant correlation with $M_*$.  This difference may arise due to the reduced range in $M_*$ of the subset of our sample with high-quality $\rm EW_{flow}$(\ion{Fe}{2}) measurements.  

Finally, we compare  $\rm EW_{2796}$ (top row) and $\rm EW_{2600}$ (bottom row)  with host galaxy SFR, $\Sigma_\mathrm{SFR, global}$ and $M_*$ in Figure~\ref{fig.galprop_ewtot}.  Here, we include galaxies both with detected winds/inflow (blue) and those without detected flows (`systemic'; black).  We find that $\rm EW_{2796}$ and $M_*$ are correlated at $5.8\sigma$ significance (with $r_\mathrm{S} = 0.63$), and hence that they exhibit the tightest relation of any of the pairs of quantities we have discussed.  We also note that the `systemic' and `winds/inflow' subsamples overlap substantially in this parameter space, with `systemic' galaxies exhibiting some of the largest $\rm EW_{2796}$ values ($\gtrsim 4$ \AA) measured.  These findings suggest that \ion{Mg}{2} absorption is sensitive not only to wind kinematics, but also to the kinematics of gas distributed throughout galaxy halos, including virialized halo gas clouds and cool accreting material.  Scattered emission likely also affects the relationship between $\rm EW_{2796} $ and $M_*$, preferentially suppressing $\rm EW_{2796}$ at lower $M_*$ in galaxies hosting winds.

$\rm EW_{2600}$, on the other hand, is only weakly correlated with $M_*$ ($r_\mathrm{S} = 0.23$ at $2.0\sigma$ significance) and $\Sigma_\mathrm{SFR, global}$ ($r_\mathrm{S} = 0.28$ at $2.5\sigma$ significance), 
but is more strongly correlated with SFR ($r_\mathrm{S} = 0.36$) at $3.2\sigma$ significance.  We speculate that because \ion{Fe}{2} probes a more narrow range in ionization parameter than \ion{Mg}{2} (and hence survives only at higher gas densities), the line profiles are not significantly affected by halo gas kinematics, but instead are primarily sensitive to the kinematics of material closer to the galaxy disks.  This may explain the strength of the correlations between star formation activity and both $\rm EW_{flow}$(\ion{Fe}{2}) and $\rm EW_{2600}$.  

Figure~\ref{fig.ewMg_mstarsfr} shows the SFR-$M_*$ distribution of our sample galaxies with detected winds (open diamonds) and without detected flows (filled circles).  The point size marking the `wind' galaxies is scaled linearly with $\rm EW_{flow}$ (\ion{Mg}{2}).  
Our sample is overlaid on top of the SFR-$M_*$ distribution of a much larger sample of galaxies at $0.4 < z < 1.0$ drawn from B11 for comparison.
With the exception of a few galaxies having low SFRs and $\rm EW_{flow} \gtrsim 2$ \AA\ (labeled A-F), the largest $\rm EW_{flow}$ values are located toward the upper edge of the star-forming sequence.  This $\rm EW_{flow}$ distribution, along with the strong positive correlations between $\rm EW_{flow}$(\ion{Mg}{2}) and SFR ($3.5\sigma$) and $\rm EW_{2796}$ and $M_*$ ($5.8\sigma$), suggests that stronger wind absorption is more likely detected in both higher-$M_*$ and higher-SFR systems.  
We discuss the exceptions to this general picture in the next subsection.  

\subsection{Strong Winds in Low-SFR Galaxies}\label{sec.lowsfr}

\begin{figure}
\includegraphics[angle=0,width=\columnwidth,trim=0 0 10 0,clip]{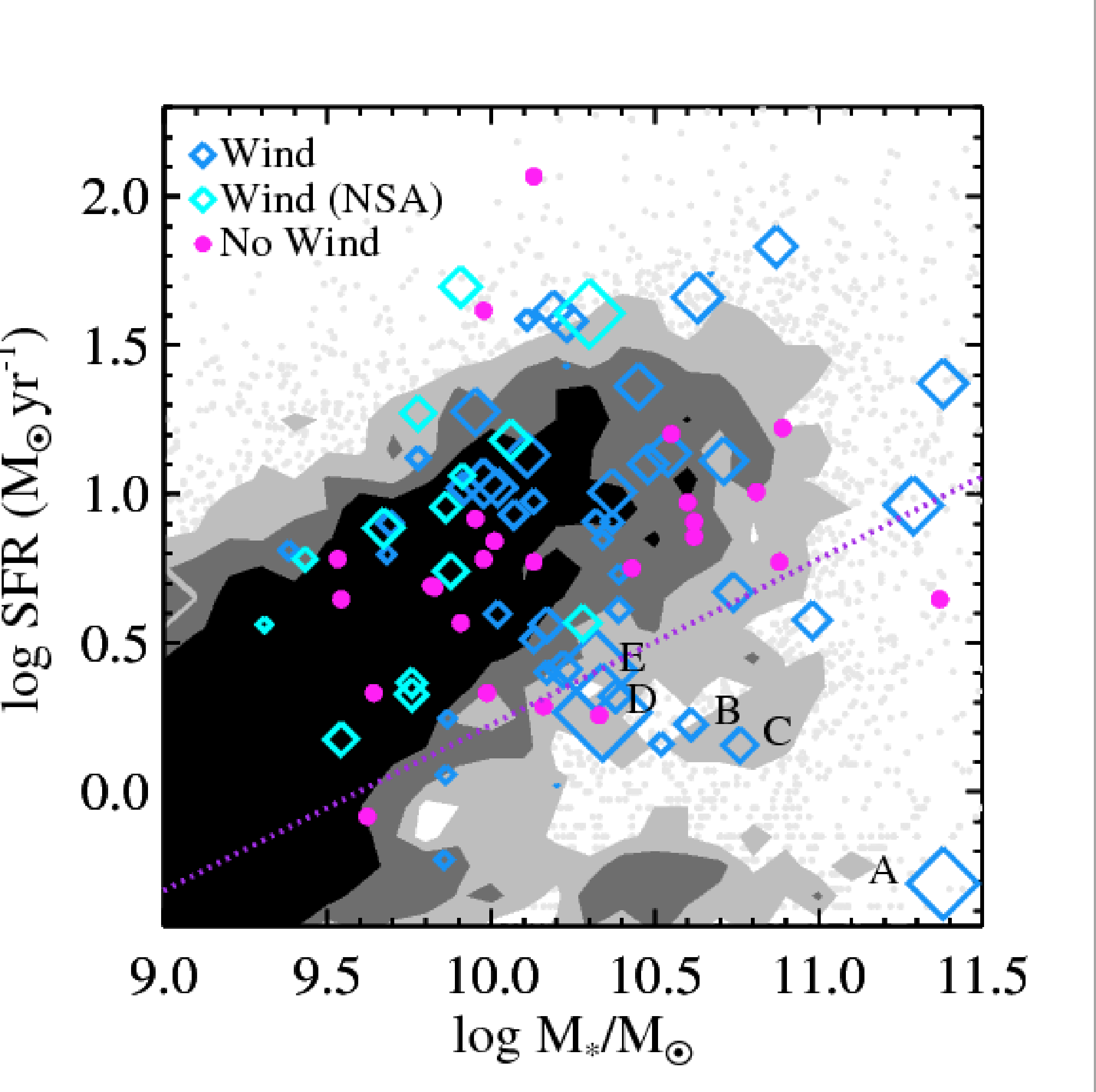}
\caption[]{Location of our sample galaxies with winds detected in \ion{Mg}{2}  (open diamonds) and without detected winds (magenta circles) on the star-forming sequence.  Galaxies having \ion{Mg}{2} profiles exhibiting absorption at $v\sim0\mkms$ are marked in blue, and galaxies having \ion{Mg}{2} profiles without systemic absorption (`NSA') are marked in cyan.  For objects with detected winds, the point size is scaled linearly with $\rm EW_{flow}$(\ion{Mg}{2}).  For reference, the galaxy marked with `A' has $\rm EW_{flow} = 4.4$ \AA, and the galaxy marked with `B' has $\rm EW_{flow} = 1.9$ \AA.
Contours and small gray points show the SFR-$M_*$ distribution of the B11 sample at $0.4 < z < 1.0$ converted to a Chabrier IMF.  While $\rm EW_{flow}$ values tend to increase toward higher SFR and $M_*$, there are a few galaxies with $\rm EW_{flow} \gtrsim 2$ \AA\ below or near the lower edge of the main star-forming sequence of B11.  We discuss these objects in \S\ref{sec.lowsfr}.  
The purple dotted line indicates the \citet{Murray2011} threshold for driving winds; see \S\ref{sec.ubiquity} for details.
     \label{fig.ewMg_mstarsfr}}
\end{figure}

In Figure~\ref{fig.ewMg_mstarsfr} we mark several objects which sit below the main star-forming sequence of the B11 sample galaxies (shown with gray contours), and yet which have large values of $\rm EW_{flow}$(\ion{Mg}{2}) ($\gtrsim 2$ \AA).  Motivated by the strong Balmer absorption lines evident in a few of these galaxy spectra and previous detections of winds hosted by post-starburst galaxies \citep{Sato2009,Coil2011},
we investigate the relative strengths of recent past and current star-formation activity and search for post-starburst signatures in these systems.  As discussed in, e.g., \citet{DresslerGunn1983}, \citet{Goto2003}, \citet{Yan2006}, and \citet{Goto2007}, the presence of strong Balmer absorption signals a recent burst of star formation, and the simultaneous absence of nebular emission lines (e.g., [\ion{O}{2}], H$\alpha$) indicates that the burst has been truncated suddenly.  To quantify the Balmer absorption strength, we measure the EW of H$\delta$ absorption in our spectra as described in \citet{Goto2003}, using a linear fit to the flux in the continuum windows listed in that work to normalize the spectra, and using a boxcar sum over the fixed rest wavelength range $4088 - 4116$ \AA.  We measure the EW of the blended [\ion{O}{2}] emission line doublet complex using a similar technique to assess ongoing star formation activity (also as described in \citealt{Goto2003}).  The criteria used to select `post-starburst' objects varies from study to study; H$\delta$ EWs must typically be greater than $3 - 5.5$ \AA, while [\ion{O}{2}] EWs must be $> -(2.5 - 5.0)$ \AA.  Only two objects in our sample whose spectra cover these transitions fully satisfy these criteria; we note that both of these objects exhibit strong winds (`B' and `C').  However, we refer to these criteria in discussing the salient properties of all of the low-SFR galaxies exhibiting winds below:


\begin{itemize}
\item {\bf A} -- EGS13041646: This galaxy has an `E/S0' morphology, and with a rest-frame $U-B$ color $\approx 1.1$ mag lies directly on the red sequence.  It exhibits no [\ion{O}{2}] emission, and has $\rm EW(H\delta) \approx 0.8$ \AA.  Its star formation activity, therefore, likely ceased $\gtrsim2$ Gyr ago; i.e., this galaxy is `red and dead'.  
The \ion{Mg}{2} line profile for this object exhibits unusually strong absorption ($\rm EW_{2796} \sim 4.7~\AA$) extending to high positive and negative velocities relative to systemic.  Much of this absorption may be associated with older stellar atmospheres or interstellar material having a velocity close to $v=0\mkms$.  However, our line profile modeling indicates that the profile is blueshifted overall with $P_\mathrm{out} = 0.967$, and hence that the galaxy hosts a wind.  
This wind cannot be driven by ongoing or recent star formation; instead, it may have been launched by past star formation, or by ongoing AGN activity.  

\item{\bf B} -- EGS13050592:  This galaxy has a red $U-B$ color (0.99 mag) and a spiral-like morphology.  Its spectrum exhibits $\rm EW(H\delta) \approx 6.0$ \AA\ and $\rm EW([OII]) \approx -4.0$ \AA, and thus satisfies the post-starburst criteria discussed above.  The strong blueshifted absorption observed in this system may therefore trace the relic of a wind launched during a recently-truncated burst of star formation.  

\item{\bf C} -- TKRS7326: This galaxy has an early spiral-like morphology with $\rm EW(H\delta) \approx 4.2$ \AA\ and $\rm EW([OII]) \approx -2.6$ \AA.  These EWs satisfy the less stringent post-starburst criteria laid out above, suggesting that the large $\rm EW_{flow}$ measured for this system may also
arise from a relic wind launched in the past.

\item{\bf D} -- TKRS5379: This spiral galaxy has $\rm EW(H\delta) \approx 4.4$ \AA\ and $\rm EW([OII]) \approx -11.5$ \AA.   The latter is suggestive of 
ongoing star-formation activity, which could drive a strong wind.  However, we measure
a large $\rm EW_{flow}$ for this object due to the presence of an extended, shallow blue wing in the \ion{Mg}{2} line profile (see Figure~\ref{fig.allspecs1}), and not because of an overall blueshift of the full absorption profile.  This blue wing may in fact arise from a systematic error in the determination of the continuum level, rather than outflowing material at extreme ($ <-700\mkms$) velocities.  

\item{\bf E} -- J033242.32-274950.3: This galaxy has an early spiral-like morphology, with $\rm EW(H\delta) \approx 3.5$ \AA\ and $\rm EW([OII]) \approx -7.0$ \AA.  Its H$\delta$ EW is large enough to satisfy the less stringent post-starburst criterion listed above, but its [\ion{O}{2}] EW is suggestive of weak ongoing star formation.  
Our spectrum of this object has unusually low S/N ($\rm 4~pix^{-1}$) near the \ion{Mg}{2} transition, and it may only be placed in the `wind' class by virtue of exceptionally strong \ion{Mg}{2} absorption ($\rm EW_{2796} \sim 5.5~\AA$).  We find that the centroid of this absorption is blueshifted with high probability  ($P_\mathrm{out} = 0.997$).  
We have visually inspected the spectrum to confirm the success of our continuum-fitting procedure, and  do not expect slight under- or overestimates of the continuum level to induce such a significant blueshift.  Furthermore, the absence of strong sky emission features at the observed wavelength of the \ion{Mg}{2} transition  ($\lambda_\mathrm{obs} \sim4080$ \AA) make sky subtraction errors which vary systematically with wavelength unlikely.
The large $\rm EW_{flow}$ we measure for  this system may trace a wind driven by the weak ongoing star formation and/or by a more vigorous recent burst.  


\end{itemize}

These results show that 
galaxies exhibiting some of the largest $\rm EW_{flow}$ values in our sample ($\rm EW_{flow} \sim 2-6$ \AA\ with $\Delta v_\mathrm{max} < -300\mkms$)
have very weak ongoing star formation activity but experienced a large burst of star formation in the recent past.
As noted above, the classification of a subset of these objects as `wind' galaxies is somewhat subjective (galaxies `A', `D' and `E'), as their red colors and older stellar populations tend to yield low spectral S/N and/or strong stellar \ion{Mg}{2} absorption.  
Furthermore, the strong correlation between $\rm EW_{2796}$ and $M_*$ discussed in \S\ref{sec.velew} could indicate that 
the strong \ion{Mg}{2} profiles (and large $\rm EW_{flow}$ values) detected in these massive systems trace virial CGM motions in addition to winds.
However, these cautionary notes aside, 
the results described above imply that 
the winds we detect 
could have been launched during past starburst activity, and furthermore
that measured wind properties may be physically linked to the star formation history of a given galaxy over a timespan of several Gyr.
Indeed, one third of the objects with post-starburst spectral signatures studied in \citet{Coil2011} were found to drive winds with typical velocities of $\sim200\mkms$.  
We also note that these five galaxies have low SFR surface densities ($\log \Sigma_\mathrm{SFR, global} [M_{\odot}~\rm yr^{-1}~kpc^{-2}] <  -1.25$) which place them in the lowest $\Sigma_\mathrm{SFR}$ bin shown in the upper panel of Figure~\ref{fig.galprop_hist}c.  Removing them from the sample results in a slightly reduced wind detection rate of $0.47\pm0.12$, still consistent with the original rate measured for this bin ($0.59\pm0.10$).   Similarly, excluding them from our measurement of the correlation between $\Delta v_\mathrm{max}$ and $\Sigma_\mathrm{SFR}$ does not significantly affect the rank correlation coefficient (which decreases from $r_\mathrm{S} = -0.02$ to $r_\mathrm{S} = -0.17$).

\subsection{Trends in Wind Velocities and EWs with Galaxy Orientation}\label{sec.velew_galorient}

\begin{figure}
\includegraphics[angle=90,width=\columnwidth]{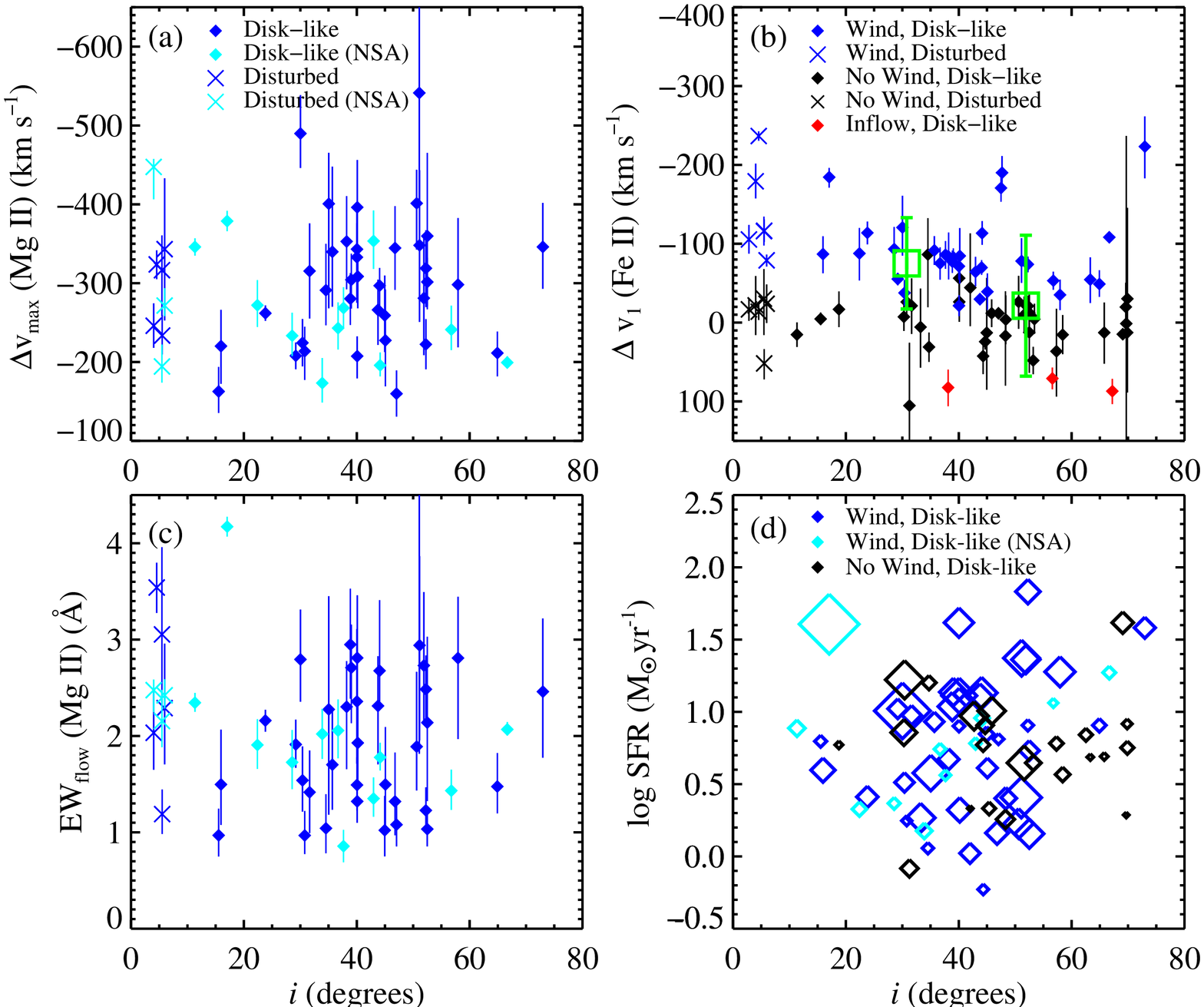}
\caption[]{\emph{(a)} $\Delta v_\mathrm{max}$ measured in the \ion{Mg}{2} transition vs.\ galaxy inclination ($i$).  Disk-like (spiral) and compact galaxies are marked with blue diamonds (for line profiles with significant absorption at $v\sim0\mkms$) and cyan diamonds (for line profiles without systemic absorption, denoted `NSA').  Galaxies with disturbed morphologies are marked with blue crosses (with systemic absorption) and cyan crosses (without systemic absorption). 
\emph{(b)} $\Delta v_1$ measured in the \ion{Fe}{2} transition vs.\ $i$.  Morphologies are indicated with diamonds and crosses as in panel \emph{(a)}.  Galaxies in the `wind' class are marked in blue; objects without detected winds or inflows are marked in black; and `inflow' galaxies are marked in red.  
Large green squares and error bars show the median and scatter in $\Delta v_1$ values  at $i< 40^{\circ}$ and $i>40^{\circ}$.  
\emph{(c)} $\rm EW_{flow}$ measured in the \ion{Mg}{2} transition vs.\ $i$.  Symbols are as described in panel \emph{(a)}.   For panels (\emph{a-c}), objects with disturbed morphologies have been plotted at $i\sim5^{\circ}$.
\emph{(d)} $\log \rm SFR$ vs.\ $i$ for disk-like and compact galaxies.  The point size is scaled linearly with $\rm EW_{2796}$.  Objects with detected winds are marked in blue; objects with no detected winds/inflows are marked in black; and objects with winds having no \ion{Mg}{2} absorption at systemic velocity are marked in cyan.  The dearth of objects with low SFR and high inclinations suggests that our sample may be biased toward higher SFRs in more edge-on systems; however, a K-S test indicates that the distribution of SFRs for systems with $i > 50^{\circ}$ is not significantly different from the distribution of SFRs for systems with $30^{\circ} < i < 50^{\circ}$.  
     \label{fig.ewtot_isfr}}
\end{figure}

Figures~\ref{fig.ewtot_isfr}a and \ref{fig.ewtot_isfr}c compare galaxy inclination with the maximum wind velocity ($\Delta v_\mathrm{max}$) and $\rm EW_{flow}$ measured from the \ion{Mg}{2} transition.  
We find no correlation between the inclinations of galaxies having `regular' morphologies and either $\Delta v_\mathrm{max}$ or $\rm EW_{flow}$, and in fact measure the highest wind velocities in systems with intermediate inclinations ($i \sim 40^{\circ}$).  
To explore this issue in greater depth, 
we also show velocities obtained from our one-component model fits of the \ion{Fe}{2} transitions ($\Delta v_1$) for `wind', `systemic', and `inflow' galaxies in Figure~\ref{fig.ewtot_isfr}b.  These velocities are weakly correlated with galaxy inclination, yielding $r_\mathrm{S} = 0.32$ at $2.6\sigma$ significance.  
The large green squares in the Figure show the median and scatter in $\Delta v_1$ values for the systems at $i < 40^{\circ}$ and $i > 40^{\circ}$.  

\citet{Martin2012} found that the fraction of galaxies driving detected winds faster than a given threshold velocity decreases as the threshold velocity is increased.\footnote{The distribution of $\Delta v_1$(\ion{Fe}{2}) measurements for our sample yields the same empirical result.}
These authors interpreted this finding as evidence for a smaller wind opening angle at higher gas velocities.
The rise in the median $\Delta v_1$ value toward lower inclinations shown in Figure~\ref{fig.ewtot_isfr}b 
is also consistent with such an interpretation; i.e., it 
may be driven by the detection of higher-velocity flows as galaxies are viewed closer to face-on.  This rise may alternatively result from
an increase in the velocity dispersion of interstellar gas clouds along the line-of-sight
 as galaxies are viewed closer to edge-on (although this is not the most likely explanation given our discussion below).  In any case, the scatter in $\Delta v_1$ is large, such that the difference in the median values at high and low inclinations is not statistically significant.  Given that we also find no evidence for a correlation between $i$ and $\Delta v_\mathrm{max}$ or $\rm EW_{flow}$, we conclude that 
while galaxy orientation plays a crucial role in whether or not a wind is detected in a given system, the strength and significance
of its relationship with
 the maximum measured wind velocities remains to be confirmed.  

Figure~\ref{fig.ewtot_isfr}d shows the distribution of SFR as a function of inclination for the disk-like and compact systems in our sample.  Galaxies with detected flows are shown in blue or cyan, and galaxies with `systemic' absorption are marked in black.  The point size is scaled linearly with $\rm EW_{2796}$.  We note a dearth of galaxies with $\log \mathrm{SFR}~ [M_{\odot}~\rm yr^{-1}]\sim 0.0 - 0.5$ at $i > 60^{\circ}$, which may result from our magnitude-limited sample selection: edge-on systems generally suffer significant extinction, and thus may require slightly higher SFRs to satisfy our selection criteria.  However, a K-S test indicates that the distribution of SFRs for galaxies with $i > 50^{\circ}$ is drawn from the same parent distribution as SFRs for galaxies with $30^{\circ} < i < 50^{\circ}$ with 25\% probability.  This suggests that our sample is not significantly biased against objects with high inclinations; however, a larger sample size over the full range of inclinations would be valuable for confirming these results.
Figure~\ref{fig.ewtot_isfr}d also shows that $\rm EW_{2796}$ not correlated with inclination for `systemic' galaxies, yielding $r_\mathrm{S} = -0.33$ at $1.5\sigma$ significance.  The line-of-sight velocity width of absorbing interstellar gas embedded in a rotating disk is maximized in edge-on systems; however, the small $\rm EW_{2796}$ values measured in highly-inclined galaxies \emph{without} detected winds suggests that the magnitude of $\rm EW_{2796}$ is not significantly affected by gaseous disk rotation.

\section{Discussion}\label{sec.discussion}

\subsection{The Ubiquity of Outflows on the Star-Forming Sequence at $z \sim 0.5$}\label{sec.ubiquity}

Empirical studies of star-formation driven outflows have traditionally focused on the physics of winds around systems with extreme characteristics; i.e., exceptionally high spatial concentrations of star formation ($\Sigma_\mathrm{SFR} \gtrsim 0.1~M_{\odot}~\rm yr^{-1}~kpc^{-2}$; \citealt{SchwartzMartin2004}; \citealt{Tremonti2007}; \citealt{Diamond-Stanic2012}) and/or ongoing merger activity \citep[e.g.,][]{Heckman2000, Martin2005}.  Recently, however, deep surveys have begun to explore 
the wind properties of more typical star-forming systems \citep[e.g.,][]{Sato2009}.
\citet{Martin2012} obtained sensitive observations of $\sim200$ star-forming galaxies with masses $9.4 \lesssim \log M_*/M_{\odot} \lesssim 11.5$ at $z\sim1$ (a sample with very similar properties to the galaxies discussed here), detecting winds in 
$\sim45\%$ of their individual galaxy spectra.  We note that the difference between their lower outflow detection rate and the $66\pm5$\% detection rate discussed in \S\ref{sec.fitresults} 
 likely arises from a combination of factors: e.g., the somewhat higher velocity resolution of our spectroscopy, and/or our use of the \ion{Mg}{2} transition in addition to near-UV \ion{Fe}{2} absorption profiles in our search for winds.  
The \ion{Mg}{2} doublet is a more sensitive transition than any of the \ion{Fe}{2} transitions used, but is also 
affected by resonantly-scattered wind emission which we expect to fill in the absorption profiles redward of $v=0\mkms$ as discussed in \S\ref{sec.bayes_constraints}.
Because it pushes to fainter magnitudes (and to higher redshift), the \citet{Martin2012} sample may additionally include a higher fraction of edge-on systems, making winds more difficult to detect.  
These differences aside, however, they interpreted the apparent lack of dependence of outflow detection rate on the SFR or $M_*$ of their galaxy sample as suggestive that winds are a common characteristic of typical star-forming galaxies at $z\sim1$.

In the local universe, \citet{ChenNaI2010} have searched for winds from galaxies having $10.4 < \log M_*/M_{\odot} \lesssim 11.0$ and $-1.6 \lesssim \log \Sigma_\mathrm{SFR}/ M_{\odot}~\rm yr^{-1}~kpc^{-2} \lesssim -0.4$.  They detected outflows traced by \ion{Na}{1} absorption in coadded spectra to velocities of $120 - 150\mkms$ over their entire galaxy sample, demonstrating for the first time that winds are common among the majority of nearby massive, star-forming galaxies.  In addition, several studies have uncovered evidence for cool gas outflow from the nuclear region of the Milky Way, in spite of its low present-day SFR ($\sim 1~M_{\odot}~\rm yr^{-1}$; \citealt{Robitaille2010}).   These include the detection of dust emission with a bipolar structure at the Galactic center \citep{BHCohen2003} and the presence of high-velocity clouds of metal-enriched gas 
directly above and below the Galactic center \citep{Keeney2006}.  

Furthermore, the analytical arguments of \citet{Murray2011} predict that winds will arise from galaxies having a broad range of SFR and $M_*$.  These authors posit that winds are initially launched by radiation pressure generated by massive star clusters.  This pressure disrupts the gas and dust in the surrounding giant molecular clouds (GMCs) when the stellar mass in a given cluster approaches a critical fraction of the total associated GMC mass \citep{Murray2010}.    The velocity with which a cluster ejects the cloud remnants 
 is similar to the cluster's escape velocity.  Arguing that the typical GMC and star cluster masses scale with both the gas surface density and the size of galactic disks, and applying the Kennicutt-Schmidt law to relate gas surface density to $\Sigma_\mathrm{SFR}$, they determine a `critical' SFR for launching a large-scale galactic wind:

\begin{eqnarray}
	\mathrm{SFR}^\mathrm{crit} \approx 5 \left(\frac{v_c}{200\mkms}\right)^{5/2} M_{\odot}~\mathrm{yr}^{-1},
\end{eqnarray}

\noindent where $v_c$ is the circular velocity at the edge of the galactic disk.  We may express this threshold as a function of $M_*$ by invoking the stellar-mass Tully-Fisher relation measured from rotation curve studies.  
Adopting the $z\sim0$ relation from \citet{BelldeJong2001}, $\log M_*/M_{\odot} = 0.52 + 4.49 \log V_\mathrm{max}$, where $V_\mathrm{max}$ is the maximum measured rotation velocity in $\mkms$, we rewrite $\rm SFR^{crit}$ as 

\begin{eqnarray}
	\mathrm{SFR}^\mathrm{crit} \approx 1.7 \left(\frac{M_*}{10^{10} M_{\odot}}\right)^{0.56}  M_{\odot}~\mathrm{yr}^{-1}
\end{eqnarray}

\noindent assuming $v_c \approx V_\mathrm{max}$.  \citet{Conselice2005} demonstrated that there is no significant evolution in the slope and offset of the Tully-Fisher relation out to $z>0.7$, allowing us to compare this threshold to the SFRs of our galaxies over the full redshift range sampled.  As shown in Figure~\ref{fig.ewMg_mstarsfr} (purple dotted line), $\rm SFR^{crit}$ lies below the SFRs of the overwhelming majority of star-forming systems at $z\sim0.5$, suggesting that nearly all galaxies on the star-forming sequence at intermediate redshift generate the radiation pressure required to launch a wind.  

Our measurements are in broad agreement with this prediction.  As described in \S\ref{sec.detrate}, our wind detection rate does not vary significantly with the $M_*$ or SFR of our sample galaxies.  Furthermore, we detect winds over the full range in SFR and $M_*$ spanned by the star-forming sequence at $z\sim0.5$, down to a stellar mass limit of $\log M_*/M_{\odot} \gtrsim 9.5$.  
We find a comparatively strong dependence of wind detection rate on galaxy orientation (discussed in \S\ref{sec.detrate_orientation}), indicating that the detected outflows typically have a collimated morphology that covers much of the galactic disk (i.e., with a large opening angle assuming a bipolar flow;
see \S\ref{sec.opening} below).  Taken together, these two results strongly suggest that collimated winds are indeed ubiquitous among $\log M_*/M_{\odot} \gtrsim 9.5$ star-forming galaxies at $z\gtrsim0.3$.  Systems with low inclinations in which no winds are detected may simply host winds which cover a small surface area (or have a smaller opening angle than is typical),
or may drive winds with relatively modest velocities or absorption strengths.  
We note that the enhancement of our wind detection rate
 in more face-on galaxies is consistent with the results of \citet{Kornei2012}, who detect higher velocity material in coadded spectra of galaxies having lower inclinations.  And at $z\sim0$, \citet{ChenNaI2010} have likewise demonstrated that the outflow velocity measured in the \ion{Na}{1} transition is higher in coadded spectra of more face-on systems.  

\subsection{The Significance of $\Sigma_\mathrm{SFR}$}\label{sec.significance_sigma}

The conclusion that every star-forming galaxy having $\log M_*/M_{\odot} \gtrsim 9.5$ at $z\gtrsim0.3$ drives a collimated outflow
additionally implies that $\Sigma_\mathrm{SFR}$ is not a crucial parameter in determining whether a galaxy launches a wind.   We detect winds in systems with $\Sigma_\mathrm{SFR}$ as low as $\sim 0.03~M_{\odot}~\rm yr^{-1}~kpc^{-2}$; i.e., well below the canonical `threshold' for driving winds \citep[$0.1 ~M_{\odot}~\rm yr^{-1}~kpc^{-2}$;][]{Heckman2002}.  We do find an enhanced wind detection rate ($84\pm7\%$) in galaxies with $\Sigma_\mathrm{SFR} \gtrsim 0.5~M_{\odot}~\rm yr^{-1}~kpc^{-2}$ (\S\ref{sec.detrate}), with only $63\pm7\%$ of galaxies having $0.05~M_{\odot}~\rm yr^{-1}~kpc^{-2} \lesssim \Sigma_\mathrm{SFR} \lesssim 0.5~M_{\odot}~\rm yr^{-1}~kpc^{-2}$ driving detected winds.   
Moreover, this finding is in qualitative agreement with \citet{Kornei2012}, who found an outflow detection rate of  $26\pm8\%$ for galaxies having low $\Sigma_\mathrm{SFR}$ ($\Sigma_\mathrm{SFR}\sim 0.18~M_{\odot}~\rm yr^{-1}~kpc^{-2}$) and a rate nearly twice as high ($48\pm9\%$) for objects having $\Sigma_\mathrm{SFR}\sim 0.93~M_{\odot}~\rm yr^{-1}~kpc^{-2}$.  
The differences in the wind detection rates measured in \citet{Kornei2012} and in the present study
likely arise from differences in the methods used to detect winds and measure $\Sigma_\mathrm{SFR}$.
Nevertheless, the consistently higher detection rates at high $\Sigma_\mathrm{SFR}$ suggest either that very high surface densities of star formation activity serve to marginally increase the probability of wind formation, or that larger $\Sigma_\mathrm{SFR}$ is associated with larger wind opening angles.  
While the results presented in Figure~\ref{fig.morph_hist}c must be interpreted with caution due to our small sample size, they appear to support the latter scenario over the former, lending further credence to our claim that wind launch occurs independently of $\Sigma_\mathrm{SFR}$, and indeed occurs in all star-forming galaxies at $z\sim0.5$.  

Still consistent with this new physical picture 
(i.e., that there is no `threshold' $\Sigma_\mathrm{SFR}$ that must be satisfied before a wind may be launched)
is the idea that higher surface densities of star formation activity may indeed enable the lofting of more material to higher velocities.
While our measurements of $\rm EW_{flow}$(\ion{Mg}{2}) are not correlated with $\Sigma_\mathrm{SFR}$, 
the rise in the maximum $\rm EW_{flow}$(\ion{Mg}{2}) values measured with increasing $\Sigma_\mathrm{SFR}$ evident in Figures~\ref{fig.galprop_ew}b and \ref{fig.galprop_ew}e as well as the weak correlation between $\rm EW_{flow}$(\ion{Fe}{2}) and $\Sigma_\mathrm{SFR}$ (with $2.8\sigma$ significance) 
 are both consistent with this hypothesis.
\citet{Kornei2012} likewise detect higher-velocity absorbing material toward galaxies with higher $\Sigma_\mathrm{SFR}$.
Indeed, the first evidence for such a relationship was presented in \citet{ChenNaI2010}, who reported a positive correlation between $\Sigma_\mathrm{SFR}$ and the EW of \ion{Na}{1} absorption tracing outflows.  
However, we also find that $\rm EW_{flow}$ is correlated with total SFR at comparable or higher significance (i.e., at $3.5\sigma$ significance for $\rm EW_{flow}$(\ion{Mg}{2})), suggesting that the weaker correlations between $\rm EW_{flow}$ and $\Sigma_\mathrm{SFR}$ are in fact driven by the former, and hence that absolute SFR has a more fundamental physical link to wind velocities and masses.  
 
We further caution that in spite of the high level of detail provided by the deep \emph{HST}/ACS imaging used in this study, these data are nevertheless a blunt tool with which to calculate the surface density of star formation activity in distant galaxies.  As noted in \citet{Meurer1997}, the distribution of massive, ionizing stars is not directly observable even in nearby galaxies, since ionizing emission does not easily escape from a host galaxy's ISM.  Local studies instead typically use H$\alpha$ or near-UV emission to trace the location of young stars.  Our analysis at $z>0.3$ relies on the spatial distribution of $\lambda_\mathrm{rest} \sim 2400 - 4550$ \AA\ emission to trace star formation activity, and crudely assumes that the `deprojected' area covered by this emission is representative of the surface area of star-forming regions.  \citet{Kornei2012}, in characterizing the spatial distribution of intense star formation activity in a similar sample of galaxies at $z\sim1$ using \emph{HST}/ACS imaging, estimated that the surface area of galaxy pixels having $\Sigma_\mathrm{SFR} > 0.1~M_{\odot}~\rm yr^{-1}~kpc^{-2}$ is on average $3.7$ times smaller than the area implied by the $V_{606}$-band Petrosian radius, demonstrating that `global'  size measurements can dramatically overestimate the area over which star formation occurs.  Starbursts may, however, have size scales as small as $\sim 0.1$ kpc \citep{Meurer1997}, which corresponds to $0.016\arcsec$ at $z\sim0.5$, or less than one \emph{HST}/ACS pixel width.  Such structure cannot be resolved in distant samples, fundamentally restricting even the most detailed analyses to estimates of lower limits on $\Sigma_\mathrm{SFR}$.  
Furthermore, the detection of outflows in this and similar studies is incomplete, as it is dependent on a variety of factors (e.g., the spectral S/N and resolution, the specific emission or absorption transition being used, and the strength of `interstellar' absorption).
These deficiencies 
make constraints on a `threshold' $\Sigma_\mathrm{SFR}$ and the fundamental relationship between $\Sigma_\mathrm{SFR}$ and the speed and mass of wind material particularly challenging.



 \subsection{Wind Morphology}\label{sec.opening}
The collimation of outflows is a basic prediction of wind models in which energy from supernovae is injected into a plane-parallel gas distribution.  Because the density of the ambient ISM in a galactic disk has a large gradient along the minor axis (at $\phi = 0^{\circ}$), a wind-driven bubble tends to become elongated in the same direction \citep{TomisakaIkeuchi1988,Heckman1990}.   
Imaging of emission from cold and shock-heated gas in winds has provided unequivocal evidence for a collimated (and specifically bipolar) outflow morphology around starburst galaxies in the local universe \citep[e.g.,][]{Walter2002,Westmoquette2008,StricklandHeckman2009}, and the detection of higher-velocity \ion{Na}{1} absorption toward more face-on SDSS galaxies at $z\sim0.1$ is fully consistent with this picture \citep{ChenNaI2010}.  In more distant galaxies, for which emission from outflows is typically prohibitively faint, outflow morphology is not yet well understood.  However, the enhanced \ion{Mg}{2} absorption detected toward background QSOs located along the minor axis of absorber host galaxies at $z \lesssim 1$ is suggestive of wind collimation \citep{Bordoloi2011,Bouche2012,Kacprzak2012}.  Further evidence for anisotropic outflows is provided by \citet{Kornei2012}, who measured higher outflow velocities toward $z\sim 1$ galaxies having $i < 45^{\circ}$.  
Moreover, \citet{Martin2012} interpreted an increase in wind detection rate with a decreasing `threshold' velocity for wind detection as evidence for an increased wind opening angle at lower gas outflow velocities.
 
The analysis presented in \S\ref{sec.detrate_orientation} provides additional, strong evidence for the ubiquity of a collimated outflow morphology around disk-like galaxies at $z\sim0.5$.  The consistently high wind detection rates for galaxies having $i \lesssim 50^{\circ}$ in both low- and high-$\Sigma_\mathrm{SFR}$ subsamples, 
combined with the 
lack of a significant  variation in detection rate with $M_*$ and SFR, implies that outflows, while active over the entire star-forming galaxy population, do not generally occur along the disk plane.   In particular, only 4 out of 14 disk-like galaxies in our sample having $i > 60^{\circ}$ exhibit outflows.  
Although we cannot constrain the detailed shape of these flows with our data, we infer that they take a roughly  biconical form, as this morphology is predicted by the simulations mentioned above, and has been observed around numerous local starburst galaxies  \citep[e.g.,][]{Walter2002,VeilleuxRupke2002,Cecil2001}.

However, in spite of this clear support for biconical flows, we do not find conclusive evidence for a relationship between outflow opening angle and wind speed as suggested by \citet{Martin2012}.  
Indeed, because the density gradient of a gaseous disk is the largest along the minor axis and decreases toward the disk plane, the highest gas velocities are predicted to occur at $\phi \sim 0^{\circ}$ \citep{TomisakaIkeuchi1988,DeYoungHeckman1994}.  
We do identify a weak correlation between galaxy orientation and velocities measured from one-component model fits to the \ion{Fe}{2} transition (Figure~\ref{fig.ewtot_isfr}b), which is suggestive of slower wind speeds at larger viewing angles ($\phi$).  This correlation is also stronger than any potential correlations between this velocity and $M_*$ or SFR (which we find are not statistically significant).
As shown in \S\ref{sec.velew_galorient}, however, we find no correlation between galaxy orientation and maximum wind speed ($\Delta v_\mathrm{max}$) or $\rm EW_{flow}$.  
In other words, as long as a wind is sufficiently powerful  along a given sightline to meet our detection criteria, we are equally likely to measure high $\rm EW_{flow}$ or $\Delta v_\mathrm{max}$ when viewing a galaxy face-on or inclined by $\sim45^{\circ}$.   
We therefore cannot make a conclusive statement on the relationship between wind speed and opening angle, and instead simply identify  $M_*$ as the galaxy property having the strongest relationship with maximum wind velocity (\S\ref{sec.velew}).




Independent of our analysis of wind detection rate as a function of galaxy inclination, our overall wind detection rate can be used to infer a `characteristic' wind opening angle for the full galaxy sample, assuming that every galaxy has the same symmetric and biconical outflow morphology.  The solid angle subtended by two cones, each with a half-opening angle of $\theta_0$, is $\Omega = 4\pi (1 - \cos \theta_0)$.  The corresponding frequency of wind detection among a galaxy sample with randomly-distributed orientations is $\Omega/4\pi$.  
Excluding the morphologically disturbed galaxies in our sample, as they are less likely to exhibit the assumed outflow symmetry, 
65\% of the remaining (compact and spiral) galaxies   
drive detected winds.  This implies $\theta_0 \approx 70^{\circ}$, or a full opening angle of $\sim 140^{\circ}$.  
Figure~\ref{fig.morph_hist}b, on the other hand, 
clearly demonstrates that the vast majority of galaxies in which winds are not detected have wind opening angles $\theta_0 < 70^{\circ}$ (under the assumption of ubiquitous outflows).  
Furthermore, one galaxy in our sample, EGS12008444 at $z= 1.28915$, has an inclination of $73^{\circ}$ and drives a detected wind, suggesting that the outflow is close to isotropic in this case.   
The assumption of a single `characteristic' opening angle is therefore not necessarily valid when describing wind properties over a large galaxy population.  

\begin{figure}
\includegraphics[angle=90,width=\columnwidth]{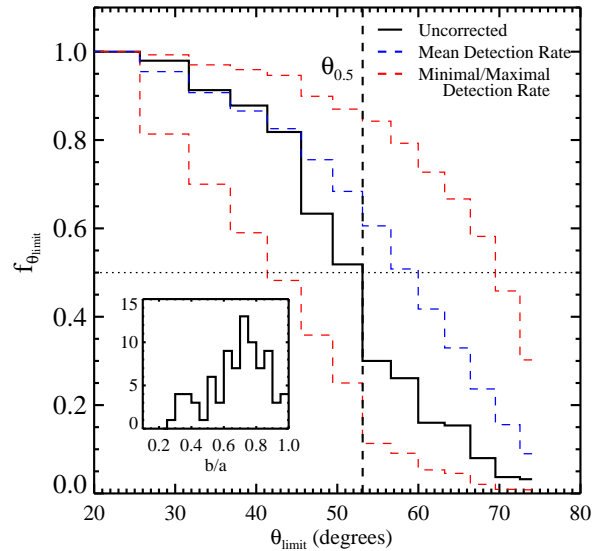}
\caption[]{The fraction of the galaxy sample ($f_{\theta_\mathrm{limit}}$) having a half wind opening angle ($\theta_0$) greater than $\theta_\mathrm{limit}$ (black).  Half of the sample has $\theta_0 > 53^{\circ}$, as indicated by the vertical dashed line.  The same measurement, corrected 
assuming that a complete sample would have a flat axis ratio distribution
(i.e., consistent with a sample of galaxies with random orientations), is shown with colored dashed histograms.  The red histograms assume that galaxies `missed' by our selection at low and high inclinations would either all drive detected winds (upper histogram) or drive no detected winds (lower histogram).  The blue histogram assumes that galaxies `missed' by our selection drive detected winds at the mean measured detection rate.  The inset plot shows the axis ratio distribution of our observed sample of compact and spiral galaxies.
     \label{fig.flim_incl}}
\end{figure}

Alternatively, we calculate the fraction ($f_{\theta_\mathrm{limit}}$) of our sample having $\theta_0$ greater than some $\theta_\mathrm{limit}$, 
\begin{eqnarray}
 	f_{\theta_\mathrm{limit}} = \frac{N_\mathrm{wind} (i > \theta_\mathrm{limit})}{N_\mathrm{wind} (i > \theta_\mathrm{limit}) + N_\mathrm{no~wind} (i \le \theta_\mathrm{limit})},
\end{eqnarray}
where $N_\mathrm{wind} (i > \theta_\mathrm{limit})$ is the number of galaxies driving detected winds and having $i > \theta_\mathrm{limit}$, 
and where $N_\mathrm{no~wind} (i \le \theta_\mathrm{limit})$ is the number of galaxies without detected winds and having $i \le \theta_\mathrm{limit}$.
We may then determine $\theta_\mathrm{limit}$ when $f_{\theta_\mathrm{limit}} = 0.5$, or $\theta_{0.5}$ (see Figure~\ref{fig.flim_incl}),   
which can be understood as the minimum $\theta_0$ exhibited by at least half of the sample galaxies.  
For the inclination distributions of galaxies with and without detected winds in our sample, $\theta_{0.5} \approx 53^{\circ}$.   
However, the quantity $\theta_{0.5}$ may be used as an indicator of the typical wind opening angle of the overall galaxy population only if it is measured from a set of galaxies whose distribution of orientations is random.  
Such a sample 
has an axis ratio ($b/a$) distribution which is approximately flat \citep[e.g.,][]{Yip2010}.
The inset plot in Figure~\ref{fig.flim_incl} therefore demonstrates that our sample is incomplete 
at both low (edge-on) and high (face-on) values of $b/a$.
To understand the effect of this incompleteness on our measurement of $\theta_{0.5}$,
 we generate a set of `supplementary' galaxies with inclinations chosen such that the distribution of axis ratios for the combined observed and supplementary samples is flat.
 We then make a variety of assumptions about the frequency of wind detection among this supplementary sample.  If we assume that none of these galaxies yield wind detections, we obtain the lower red dashed histogram for $f_{\theta_\mathrm{limit}}$. 
If we instead assume that all supplementary galaxies drive detectable winds, we obtain the upper red histogram.  These distributions suggest that under the most conservative assumptions, $\theta_{0.5}$ must lie between $42^{\circ}$ and $70^{\circ}$.  If we instead assume that the wind detection rates for the supplementary sample are consistent with the rates we measure (i.e., 50\% if $i > 45^{\circ}$ and 64\% if $i<45^{\circ}$), we obtain the blue dashed histogram for $f_{\theta_\mathrm{limit}}$, and find $\theta_{0.5} \sim 60^{\circ}$.  All together, this analysis suggests that 50\% of the star-forming galaxy population at $z\sim0.5$ has $\theta_0$ greater than $\sim42^{\circ} - 60^{\circ}$.


These opening angles are  consistent with the wind opening angle measured in the molecular gas outflow from M82 ($\theta_0 = 55^{\circ}$; \citealt{Walter2002}), the opening angle implied by the detection of \ion{Na}{1}-absorbing outflows from SDSS star-forming galaxies ($\theta_0 \sim 60^{\circ}$; \citealt{ChenNaI2010}), and
 the opening angles inferred from analysis of the azimuthal angle dependence of \ion{Mg}{2} absorption toward background QSOs ($\theta_0 \approx 50^{\circ}$; \citealt{Kacprzak2012}).  \citet{Rupke2005b} inferred somewhat smaller opening angles for \ion{Na}{1}-absorbing winds from infrared-selected starbursts and ULIRGs out to $z\sim0.5$;   however, they also noted a possible dependence of opening angle on the IR luminosity of their sample: galaxies with $10^{10} < L_\mathrm{IR} / L_{\odot} < 10^{12}$ exhibited $\theta_0 \sim 32^{\circ}$, while galaxies with $L_\mathrm{IR} / L_{\odot} > 10^{12}$ exhibited $\theta_0 \sim 46^{\circ}$. 
 These findings, together with those presented in \S\ref{sec.detrate_orientation} and Figure~\ref{fig.morph_hist}c, suggest that the distribution and strength of star formation activity have a significant effect on the morphology of outflows.  We await future studies with larger samples to confirm these trends and further disentangle the concurrent effects of stellar morphology, star formation rate, and the spatial distribution of star formation activity on the gas kinematics and opening angle of winds. 




\subsection{The Mass and Spatial Distribution of Outflows}\label{sec.massoutflowrate}

In this section, we discuss the constraints our measurements place on the amount of mass carried by the observed outflows in our sample.  If we assume that the gas 
is arranged in a smooth, continuous flow extending from the center of its host galaxy to a distance $D$, and that it travels at a constant speed $v$, the mass outflow rate can be written:
\begin{eqnarray}
	dM/dt = \frac{1}{3} \mu m_p N_\mathrm{flow} \mathrm{(H)} A_\mathrm{flow} v / D.
\end{eqnarray}

\noindent Here, $\mu m_p$ is the mean atomic weight, $N_\mathrm{flow} \rm (H)$ is the hydrogen column density of the flow, and $A_\mathrm{flow}$ is the projected surface area occupied by clouds in the wind.  In the following, we discuss our constraints on $N_\mathrm{flow}\rm (H)$, $A_\mathrm{flow}$, and $D$ in turn, and present the resulting constraints 
on $dM/dt$ in the final subsection.

\subsubsection{Outflow Column Density}

As shown in Figures~\ref{fig.ewNCfbD}a and \ref{fig.ewNCfbD}d, our two-component model fits to those spectra which exhibit outflows generally require $N_\mathrm{flow} > 10^{14.4}~\rm cm^{-2}$ for \ion{Mg}{2} and $N_\mathrm{flow} > 10^{14.5}~\rm cm^{-2}$ for \ion{Fe}{2}.  An estimate of the total hydrogen column in the flow requires knowledge of the ionization state and metallicity of the gas as well as the degree of dust depletion for these elements.  Because our data provide no constraints on any of these quantities, we adopt assumptions motivated by metallicity measurements in complementary samples and conservative conjectures on the physical conditions in the flow.   \citet{KK2004} presented gas-phase oxygen abundance measurements for a sample of galaxies at $0.3 < z < 1$ having $-18.5 > M_B > -21.5$.  For galaxies with $M_B < -19.5$, i.e., with luminosities similar to those in the present sample, nebular oxygen abundances fell in the range $8.5 \lesssim 12 + \log\rm(O/H) \lesssim 9$.  This implies that the ISM in our galaxy sample has abundances close to the solar value (8.7; \citealt{AllendePrieto2001}), with variations of $\sim 0.2-0.3$ dex.  Further, the ionization potentials of neutral Mg and Fe (7.6 eV and 7.9 eV) and  \ion{Mg}{2} and \ion{Fe}{2} (15.0 eV and 16.2 eV) are such that the singly-ionized state of both elements are likely dominant in photoionized gas having $T\sim10^4$ K \citep{Murray2007}.    In the case of Mg, the dominance of \ion{Mg}{2} persists as the ionization parameter surpasses $\log U > -2$ (where $U = n_{\gamma} / n_\mathrm{H}$), while the \ion{Fe}{2} fraction drops precipitously with the density of ionizing photons above this threshold \citep{Churchill2003,Narayanan2008}.  The assumption that both elements are fully singly-ionized (i.e., $\chi =  n(\mathrm{X}^+)/n(\mathrm{X}) = 1$) is therefore realistic in the case of Mg, and conservative in the case of Fe.  Finally, dust depletion factors 
($d(\mathrm{X})$) measured in the local Galactic ISM fall in the range $-(0.3 - 1.5)$ dex for Mg and $-(1.0 - 2.3)$ dex for Fe \citep{Jenkins2009}.  

Thus, adopting a solar abundance ratio ($\log \rm Mg/H = -4.42$; \citealt{SavageSembach1996}) and a dust depletion factor of $-0.5$ dex for Mg, our limit on $N_\mathrm{flow}$ implies 
\begin{eqnarray*}
N_\mathrm{flow} \mathrm{(H)} > \frac{10^{19.3}~\mathrm{cm}^{-2}}{\chi(\mathrm{MgII})} \frac{N_\mathrm{flow} (\mathrm{MgII})}{10^{14.4}~\mathrm{cm}^{-2}} \frac{10^{-4.42}}{10^{\log \mathrm{Mg/H}}} \frac{10^{-0.5}}{10^{d(\rm Mg)}}.
\end{eqnarray*}
Similarly, our limit on $N_\mathrm{flow}$ from \ion{Fe}{2}, assuming solar abundance ($\log \rm Fe/H = -4.49$) and a dust depletion factor of $-1.0$ dex, yields
\begin{eqnarray*}
N_\mathrm{flow}  (\mathrm{H}) > \frac{10^{20.0}~\mathrm{cm}^{-2}}{\chi(\mathrm{FeII})} \frac{N_\mathrm{flow} (\mathrm{FeII})}{10^{14.5}~\mathrm{cm}^{-2}} \frac{10^{-4.49}}{10^{\log \mathrm{Fe/H}}} \frac{10^{-1.0}}{10^{d(\rm Fe)}}.
\end{eqnarray*}
These values are close to the columns of neutral gas measured in galactic disks.  If we instead adopt the ionization correction and depletion factors assumed by \citet{Martin2012} for \ion{Fe}{2} ($\chi(\mathrm{FeII}) = 0.5$ and $d(\mathrm{Fe}) = -0.69$), we obtain a nearly equivalent constraint on $N_\mathrm{flow} \rm (H)$.
We emphasize that these limits are conservative not only by virtue of our assumed ionization correction, dust depletion, and abundance ratios, 
 but also because the
absorption in the `flow' component is almost always saturated.  These data are fully consistent with a flow column density that is a factor of 10 higher than the quoted limit.  

\subsubsection{Outflow Surface Area}\label{sec.aflow}

The projected surface area covered by the outflowing clouds, $A_\mathrm{flow}$, must be greater than or equal to the product $A_\mathrm{flow} \ge C_{f,\rm flow} \times \rm CONT_{UV}$, where $\rm CONT_{UV}$ is the surface area of continuum emission at $\lambda_\mathrm{rest} \sim 2600$ \AA\ or $\sim 2800$ \AA, and $C_{f, \rm flow}$ is the outflow covering fraction constrained by our two-component model fitting.    
$A_\mathrm{flow}$ is typically assumed to be approximately equal to the area inferred from adopting the half-light radius of the host galaxy measured from rest-frame near-UV or optical imaging \citep[e.g.,][]{Weiner2009,RubinTKRS2009,Martin2012}; however, this assumption may significantly overestimate the physical scale of the flow.  

The continuum of a single-burst stellar population is dominated by high-mass stars ($> 5~M_{\odot}$) in the near-UV \citep{Kennicutt1998}.  Integrating a Salpeter IMF in the mass range $> 5~M_{\odot}$, and assuming these stars have a lifetime of $\sim 100$ Myr, the number of such stars in a galaxy having SFR $\sim 10~M_{\odot}~\rm yr^{-1}$ is $\sim 10^7$.  If each of these stars has a radius $4R_{\odot}$, the total surface area of this population is $\sim 10^{-7}~\rm pc^{2}$.  This is the minimum $\rm CONT_{UV}$ consistent with our data, and $A_\mathrm{flow} \sim C_{f, \rm flow} \times \rm CONT_{UV}$ is appropriate if the outflow remains within the immediate vicinity of the stars/supernovae which drive it. 

Wishing to adopt  less fastidious limits, we appeal to observations of wind-driven bubbles detected in emission.  In the Milky Way, shells associated with young clusters have size scales of up to $\sim100$ pc traced by \ion{H}{1}, free-free, or $8~\mu$m PAH emission \citep{Heiles1979,RahmanMurray2010}.  The total number of these bubbles in our Galaxy is difficult to determine observationally; however, studies of such objects have sample sizes of $\sim50$.  The projected surface area of 50 spherical bubbles with radii $\sim100$ pc is $ \sim2~\rm kpc^{2}$.  
 If we instead adopt the median rest-frame near-UV semi-major axis of our sample as the appropriate length scale for UV emission (3.3 kpc), we find $\rm CONT_{UV} \sim 35~\rm kpc^{2}$.  
Our two-component model fitting procedure finds $C_{f, \rm flow} \gtrsim 0.5$ in $\sim75\%$ of cases in which $\rm EW_{flow}^{16\%} > 0.2$ \AA, with a median value of $\sim0.65$ for both the \ion{Mg}{2} and \ion{Fe}{2} transitions.  
Assuming $C_f \sim 0.65$, and that the flows recede from both sides of the galactic disks, 
these extremes  correspond to a range in $A_\mathrm{flow} \sim 2.6 -  45.4~\rm kpc^{2}$.  
Studies of winds traced by rest-frame near-UV absorption lines in galaxy spectroscopy such as this one cannot distinguish between these scenarios.  


 
 \subsubsection{The Spatial Extent of the Observed Absorbing Wind}\label{sec.spatialextent}
 Our spectroscopy is sensitive to absorbing gas at any location along the line of sight to the target galaxies, including material many tens to hundreds of kiloparsecs from the central light source.  We must therefore invoke measurements from complementary datasets in order to constrain the distance between the observed wind and its host.  \citet{Chen2010} has measured the EW of \ion{Mg}{2} absorption associated with galaxies at $0.1 < z < 0.5$ using spectroscopy of close projected background QSO sightlines.  The host galaxy sample, while at a lower average redshift than the present sample, has a comparable stellar mass range ($9.5 \lesssim \log M_* / M_{\odot} \lesssim 11.0$) and occupies the SFR-$M_*$ parameter space covered by the star-forming sequence at $z\sim0.1$ \citep{Chen2010b}.  
This analysis is sensitive to absorbers with $\rm EW_{2796} > 0.1$ \AA; i.e., systems with EWs as large as those shown in Figure~\ref{fig.galprop_ew} ($\gtrsim 1$ \AA) are well within the detection limit.  However, among 77 sightlines with impact parameters out to $\sim 170$ kpc, only 12 yielded absorbers having $\rm EW_{2796} > 1$ \AA, and these were detected only within $10~\rm kpc \lesssim \rho \lesssim 50~\rm kpc$ of the host galaxy.  Even within 50 kpc, a 1 \AA\ absorber was detected in only $\sim25$\% of the sample.  

Our analysis, on the other hand, finds $\rm EW_{flow}$ (\ion{Mg}{2}) $> 1$ \AA\ in 60 of our galaxies, or in 59\% of our sample spectra with coverage of \ion{Mg}{2}.  This suggests that the bulk of the gas observed in absorption `down-the-barrel' is not often detected in QSO-galaxy halo studies that probe impact parameters $\rho > 10$ kpc, and hence that this absorption occurs primarily at even smaller separations.  Studies searching for the galaxies associated with very strong \ion{Mg}{2} absorbers having $\rm EW_{2796} > 2$ \AA\  have identified potential counterparts at impact parameters $\rho \sim 10 - 60$ kpc \citep[e.g.,][]{Bouche2007,Nestor2011}; however, these surveys may miss the true galaxy counterparts if their emission is blended with the bright QSO emission or below the detection limit.  To constrain the frequency with which high-EW absorbers occur beyond 5-10 kpc, larger samples of halo absorption strength measurements for galaxies whose redshifts are known \emph{a priori} at impact parameters $< 50$ kpc are needed, preferably in combination with analysis of the galaxy disk orientation \citep[e.g.,][]{Bouche2012,Kacprzak2012}.  \\

 
Returning to our goal of estimating the mass flux in the observed wind, we may rewrite equation (5) as follows:
\begin{eqnarray*}
	dM/dt \approx 1~ M_{\odot}~ \mathrm{yr^{-1}} \frac{N_\mathrm{flow} \mathrm{(H)}}{10^{20} \mathrm{cm^{-2}}} \frac{A_\mathrm{flow}}{45~\mathrm{kpc^2}} \frac{v}{300~\mkms} \frac{5~\mathrm{kpc}}{D}
\end{eqnarray*}

 Here we have adopted a value of $A_\mathrm{flow}$ on the high end of the range discussed in \S\ref{sec.aflow} and a conservative value of $D$.  We note that the more standard assumption that the wind is distributed in a spherical shell having radius $R \approx D$ would result in $A_\mathrm{flow} \approx 4\pi R^2$, or 314 $\rm kpc^2$, and a mass flux of $\sim7~M_{\odot}~\rm yr^{-1}$ \citep[e.g.,][]{Weiner2009,RubinTKRS2009,Martin2012}.  
 Further, the recent detection of spatially-extended, scattered \ion{Mg}{2} emission from an outflow around a strongly star-forming galaxy at $z\sim0.94$ suggests a mass outflow rate as large as $330-500~M_{\odot}~\rm yr^{-1}$, assuming a very low fraction of Mg is singly-ionized \citep{Martin2013}.
 These disparities illustrate the importance of additional observational constraints on the morphology of winds from studies tracing the location of multiple gas phases in emission.  
 Nevertheless, our conservative estimate  
of the mass outflow rate is well below the SFR of most of our sample, and limits the mass loading factor for the wind, $\frac{dM/dt}{\rm SFR}$, to $\gtrsim 0.02 - 0.6$. 
 

\begin{figure}
\includegraphics[angle=90,width=\columnwidth]{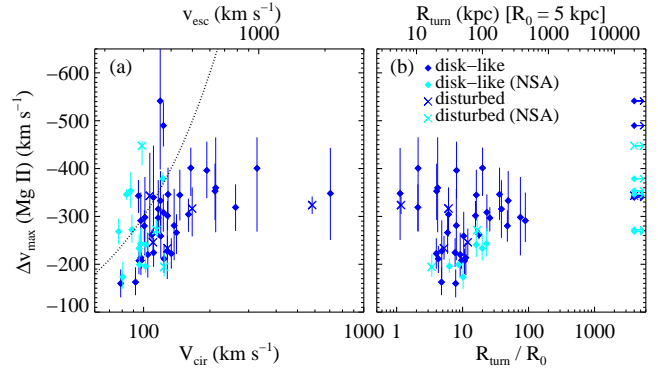}
\caption[]{\emph{(a)} $\Delta v_\mathrm{max}$ measured in the \ion{Mg}{2} transition vs.\ galaxy `circular velocity'.  Disk-like (spiral) and compact galaxies are marked with blue diamonds (for line profiles with significant absorption at systemic velocity) 
and cyan diamonds (for line profiles without systemic absorption, denoted `NSA').  Galaxies with disturbed morphologies are marked with blue crosses (with systemic absorption) 
and cyan crosses (without systemic absorption).  The galaxy escape velocity ($v_\mathrm{esc}$) is shown on the upper x-axis.  The dotted line shows a 1:1 relation between $\Delta v_\mathrm{max}$ and $v_\mathrm{esc}$.  
\emph{(b)} $\Delta v_\mathrm{max}$ measured in the \ion{Mg}{2} transition vs.\ $R_\mathrm{turn} / R_0$, the ratio of the turnaround radius to the initial radius of the wind material.  See \S\ref{sec.fate} for further description of this quantity.  Those systems with a wind speed exceeding the galaxy $v_\mathrm{esc}$ are indicated with arrows.
     \label{fig.vel_charvel}}%
\end{figure}

 \subsection{The Fate of Outflowing Gas and its Relation to the CGM}\label{sec.fate}
 
Our conclusion that nearly all massive, star-forming galaxies at $0.3\lesssim z \lesssim 1.4$ launch large-scale winds does not necessarily imply that such galaxies expel a significant amount of this cool material from their host halos.  
To compare the dynamics of our host galaxies with their
 cool gas kinematics, we first estimate the dark matter halo masses ($M_h$) of our sample by referring to the multiple-epoch halo abundance matching analysis of \citet{Moster2012}.  We invert the relation given in their Equation 2 to calculate $M_h$ for the redshift and $M_*$ of each galaxy, finding that the sample halos have masses falling within the range $11.1 \lesssim \log M_h/M_{\odot} \lesssim 14.0$.  We then use the relations between $M_h$ and halo maximum rotation velocity given in \citet{MB2004} to 
calculate $V_\mathrm{cir}$ for each galaxy, assuming a singular isothermal spherical halo density profile.  This is compared to $\Delta v_\mathrm{max}$ in Figure~\ref{fig.vel_charvel}a.  
To crudely estimate the galaxy escape velocities, we simplify the relation $v_\mathrm{esc}(R)^2 = 2 V^2_\mathrm{cir} \ln(1 + r_*/R)$, appropriate for a halo with radius $r_*$ and having a flat rotation curve \citep{BinneyTremaine1987}, to $v_\mathrm{esc} \approx 3 V_\mathrm{cir}$ \citep[e.g.,][]{Weiner2009}.  This is shown on the upper x-axis of the Figure, with 
the dotted line showing a 1:1 relation between $\Delta v_\mathrm{max}$ and $-1 \times v_\mathrm{esc}$.  
Only a handful ($\sim10$) of our galaxies have a measured $\Delta v_\mathrm{max}$ value which exceeds this escape velocity estimate; on the contrary, in most of the sample, $v_\mathrm{esc}$ is $\gtrsim 50 \mkms$ larger than $\Delta v_\mathrm{max}$.  
Our \ion{Mg}{2} line profiles indicate the presence of absorption at yet higher velocities than $\Delta v_\mathrm{max}$ (e.g., Figure~\ref{fig.allspecs1}), suggesting that a fraction of the material carried by the wind may indeed have the energy to escape these systems.  However, Figure~\ref{fig.vel_charvel} indicates that $\gtrsim 84\%$ of the optical depth of our fitted `flow' components (i.e., the fractional area of the `flow' components redward of $\Delta v_\mathrm{max}$) lies at velocities $< v_\mathrm{esc}$.  
We caution that our estimate of $v_\mathrm{esc}$ suffers numerous uncertainties, including the unconstrained halo potential profile, the unknown location of the wind in the halo, and the possible presence of ambient material in the halo.  However, this simplified analysis suggests that substantial wind material may escape from only the lowest-$M_*$ systems in our sample.


Furthermore, 
for the cases in which the gas remains bound to the halos, 
we may estimate the maximum distance that it can achieve in its path through its host environs.  We again approximate the total mass distribution in each galaxy halo as a singular isothermal sphere having $M(r) = \frac{2 V_\mathrm{cir}^2 r}{G}$.  
Neglecting the effects of drag from interaction of this gas with ambient material in the halo (the CGM), and assuming the gas is not further accelerated by ram or radiation pressure produced by hot stars or supernovae, 
we may write the equation of motion for the gas as follows:
\begin{eqnarray}
	\frac{dv}{dt} = - \frac{V^2_\mathrm{cir}}{r}.
\end{eqnarray}

Integrating this equation, and assuming the gas has initial velocity $v_0$ at distance $R_0$, we find that the `turnaround' radius for the gas cloud, $R_\mathrm{turn}$, is 
\begin{eqnarray}
	R_\mathrm{turn} = R_0 \exp \left (\frac{v_0^2}{2 V_\mathrm{cir}^2} \right ).
\end{eqnarray}
We set $v_0 = \Delta v_\mathrm{max}$ (\ion{Mg}{2}), and calculate the ratio $R_\mathrm{turn} / R_0$.  This quantity is compared with $\Delta v_\mathrm{max}$ in Figure~\ref{fig.vel_charvel}b.  

For nearly all of the systems in which the gas remains bound, it does not have sufficient momentum to travel beyond 50 kpc from its host, assuming that it is initially at a distance of 1 kpc.  
However, if the gas has an initial distance of at least 5 kpc (see top axis in Figure~\ref{fig.vel_charvel}b), it may reach $R_\mathrm{turn} > 50$ kpc in nearly half the systems shown, suggesting that while this gas may not have the energy to escape from its halo gravitational potential well altogether, it may indeed travel far enough to become a \emph{bona fide} component of the CGM.  
Given the growing body of evidence showing that the halos of these galaxies are likely already enriched with a significant amount of material \citep[e.g.,][]{Chen2010,Steidel2010,Cooksey2012,Rudie2012}, our neglect of drag forces is not well justified, and their inclusion may substantially modify the distances to which the observed gas is expected to travel.  Nevertheless, these results suggest that winds observed at $z\sim0.5$ may indeed contribute to the gas reservoir observed around bright galaxies 
at $z\sim0$ \citep[e.g.,][]{Tumlinson2011}.  Adopting our fiducial mass outflow rate of $\sim1~M_{\odot}~\rm yr^{-1}$ (\S\ref{sec.massoutflowrate}), the observed winds could add $\sim 10^{10}~M_{\odot}$ of cool material to the CGM if they were to persist at a constant rate between $z\sim1$ and today.  This is a substantial fraction of the total mass in cool photoionized gas detected around nearby $L^*$ galaxies  (Werk et al.\ 2012), suggesting that the observed outflows do indeed play a significant role in redistributing cool material from the ISM of the host galaxies to their surroundings.  Furthermore, because much of this material likely remains bound to its halo, it may eventually be re-accreted into its host galactic disk.  While evidence for such `recycling' remains sparse \citep{Rubin2012,Martin2012}, gas that has previously been ejected in a wind is predicted to provide the primary source of fuel for star formation in halos with masses above $10^{11.2}~M_{\odot}$ at $z\sim0$ \citep{Oppenheimer2010}.  The energetics of the cool outflows observed in our sample are fully consistent with this prediction.


\subsection{Outflow-Galaxy Scaling Laws}

Simulations of galaxy formation in a cosmological context have relied on feedback `recipes', or scaling laws which relate outflow kinematics and energetics to intrinsic galaxy properties (e.g., velocity dispersion, SFR),
 in order to sufficiently suppress star formation to reproduce the observed
 galaxy stellar mass function at $z\sim 0$ \citep{Oppenheimer2010,Guo2011}.   Such scalings, while motivated by simple physical arguments \citep[e.g.,][]{Martin2005, Murray2005}, have been based primarily on empirical constraints on the kinematics of outflows from local starbursts.  Even in nearby galaxies, furthermore, observational constraints on the amount of material in outflows have for the most part provided only lower limits \citep[e.g.,][]{Heckman1990,Martin2005,Rupke2005b}, although recently, valuable upper limits on the mass of outflows from blue cloud galaxies at $z\sim1$ have been reported by \citet{Martin2012}.  
 Finally, simple scalings (by construction) fail to capture the large dispersion in gas velocities and absorption strengths among galaxies with similar SFRs and/or masses observed in nearly all empirical studies of outflows, both in the local universe and in distant objects \citep[e.g.,][]{Martin2005,Rupke2005b,Kornei2012,Martin2012}.
 
The present study is no exception.  As described in \S\ref{sec.velew}, we find no correlation between outflow velocity and SFR or $\Sigma_\mathrm{SFR}$ over the ranges $1~M_{\odot}~\mathrm{yr}^{-1} \lesssim \mathrm{SFR} \lesssim 63~M_{\odot}~\mathrm{yr}^{-1}$  and $0.03~M_{\odot}~\mathrm{yr}^{-1}~\mathrm{kpc}^{-2} \lesssim \Sigma_\mathrm{SFR} \lesssim 3~M_{\odot}~\mathrm{yr}^{-1}~\mathrm{kpc}^{-2} $.  We instead measure a dispersion in $\Delta v_\mathrm{max}$ of $\sim80\mkms$ over the full range of SFR and $\Sigma_\mathrm{SFR}$ occupied by our sample.  
This is fully consistent with most previous work on outflows in galaxies with similar SFRs \citep[e.g.,][]{Rupke2005b,Grimes2009,ChenNaI2010,Heckman2011,Martin2012}, with two notable exceptions (discussed below).  
The concomitant, $3.4\sigma$-significant correlation between $\Delta v_\mathrm{max}$ and $M_*$ discussed in \S\ref{sec.velew} instead suggests that the maximum velocity achieved by cool wind material is most strongly dependent on galaxy dynamics, rather than star formation activity.  We speculate that this trend is driven in part by absorption from virialized CGM gas along the line of sight, the kinematics of which may be closely tied to the halo mass of the host galaxy.
We emphasize that this hypothesis is not substantiated by the analysis presented here, and must be tested via empirical constraints on the spatial extent of outflows and in cosmological `zoom-in' simulations of galaxy formation in individual halos \citep[e.g.,][]{Shen2012,Stinson2012}.
\citet{Martin2012} also found a positive correlation between $M_*$ and the maximum negative velocity at which the \ion{Mg}{2} absorption profile meets the continuum, though 
 the trend was only marginally significant ($1.8\sigma$).  We speculate that the weakness of this trend may be due to the resolution-dependent nature of their velocity measurement, combined with the broad range in spectral resolution of their sample (FWHM $\sim282-435\mkms$).  
\citet{Weiner2009} additionally provide some evidence for an increase in the maximum velocities achieved by cool gas with increasing $M_*$ \emph{and} SFR over the range $10~M_{\odot}~\mathrm{yr}^{-1} \lesssim \mathrm{SFR} \lesssim 65~M_{\odot}~\mathrm{yr}^{-1}$.  However, because the \citet{Weiner2009} sample is selected based on galaxy luminosity at 2800 \AA\ in the rest frame, the SFR and $M_*$ of the galaxies are highly covariant, making it difficult to disentangle the effects of these two quantities on wind velocity.  Our thorough sampling of the SFR-$M_*$ parameter space occupied by galaxies over a broad range in redshift (Figure~\ref{fig.ewMg_mstarsfr}) enables us to isolate these effects for the first time in a distant galaxy sample.  

We do, however, present evidence for a relationship between $\rm EW_\mathrm{flow}$ measured in \ion{Mg}{2} and host galaxy SFR, which are correlated at a $3.5\sigma$ significance level (Figure~\ref{fig.galprop_vel}).  As shown in Figure~\ref{fig.ewNCfbD}, higher $\rm EW_{flow}$(\ion{Mg}{2}) values are associated with higher limits on $N_\mathrm{flow}$ and larger flow velocity widths ($b_{D, \rm flow}$).  This association suggests 
that hosts with higher SFRs may launch more material into a wind, although the trend could also be driven primarily by a rise in $b_{D,\rm flow}$ with SFR.  In either case, this is indicative of a physical link between star formation activity and outflow kinematics and/or column density which was not evident from our analysis of outflow velocities.  \citet{Weiner2009} reported a similarly suggestive correlation between \ion{Mg}{2} outflow EW and SFR (and $M_*$) at $z\sim 1.4$, and \citet{ChenNaI2010} noted a strong correlation between wind absorption EW measured in \ion{Na}{1} and $\Sigma_\mathrm{SFR}$ in star-forming galaxies at $z\lesssim0.2$.
Finally, \citet{Kornei2012} presented evidence for an increase in outflow velocity measured from single-component model fits to \ion{Fe}{2} with $\Sigma_\mathrm{SFR}$ over the range $0.1~M_{\odot}~\mathrm{yr}^{-1}~\mathrm{kpc}^{-2} \lesssim \Sigma_\mathrm{SFR} \lesssim 1.5~M_{\odot}~\mathrm{yr}^{-1}~\mathrm{kpc}^{-2}$.  At face value, this appears inconsistent with our finding that $\Delta v_\mathrm{max}$ is independent of both SFR and $\Sigma_\mathrm{SFR}$.  However, 
this discrepancy may be due to the different techniques used to characterize outflow velocities: while our $\Delta v_1$ measurement is independent of $\Delta v_\mathrm{max}$ for the \ion{Fe}{2} transition (yielding a sum-squared difference of ranks only $0.4\sigma$ from the null hypothesis), $\Delta v_1$ is weakly correlated with $\rm EW_{flow}$ (at $1.5\sigma$ significance).  
Furthermore, we find a marginally significant increase in our detection rate of winds (as constrained by one-component model fits similar to those adopted by \citealt{Kornei2012}) with increasing $\Sigma_\mathrm{SFR}$, and suggest that this may be due to a larger wind opening angle in galaxies with higher surface densities of star formation activity.
Thus, we expect that the velocity measurement used in \citet{Kornei2012} is more closely tied to the velocity spread of the flow than the maximum velocity achieved by the wind, and suggest that their reported correlation is consistent with our finding that star formation activity and outflow incidence and absorption strength are physically associated.   

We reiterate, however, that a prevailing feature of all of these studies is the large variation in outflow properties measured for galaxies with similar intrinsic characteristics (SFR, $M_*$, $\Sigma_\mathrm{SFR}$, etc.).  While this variation may be due in part to viewing angle, our finding that maximum wind velocities are independent of galaxy inclination suggests that there are other factors at play.  The physics which relate the star formation activity in individual clusters to the properties of the extremely hot, hard X-ray emitting wind fluid produced from supernovae ejecta is only beginning to be understood \citep{StricklandHeckman2009}.  The action of this wind fluid on the surrounding cool gas, giving rise to the low-ionization absorption lines analyzed here, is yet more difficult to predict, and must depend on a great number of variables (e.g., the mass and physical conditions in the ambient interstellar material, the interaction of gas clouds accelerated by different star-forming regions, the further interaction of the wind with material in the extended galaxy halo).  Comparison of observed cool gas kinematics with results from hydrodynamic simulations of individual galaxy halos which implement a wide range of wind prescriptions is a promising avenue for constraining the physics relevant to cool outflows, or alternatively for understanding the limitations of these data in establishing such constraints.  Simulations tracking the enrichment of gaseous halos with hydrogen and metals have already been used to better understand feedback models via comparison of `observations' of the gas distribution along sightlines through the simulated halos with QSO absorption line studies \citep{Fumagalli2011,Shen2012,Stinson2012}.  Similar analysis of gas dynamics `down-the-barrel' toward simulated galaxies may be directly compared to the measurements described here, providing new insight into the origins, energetics, and fate of galactic outflows.   







\section{Summary and Conclusions}\label{sec.conclusions}

With the aim of characterizing the frequency, velocities, and absorption strength of cool, large-scale gaseous outflows, we have analyzed the absorption line profiles for the \ion{Mg}{2} $\lambda \lambda2796, 2803$ and \ion{Fe}{2} $\lambda \lambda 2586, 2600$ transitions in individual spectra of 105 star-forming galaxies at $0.3 < z < 1.4$ selected from the GOODS fields and the Extended Groth Strip.  This sample is magnitude-selected (to $B_\mathrm{AB} < 23$), and fully covers the star-forming sequence at $0.3 \lesssim z \lesssim 0.7$.  We identified outflows via the blueshift of these absorption transitions with respect to the galaxy systemic velocities, and modeled each line profile with two absorption `components' to constrain the maximum velocity ($\Delta v_\mathrm{max}$), equivalent width ($\rm EW$), column density, velocity width, and covering fraction of the flow.  Our spectroscopy and fitting procedure are sensitive to winds for galaxies driving saturated, outflowing gas clouds to velocities of $\sim250\mkms$ or greater.
Our analysis reveals the following:
\begin{itemize}
	\item We detect outflows in $66 \pm 5$\% of our sample.  These flows are detected over the full ranges of $M_*$, SFR, and SFR surface density ($\Sigma_\mathrm{SFR}$) occupied by the sample galaxies.  The detection rate is independent of host galaxy SFR and $M_*$, and is weakly dependent on the host SFR surface density ($\Sigma_\mathrm{SFR}$).  We find no evidence for a `threshold' $\Sigma_\mathrm{SFR}$ below which winds are not driven, detecting winds in galaxies with $\Sigma_\mathrm{SFR}$ as low as $0.03~M_{\odot}~\mathrm{yr}^{-1}~\mathrm{kpc}^{-2}$.   
	\item The detection rate of winds is strongly dependent on the host galaxy orientation.  That is, winds are detected in $\sim89\%$ of face-on galaxies (with inclinations $< 30^{\circ}$), but are detected in only $\sim45\%$ of edge-on galaxies 
(with inclinations $> 50^{\circ}$).  These results, in combination with the lack of a strong dependence of detection rate on intrinsic galaxy properties, suggest that biconical outflows are ubiquitous on the star-forming sequence at $z\sim0.5$.  Our analysis of outflow detection rates as a function of galaxy inclination indicates that over half of the sample galaxies have full wind cone opening angles of at least $2\theta_0 \sim100^{\circ}$ (with an isotropic wind having $2\theta_0 \sim 180^{\circ}$).  
Finally, we suggest that the weak dependence of outflow detection rate on $\Sigma_\mathrm{SFR}$ may be driven by larger cone opening angles in galaxies with higher $\Sigma_\mathrm{SFR}$.  
	\item We find that maximum wind velocity is most strongly correlated (at $3.4\sigma$ significance) with host galaxy stellar mass, rather than with the host's star-forming properties, which may suggest that the kinematics of cool wind material are dominated by the dynamics of the host galaxy halo.  The EW of the flow, however, is most significantly correlated with SFR (at the $3.5\sigma$ level), suggesting that hosts with higher SFR may launch more material into a wind and/or generate a larger velocity spread for the absorbing gas clouds. 
	\item Comparison of the large outflow EWs measured for those galaxies in our sample driving winds to the EWs of \ion{Mg}{2} absorbers observed along QSO sightlines through foreground galaxy halos suggests that the bulk of the outflowing gas is at most $\sim50$ kpc from the host galaxies.  However, we estimate that the gas velocities, while typically insufficient to enable escape from the gravitational potential of the host halos, could carry the gas to distances of $\gtrsim 100$ kpc.  Combined with a mass outflow rate of 
	at least $\sim1~M_{\odot}~\rm yr^{-1}$ implied by our absorption-line fitting results, these energetics suggest that the detected 
	 outflows are a viable source of cool material for replenishment of the circumgalactic medium observed around $z\sim0$ galaxies \citep{Tumlinson2011,Prochaska2011b}.  
	 
\end{itemize} 

Cool outflows play an integral role in redistributing the gas supply in star-forming galactic disks to 
their surrounding gaseous environments from $z\sim1$ to today.  While this redistribution 
is likely crucial to modulating the growth of galaxies, it is only one of many processes 
that affect the buildup of stellar mass and the accumulation of metals in the CGM.  
The rate of re-accretion, or recycling, of this material 
must play an equally significant part in regulating the formation of new stars \citep[e.g.,][]{Oppenheimer2010}, 
and yet the hydrodynamics of the interactions between outflowing, accreting, and 
ambient CGM material remain poorly constrained.  
Comparisons between hydrodynamic simulations 
which resolve these gas flows around individual galaxies \citep[e.g.,][]{Shen2012,Stinson2012} and the kinematics of cool material measured at 
higher spectral resolution will be critical to understanding this interplay.  By combining these comparisons with constraints on the energetics of outflows obtained from 
studies of resonantly-scattered and shock-excited wind emission \citep[e.g.,][]{Westmoquette2008,Rubin2011,Erb2012}, we will 
achieve new insight into the function and fate of star-formation driven outflows over cosmic time. \\





\acknowledgements

The authors are grateful for support for this project from NSF grants AST-0808133, AST-0507483, AST-0548180, and AST-1109288.
KHRR and JXP acknowledge support from the Alexander von Humboldt foundation in the form of the 
Humboldt Postdoctoral Fellowship and a visitor fellowship to MPIA, respectively.  
The Humboldt foundation is funded by the German Federal Ministry for Education and Research.  

Much of the data presented herein were obtained at the W.\ M.\ Keck Observatory, which
is operated as a scientific partnership among the California Institute of Technology, the
University of California, and the National Aeronautics and Space Administration. 
The Observatory was made possible by the generous financial support of the W.\ M.\ Keck Foundation.

The authors wish to thank the TKRS, AEGIS/DEEP2, and GOODS teams for making their data and catalogs publicly available, and Jennifer Lotz for 
providing SExtractor data products.  We thank Taro Sato and Phil Marshall for helpful discussions on the use of 
Bayesian statistical methods, Robert da Silva for discussions on the IMF and star formation in clusters, and Arjen van der Wel for 
help with the interpretation of morphological measurements.  
We wish to acknowledge Elisabete da Cunha for her generous assistance with the implementation of MAGPHYS SED fitting.
We thank Alison Coil and Girish Kulkarni for valuable comments on this analysis, 
and we are grateful to Neil Crighton and Joseph Hennawi for numerous discussions of these results and for their reading of 
an earlier version of this manuscript.




Finally, the authors wish to recognize and acknowledge the very significant cultural role and 
reverence that the summit of Mauna Kea has always had within the indigenous Hawaiian community. 
We are most fortunate to have the opportunity to conduct observations from this 
mountain.

\clearpage
{\footnotesize
\begin{longtable}{lcccccccccc}
\caption{Derived Galaxy Properties\label{tab.gals}}\\
\hline \hline  \\[-2ex]
Galaxy Name$^a$ & z & Morphology & $M_B$ & $U-B$ & $R$ & $i$ & SFR & $\log M_*/M_{\odot}$ & $\Sigma_\mathrm{SFR,global}$ & $\Sigma_\mathrm{SFR,max}$\\ [0.5ex]
 & & & mag & mag & kpc & degrees & $M_{\odot}~\rm yr^{-1}$ & & \multicolumn{2}{c}{$M_{\odot}~\rm yr^{-1}~kpc^{-2}$}\\
\hline \hline
\endfirsthead
\multicolumn{11}{c}{{\tablename} \thetable{} -- Continued} \\ [0.5ex]
\hline \hline
Galaxy Name$^a$ & z & Morphology & $M_B$ & $U-B$ & $R$ & $i$ & SFR & $\log M_*/M_{\odot}$ & $\Sigma_\mathrm{SFR,global}$ & $\Sigma_\mathrm{SFR,max}$\\ [0.5ex]
 & & & mag & mag & kpc & degrees & $M_{\odot}~\rm yr^{-1}$ & & \multicolumn{2}{c}{$M_{\odot}~\rm yr^{-1}~kpc^{-2}$}\\
\hline \\[-1.8ex]
\endhead
\\[-1.8ex] \hline \hline
\multicolumn{11}{c}{{Continued on Next Page\ldots}} \\
\endfoot
\\[-1.8ex] \hline \hline
\multicolumn{11}{l}{$^a$ Galaxies are named according to ID numbers listed in \citet[][TKRS]{Wirth2004}, the AEGIS survey \citep[][EGS]{Davis2007}, and}\\
\multicolumn{11}{l}{$~~$ \citet[]{Giavalisco2004}. IDs beginning with numbers mark objects from \citet[]{Giavalisco2004}, and in all other instances are}\\
\multicolumn{11}{l}{$~~$ preceded with ``J0332".}\\
\multicolumn{11}{l}{$^*$ FIREWORKS photometry is not available for this object.}\\
\multicolumn{11}{l}{$^{**}$ MOIRCS Deep Survey photometry is not available for this object.}\\
\endlastfoot
 EGS12004441 &  0.37830 & S & -19.58 &   0.59 &   3.15 &   56.0 & $   0.7^{+   0.4}_{  -0.1} $ & $  9.31^{+  0.05}_{ -0.06} $ & $ 0.014^{+ 0.007}_{-0.002} $ & $ 0.042^{+ 0.021}_{-0.006} $\\
 EGS12004471 &  0.41863 & S & -20.83 &   0.64 &   3.05 &   23.8 & $   2.6^{+   0.6}_{  -0.1} $ & $ 10.22^{+  0.03}_{ -0.11} $ & $ 0.083^{+ 0.019}_{-0.003} $ & $ 0.095^{+ 0.022}_{-0.003} $\\
 EGS12008005 &  0.76689 & S & -21.23 &   0.52 &   1.98 &   44.7 & $   8.1^{+   0.0}_{  -0.1} $ & $ 10.62^{+  0.13}_{ -0.01} $ & $ 0.608^{+ 0.000}_{-0.007} $ & $ 1.921^{+ 0.000}_{-0.022} $\\
 EGS12008157 &  1.27869 & merger & -22.48 &   0.52 &   2.54 &   62.7 & $ 116.7^{+   2.7}_{ -22.9} $ & $ 10.13^{+  0.08}_{ -0.06} $ & $ 3.248^{+ 0.076}_{-0.638} $ & $17.075^{+ 0.398}_{-3.355} $\\
 EGS12008197 &  0.98039 & S & -21.76 &   0.55 &   3.37 &   40.0 & $  41.4^{+  43.1}_{ -11.1} $ & $ 10.19^{+  0.10}_{ -0.04} $ & $ 0.887^{+ 0.924}_{-0.237} $ & $ 1.238^{+ 1.290}_{-0.331} $\\
 EGS12008247 &  0.48588 & S & -20.25 &   0.73 &   3.53 &   32.2 & $   1.0^{+   0.4}_{   0.0} $ & $  9.74^{+  0.03}_{  0.00} $ & $ 0.019^{+ 0.008}_{ 0.000} $ & $ 0.036^{+ 0.015}_{ 0.000} $\\
 EGS12008269 &  0.35343 & S & -20.70 &   0.63 &   2.30 &   22.9 & $   3.1^{+   0.6}_{  -0.4} $ & $ 10.03^{+  0.05}_{ -0.06} $ & $ 0.182^{+ 0.037}_{-0.025} $ & $ 0.381^{+ 0.077}_{-0.053} $\\
 EGS12008427 &  0.64346 & S & -21.21 &   0.67 &   4.09 &   64.9 & $   8.1^{+   2.2}_{  -1.0} $ & $ 10.32^{+  0.04}_{ -0.05} $ & $ 0.076^{+ 0.021}_{-0.009} $ & $ 0.393^{+ 0.107}_{-0.047} $\\
\end{longtable}}
\clearpage

\clearpage
{\footnotesize
\begin{landscape}
\begin{longtable}{llccccccccccccc}
\caption{\ion{Mg}{2} Absorption Line Measurements\label{tab.mglines}}\\
\hline \hline  \\[-2ex]
Galaxy Name$^a$ & Category & $\rm S/N $ & $\rm EW_{2796}$$^b$ & $\rm EW_{2803}$$^b$ & $P_\mathrm{out,1}$$^c$ & $\Delta v_1$$^d$ & $C_{f,1}$$^e$ & $P_\mathrm{out,2}$$^f$ & $\Delta v_\mathrm{max}$$^g$ & $C_{f,\rm flow}$$^h$ & $v_{\rm flow}$$^i$ & $\rm EW_{flow}$$^j$ &  $\log N_\mathrm{flow}$$^k$ & $\rm EW_{sys}$$^l$\\ [0.5ex]
 & & $\rm pix^{-1}$ & \AA & \AA &  & $\mkms$ & & & $\mkms$ & & $\mkms$ & \AA & $\rm cm^{-2}$ & \AA\\
\hline \hline
\endfirsthead
\multicolumn{15}{c}{{\tablename} \thetable{} -- Continued} \\ [0.5ex]
\hline \hline
Galaxy Name$^a$ & Category & $\rm S/N $ & $\rm EW_{2796}$$^b$ & $\rm EW_{2803}$$^b$ & $P_\mathrm{out,1}$$^c$ & $\Delta v_1$$^d$ & $C_{f,1}$$^e$ & $P_\mathrm{out,2}$$^f$ & $\Delta v_\mathrm{max}$$^g$ & $C_{f,\rm flow}$$^h$ & $v_{\rm flow} $$^i$ & $\rm EW_{flow}$$^j$ &  $\log N_\mathrm{flow}$$^k$ & $\rm EW_{sys}$$^l$\\ [0.5ex]
 & & $\rm pix^{-1}$ & \AA & \AA & & $\mkms$ & & & $\mkms$ & & $\mkms$ & \AA & $\rm cm^{-2}$ & \AA\\
\hline \\[-1.8ex]
\endhead
\\[-1.8ex] \hline \hline
\multicolumn{15}{c}{{Continued on Next Page\ldots}} \\
\endfoot
\\[-1.8ex] \hline \hline
\multicolumn{15}{l}{$^a$ Galaxies are named according to ID numbers listed in \citet[][TKRS]{Wirth2004}, the AEGIS survey \citep[][EGS]{Davis2007}, and}\\
\multicolumn{15}{l}{$~~$ \citet[]{Giavalisco2004}. IDs beginning with numbers mark objects from \citet[]{Giavalisco2004}, and in all other instances are preceded with ``J0332".}\\
\multicolumn{15}{l}{$^b$ In cases for which a line is not detected, the 3$\sigma$ EW upper limit is listed.}\\
\multicolumn{15}{l}{$^c$ Fraction of the marginalized PPDF for the one-component model having $v < 0\mkms$. }\\
\multicolumn{15}{l}{$^d$ Central velocity for the one-component model.}\\
\multicolumn{15}{l}{$^e$ Covering fraction for the one-component model.}\\
\multicolumn{15}{l}{$^f$ Fraction of the marginalized PPDF for the flow component in the two-component model having $v < 0\mkms$.}\\
\multicolumn{15}{l}{$^g$ Maximum outflow velocity, defined as $\Delta v_\mathrm{max} =  v_\mathrm{flow} - b_{D,\rm flow}/\sqrt{2}$. }\\
\multicolumn{15}{l}{$^h$ Covering fraction of the flow component in the two-component model.}\\
\multicolumn{15}{l}{$^i$ Central velocity of the flow component in the two-component model.}\\
\multicolumn{15}{l}{$^j$ Equivalent width of the flow component in the two-component model.}\\
\multicolumn{15}{l}{$^k$ Column density of the flow component in the two-component model.  Lower limits include 95\% of the marginalized PPDF.}\\
\multicolumn{15}{l}{$^l$ Equivalent width of the systemic component in the two-component model.}\\
\endlastfoot
 EGS12004441 & low S/N & $   4.4$ & $ <    1.0$ & $ <    1.0$ & \nodata & \nodata & \nodata & \nodata & \nodata & \nodata & \nodata & \nodata & \nodata & \nodata\\
 EGS12004471 & wind (Mg,Fe) & $  20.2$ & $   2.9\pm   0.1$ & $   2.7\pm   0.1$ & $0.968$ & $  -184^{+     9}_{    -8} $ & $0.89^{+0.07}_{-0.07} $ & $1.000$ & $  -261^{+     9}_{    -9} $ & $0.92^{+0.05}_{-0.08} $ & $  -229^{+     8}_{    -8} $ & $  2.16^{+  0.11}_{ -0.12} $ & $ >   15.4$ & $  0.82^{+  0.06}_{ -0.04} $\\
 EGS12008005 & systemic & $  12.7$ & $   2.2\pm   0.2$ & $   2.5\pm   0.2$ & $0.066$ & $    21^{+    13}_{   -14} $ & $0.60^{+0.07}_{-0.06} $ & \nodata & \nodata & \nodata & \nodata & \nodata & \nodata & \nodata\\
 EGS12008157 & systemic & $   9.1$ & $   2.2\pm   0.2$ & $   2.4\pm   0.2$ & $0.038$ & $    34^{+    10}_{   -13} $ & $0.92^{+0.06}_{-0.09} $ & \nodata & \nodata & \nodata & \nodata & \nodata & \nodata & \nodata\\
 EGS12008197 & wind (Mg,Fe) & $   4.8$ & $   4.0\pm   0.3$ & $   3.0\pm   0.3$ & $0.997$ & $  -129^{+    24}_{   -26} $ & $0.66^{+0.10}_{-0.08} $ & $1.000$ & $  -332^{+    57}_{   -61} $ & $0.60^{+0.18}_{-0.14} $ & $  -232^{+    58}_{   -55} $ & $  2.36^{+  0.75}_{ -0.65} $ & $ >   14.0$ & $  1.51^{+  0.39}_{ -0.33} $\\
 EGS12008247 & low S/N & $   3.6$ & $ <    1.7$ & $ <    1.8$ & \nodata & \nodata & \nodata & \nodata & \nodata & \nodata & \nodata & \nodata & \nodata & \nodata\\
 EGS12008427 & wind (Mg,Fe) & $  10.7$ & $   1.8\pm   0.2$ & $   2.4\pm   0.2$ & $0.967$ & $  -152^{+    34}_{   -83} $ & $0.48^{+0.27}_{-0.14} $ & $0.998$ & $  -211^{+    29}_{   -27} $ & $0.63^{+0.19}_{-0.12} $ & $  -172^{+    30}_{   -26} $ & $  1.48^{+  0.35}_{ -0.28} $ & $ >   14.4$ & $  1.00^{+  0.16}_{ -0.13} $\\
\end{longtable}
\clearpage
\end{landscape}}

\clearpage
{\footnotesize
\begin{landscape}
\begin{longtable}{llccccccccccccc}
\caption{\ion{Fe}{2} Absorption Line Measurements\label{tab.felines}}\\
\hline \hline  \\[-2ex]
Galaxy Name$^a$ & Category & $\rm S/N $ & $\rm EW_{2586}$$^b$ & $\rm EW_{2600}$$^b$ & $P_\mathrm{out,1}$$^c$ & $\Delta v_1$$^d$ & $C_{f,1}$$^e$ & $P_\mathrm{out,2}$$^f$ & $\Delta v_\mathrm{max}$$^g$ & $C_{f,\rm flow}$$^h$ & $ v_\mathrm{flow}$$^i$ & $\rm EW_{flow}$$^j$ &  $\log N_\mathrm{flow}$$^k$ & $\rm EW_{sys}$$^l$\\ [0.5ex]
 & & $\rm pix^{-1}$ & \AA & \AA &  & $\mkms$ & & & $\mkms$ & & $\mkms$ & \AA & $\rm cm^{-2}$ & \AA\\
\hline \hline
\endfirsthead
\multicolumn{15}{c}{{\tablename} \thetable{} -- Continued} \\ [0.5ex]
\hline \hline
Galaxy Name$^a$ & Category & $\rm S/N $ & $\rm EW_{2586}$$^b$ & $\rm EW_{2600}$$^b$ & $P_\mathrm{out,1}$$^c$ & $\Delta v_1$$^d$ & $C_{f,1}$$^e$ & $P_\mathrm{out,2}$$^f$ & $\Delta v_\mathrm{max}$$^g$ & $C_{f,\rm flow}$$^h$ & $v_\mathrm{flow}$$^i$ & $\rm EW_{flow}$$^j$ &  $\log N_\mathrm{flow}$$^k$ & $\rm EW_{sys}$$^l$\\ [0.5ex]
 & & $\rm pix^{-1}$ & \AA & \AA &  & $\mkms$ & & & $\mkms$ & & $\mkms$ & \AA & $\rm cm^{-2}$ & \AA\\
\hline \\[-1.8ex]
\endhead
\\[-1.8ex] \hline \hline
\multicolumn{15}{c}{{Continued on Next Page\ldots}} \\
\endfoot
\\[-1.8ex] \hline \hline
\multicolumn{15}{l}{$^a$ Galaxies are named according to ID numbers listed in \citet[][TKRS]{Wirth2004}, the AEGIS survey \citep[][EGS]{Davis2007}, and}\\
\multicolumn{15}{l}{$~~$ \citet[]{Giavalisco2004}. IDs beginning with numbers mark objects from \citet[]{Giavalisco2004}, and in all other instances are preceded with ``J0332".}\\
\multicolumn{15}{l}{$^b$ In cases for which a line is not detected, the 3$\sigma$ EW upper limit is listed.}\\
\multicolumn{15}{l}{$^c$ Fraction of the marginalized PPDF for the one-component model having $v < 0\mkms$. }\\
\multicolumn{15}{l}{$^d$ Central velocity for the one-component model.}\\
\multicolumn{15}{l}{$^e$ Covering fraction for the one-component model.}\\
\multicolumn{15}{l}{$^f$ Fraction of the marginalized PPDF for the flow component in the two-component model having $v < 0\mkms$.}\\
\multicolumn{15}{l}{$^g$ Maximum outflow velocity, defined as $\Delta v_\mathrm{max} =  v_\mathrm{flow} - b_{D,\rm flow}/\sqrt{2}$. }\\
\multicolumn{15}{l}{$^h$ Covering fraction of the flow component in the two-component model.}\\
\multicolumn{15}{l}{$^i$ Central velocity of the flow component in the two-component model.}\\
\multicolumn{15}{l}{$^j$ Equivalent width of the flow component in the two-component model.}\\
\multicolumn{15}{l}{$^k$ Column density of the flow component in the two-component model.  Lower limits include 95\% of the marginalized PPDF.}\\
\multicolumn{15}{l}{$^l$ Equivalent width of the systemic component in the two-component model.}\\
\endlastfoot
 EGS12004441 & low S/N & $   3.0$ & $ <    2.1$ & $ <    1.7$ & \nodata & \nodata & \nodata & \nodata & \nodata & \nodata & \nodata & \nodata & \nodata & \nodata\\
 EGS12004471 & wind (Mg,Fe) & $  13.0$ & $   3.0\pm   0.2$ & $   5.3\pm   0.3$ & $0.994$ & $  -113^{+    14}_{   -14} $ & $0.75^{+0.13}_{-0.11} $ & $1.000$ & $  -277^{+    39}_{   -47} $ & $0.67^{+0.18}_{-0.14} $ & $  -200^{+    32}_{   -35} $ & $  0.96^{+  0.28}_{ -0.24} $ & $ >   14.3$ & $  1.03^{+  0.18}_{ -0.16} $\\
 EGS12008005 & systemic & $  12.8$ & $   2.4\pm   0.2$ & $   3.2\pm   0.2$ & $0.011$ & $    24^{+     9}_{    -9} $ & $0.82^{+0.07}_{-0.07} $ & \nodata & \nodata & \nodata & \nodata & \nodata & \nodata & \nodata\\
 EGS12008157 & systemic & $   8.7$ & $   1.6\pm   0.3$ & $   2.4\pm   0.2$ & $0.835$ & $   -23^{+    24}_{   -24} $ & $0.61^{+0.11}_{-0.09} $ & \nodata & \nodata & \nodata & \nodata & \nodata & \nodata & \nodata\\
 EGS12008197 & wind (Mg,Fe) & $   8.9$ & $   2.2\pm   0.2$ & $   2.7\pm   0.3$ & $0.996$ & $   -71^{+    17}_{   -16} $ & $0.64^{+0.08}_{-0.07} $ & $0.996$ & $  -260^{+    60}_{   -60} $ & $0.46^{+0.19}_{-0.15} $ & $  -193^{+    71}_{   -64} $ & $  0.98^{+  0.53}_{ -0.34} $ & $ >   14.4$ & $  1.38^{+  0.26}_{ -0.27} $\\
 EGS12008247 & low S/N & \nodata & \nodata & \nodata & \nodata & \nodata & \nodata & \nodata & \nodata & \nodata & \nodata & \nodata & \nodata & \nodata\\
 EGS12008427 & wind (Mg,Fe) & $   9.8$ & $   2.4\pm   0.2$ & $   1.7\pm   0.2$ & $0.990$ & $   -48^{+    16}_{   -17} $ & $0.83^{+0.12}_{-0.18} $ & $0.930$ & $  -288^{+   144}_{  -226} $ & $0.46^{+0.34}_{-0.32} $ & $  -156^{+   102}_{  -171} $ & $  0.00^{+  0.32}_{ -0.00} $ & $ >    9.3$ & $  1.51^{+  0.19}_{ -0.22} $\\
\end{longtable}
\clearpage
\end{landscape}}

\appendix
\section{Redshifts}\label{sec.appen_redshifts}

Previous studies of cool absorbing outflows have exercised care in selecting the best estimate of the host galaxy systemic velocity for comparison with outflow kinematics.  For instance, \citet{Heckman2000} determined the systemic velocities of their sample of infrared-bright galaxies using either spatially-resolved rotation curves, CO emission line profiles, nuclear stellar velocities, \ion{H}{1} emission line profiles, or nuclear emission line profiles, in order of preference.  \citet{Martin2005} used CO emission profiles to trace the systemic velocities of her ULIRG sample, arguing that in galaxy mergers, the molecular ISM rapidly sinks to the dynamical center of the system.  In the \citet{Rupke2005a} study of \ion{Na}{1} outflows from IR-luminous starbursts, the authors preferred stellar absorption lines for tracing the galaxy systemic velocities; however where high-quality centroids for these lines were not available, they instead used nebular emission rotation curves, \ion{H}{1} emission or absorption, or nebular emission centroids.  \citet{Schwartz2006} also preferred stellar absorption lines in spectroscopy of nearby dwarf galaxies for tracing the systemic velocity, and resorted to CO and \ion{H}{1} velocities when absorption measurements were not available.

While at $z > 0.3$ we lack many of the dynamical tracers available for nearby galaxies, we do
 obtain measurements of stellar absorption redshifts independently of the velocities of nebular emission where possible.
As in previous studies, we assume that absorption lines in stellar atmospheres provide the best tracer of the velocity of the stellar population, as opposed to nebular emission lines, which trace the gas velocities around young stars.  We also measure redshifts for the red- and blue-side LRIS spectra separately, preferring to adopt the fitted redshift for the side which covers \ion{Mg}{2} and \ion{Fe}{2} transitions.  Finally, 
while the [\ion{O}{2}] $\lambda \lambda 3727, 3729$ doublet is for some spectra the only transition available for tracing the host galaxy systemic velocity, because we do not resolve the doublet lines (separated by $\sim 222\mkms$) in our data, we adopt the redshift measured from a previously-obtained DEIMOS spectrum of the same target in such cases if available.  The DEIMOS spectra in the GOODS fields have slightly higher velocity resolution \citep{Wirth2004}, and the DEEP2 spectra have a resolution at least twice as high \citep{Davis2007}, such that uncertainties in the doublet line strength ratio have a less significant effect on the fitted redshift.  

Redshift fitting is done using IDL code included in the LowRedux package adapted for use with LRIS spectra from the publicly available programs developed for the SDSS (IDLSPEC2D)\footnote{http://spectro.princeton.edu/idlspec2d$\_$install.html}.  
This code calculates $\chi^2$ values 
as a function of the lag between 
an observed spectrum and a linear combination of eigenspectra.  These eigenspectra are derived from SDSS spectra and are available with the IDLSPEC2D package.  They  cover 2700 \AA \ - 9000 \AA; however, we mask spectral regions in the data below 2860 \AA \ in the rest frame to exclude \ion{Mg}{2} $\lambda \lambda 2796, 2803$, \ion{Fe}{2} lines further to the blue, and \ion{Mg}{1} $\lambda 2853$ from the fit.  
We note that \ion{Ca}{2} H \& K $\lambda \lambda 3934, 3969$ absorption may arise both in stellar atmospheres and from interstellar or wind material, such that fitting to this spectral region could in principle bias our results.  However, we find that exclusion of this transition does not significantly affect 
our measured redshifts (e.g., the resulting mean offsets for categories (1) and (3) below are $< 5\mkms$).  
The $\chi^2$ values surrounding a global minimum are fit by a Gaussian or quadratic to determine the best-fit lag, and the lag (or redshift) uncertainty includes 
both the formal $1\sigma$ uncertainties of the fitted center of the minimum, as well as the range in lag values over which $\chi^2$ increases by 1.  Four spectra exhibiting the broad emission signatures of Type 1 AGN are flagged and fit with an SDSS QSO template, and excluded from the remaining analysis.  ``First-pass" redshifts ($z_\mathrm{FP}$) are measured by fitting the full red-side galaxy spectra, including both stellar absorption and nebular emission lines, and provide values accurate enough to mask emission lines in 
subsequent fits.

\begin{figure}
\centering \includegraphics[angle=90,width=5in]{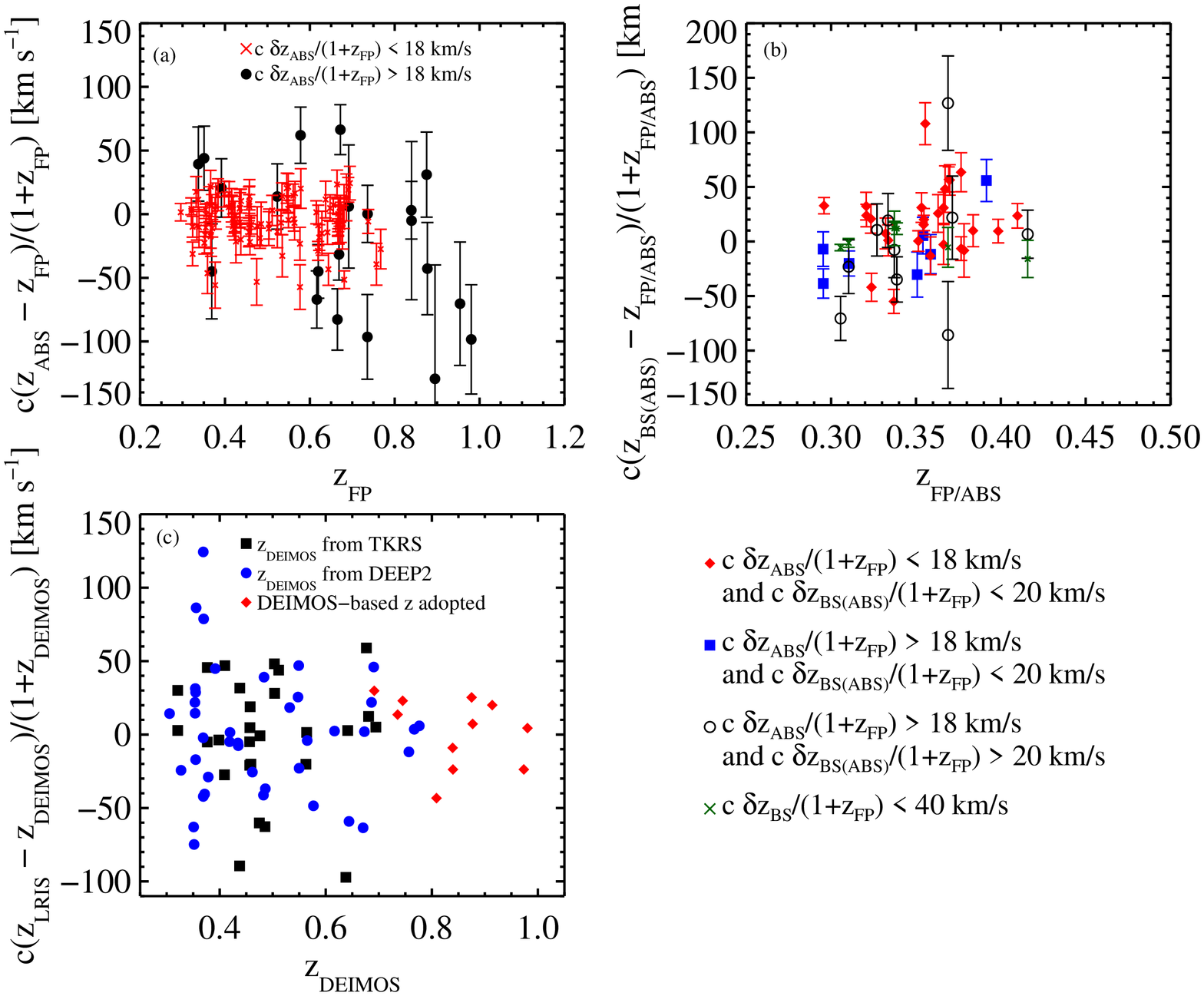}
\caption[]{\emph{(a)}  Offset between redshifts measured using the full galaxy spectrum ($z_\mathrm{FP}$) and those measured after masking emission lines in the data ($z_\mathrm{ABS}$) for spectra with low (red crosses) and high (black circles) uncertainties in $z_\mathrm{ABS}$.  For most spectra with high-quality $z_\mathrm{ABS}$ measurements, the offset is $\lesssim 30\mkms$; however, a handful of spectra yield offsets $\sim -50\mkms$.  \emph{(b)}  Offset between $z_\mathrm{BS(ABS)}$ (absorption-line redshifts measured with blue-side spectra) and $z_\mathrm{ABS}$ or $z_\mathrm{FP}$ measured from the red-side spectra.  Objects with both high-quality $z_\mathrm{BS(ABS)}$ and high-quality $z_\mathrm{ABS}$ measurements are indicated with red diamonds; objects with high-quality $z_\mathrm{BS(ABS)}$ measurements for which $z_\mathrm{FP}$ is adopted on the red side are indicated with blue squares, and objects with both poor-quality $z_\mathrm{BS(ABS)}$ and $z_\mathrm{ABS}$ measurements are shown with open black circles.  Green crosses show the offset between high-quality $z_\mathrm{BS}$ measurements and $z_\mathrm{FP}$.  Taken together, the dispersion in the colored points indicates the uncertainty introduced by the separate wavelength and flexure calibration of the blue and red sides.   \emph{(c)}  Offset between our preferred redshift measurement from our LRIS spectra ($z_\mathrm{LRIS}$) and the redshift for the same galaxy measured in Keck/DEIMOS spectra ($z_\mathrm{DEIMOS}$) from \citet{Wirth2004} (black squares) and \citet{Davis2007} (blue circles).  Galaxies for which we ultimately adopt the DEIMOS-based redshift   
 are shown with red diamonds.  The median offset for the blue and black points is $1.4\mkms$, with a dispersion of $41\mkms$, indicating good agreement between redshift measurements from these different studies.
     \label{fig.redshifts}}
\end{figure}

We then assign redshifts with the following preference:
\begin{enumerate}
\item Redshifts obtained from fitting blue-side spectra ($z_\mathrm{BS(ABS)}$) with coverage that extends redward of \ion{Ca}{2} K $\lambda 3934$, and with the following emission lines masked over $\pm 280 \mkms$: [\ion{O}{2}], [\ion{Ne}{3}] $\lambda 3869$, 
and the Balmer series 
through Hf.

\item Redshifts fitted to blue-side spectra with coverage that extends redward of \ion{Ca}{2} K $\lambda 3934$ without emission-line masking ($z_\mathrm{BS}$).  

\item Redshifts fitted to red-side spectra with masking of the emission lines listed in (1), as well as of [\ion{O}{3}] $\lambda \lambda 4960, 5008$ ($z_\mathrm{ABS}$).  

\item $z_\mathrm{FP}$, if the red-side spectrum extends to at least H$\gamma$ at 4341 \AA.

\item Redshift determined from DEIMOS spectrum ($z_\mathrm{DEIMOS}$). 

\end{enumerate}

Where $z_\mathrm{DEIMOS}$ is not available for the remaining objects, we adopt $z_\mathrm{FP}$ if our LRIS spectra cover [\ion{O}{2}].  We exclude 3 objects which lack both [\ion{O}{2}] coverage and DEIMOS spectroscopy from our analysis. 

Every redshift measurement described above must meet specific quality standards before it is accepted.  For instance, $z_\mathrm{BS(ABS)}$ is adopted only if it is within $200\mkms$ of $z_\mathrm{FP}$, and if the error in the fit is $< 20\mkms$.  We note that these formal uncertainties are likely underestimated by the fitting code, and further quantify our true measurement uncertainties below.  However, this prevents the use of $z_\mathrm{BS(ABS)}$ in cases of spectra with low S/N in the stellar continuum on the blue side.  For the remaining spectra, $z_\mathrm{BS}$ is adopted only if it is within $40\mkms$ of $z_\mathrm{FP}$.  This latter cut prevents us from adopting redshifts which are dominated by [\ion{O}{2}] emission; the remaining $z_\mathrm{BS}$ values are all within $< 23\mkms$ of the corresponding DEIMOS redshifts. 

Moving over to the red side, several of the remaining spectra (18) do not yield successful $z_\mathrm{ABS}$ measurements; i.e., $z_\mathrm{ABS}$ is offset from $z_\mathrm{FP}$ by more than $200\mkms$.  These spectra lack strong absorption lines and/or exhibit a low continuum S/N.  
Even among the remaining ``successful" $z_\mathrm{ABS}$ measurements, however, the difference between the two redshift determinations exceeds $40 \mkms$ (the approximate size of a pixel) in several cases.  This is evident in Figure~\ref{fig.redshifts}a, which compares $z_\mathrm{ABS}$ and $z_\mathrm{FP}$ for {\it all} galaxies, including those with successful blue-side redshifts.  

\begin{figure}
\centering \includegraphics[angle=90,width=4in]{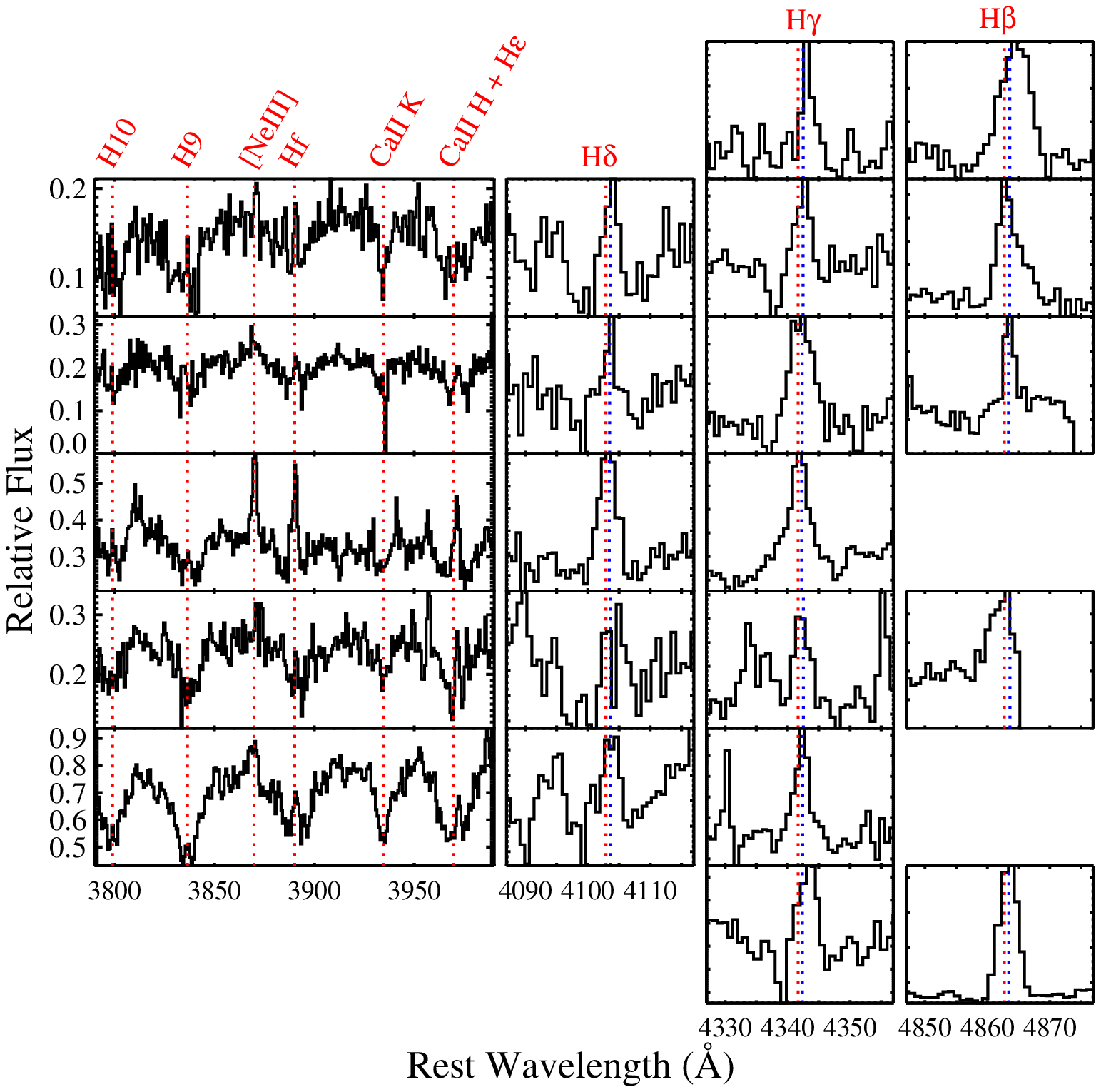}
\caption[]{Seven spectra for which the error on $z_\mathrm{ABS}$ is $< 18 \mkms$ and the offset between $z_\mathrm{FP}$ and $z_\mathrm{ABS}$ is greater than $36 \mkms$.  Each row shows a different object.  The red dotted lines mark the systemic velocity of each transition as determined by $z_\mathrm{ABS}$, while the blue dotted lines mark the systemic velocity implied by $z_\mathrm{FP}$.
     \label{fig.zoffsets}}
\end{figure}

We show a subset of the spectra with successful $z_\mathrm{ABS}$ measurements but large offsets between $z_\mathrm{ABS}$ and $z_\mathrm{FP}$ in Figure~\ref{fig.zoffsets}.  For each of these spectra our measurements yield a relatively low uncertainty on $z_\mathrm{ABS}$ ($< 18 \mkms$) but an offset between $z_\mathrm{FP}$ and $z_\mathrm{ABS}$ of at least 36 \kms.  In each of these cases, $z_\mathrm{ABS}$ is blueward of $z_\mathrm{FP}$; however, we do not detect a systematic blueshift between $z_\mathrm{ABS}$ and $z_\mathrm{FP}$ (e.g., the mean offset is $-5 \mkms$ for spectra with $z_\mathrm{ABS}$ uncertainties $< 18\mkms$).  A by-eye examination of this Figure indicates that $z_\mathrm{ABS}$ may indeed provide a better indication of the velocities of absorption lines in these spectra.  In nearly every case, the H$\delta$ and H$\gamma$ emission lines sit slightly to the red within the broader stellar absorption profile.  This is evident in the top two and bottom H$\beta$ panels as well.  See \citet{Rodrigues2012} for further discussion of this effect.

While we would ideally adopt $z_\mathrm{ABS}$ for every galaxy remaining in our sample, we also wish to minimize the uncertainty in our redshift measurements, and many of the $z_\mathrm{ABS}$ values exhibit formal uncertainties greater than 20 \kms or outright fit failures.  We find that spectra with $z_\mathrm{ABS}$ measurement uncertainties less than 18 \kms exhibit acceptable offsets between $z_\mathrm{ABS}$ and $z_\mathrm{FP}$ (as determined by eye; see red crosses in Figure~\ref{fig.redshifts}a), while $z_\mathrm{ABS}$ values with larger uncertainties do not necessarily yield acceptable fits.  We therefore adopt $z_\mathrm{ABS}$ as the systemic redshift whenever its formal uncertainty is $< 18 \mkms$.  This applies for 75 objects which lack acceptable blue-side redshifts.  Nineteen of the remaining spectra have coverage extending redward of H$\gamma$, and 3 additional spectra do not have this coverage but lack a measurement of $z_\mathrm{DEIMOS}$; for these we adopt $z_\mathrm{FP}$.  For the remaining 19 spectra, we adopt $z_\mathrm{DEIMOS}$.  

Figure~\ref{fig.redshifts}b compares our $z_\mathrm{BS(ABS)}$ measurements with either $z_\mathrm{ABS}$ or $z_\mathrm{FP}$.  Open points indicate $z_\mathrm{BS(ABS)}$ measurements which fail to meet our selection criteria as described above, while solid points indicate galaxies for which $z_\mathrm{BS(ABS)}$ is adopted.  Points for which the criterion for acceptance of $z_\mathrm{ABS}$ is met are red and $z_\mathrm{ABS}$ is plotted on the horizontal axis; otherwise, $z_\mathrm{FP}$ is used (blue).    
For completeness, green crosses show objects for which $z_\mathrm{BS}$ is adopted.
This plot provides an indication of the uncertainties introduced by the separate wavelength and flexure calibration of the blue and red sides.  The RMS of the quantity $c (z_\mathrm{BS(ABS)} - z_\mathrm{ABS})/(1+z_\mathrm{ABS})$ for the red points is $33 \mkms$, while the median offset between $z_\mathrm{ABS}$ and $z_\mathrm{BS(ABS)}$ is $21 \mkms$.  This suggests there is a systematic offset between the wavelength solutions on the blue and red sides as large as $\sim 20\mkms$, although this shift also incorporates uncertainties due to measurement of the velocities of different absorption lines on different sides of the spectrograph.  

We now discuss the overall uncertainties on our measurements.  First, 8 galaxies in our sample with successful LRIS-based redshifts were observed twice, on two different slitmasks.  We find that the mean offset in the redshifts calculated for each set of two spectra is $19 \mkms$, with a maximum offset of $32 \mkms$, and an RMS value of $22\mkms$.  For six of these objects, the same type of redshift was assigned to each, while for two objects, redshifts of different types were assigned.  

We also compare our results with those of the TKRS and EGS surveys in Figure~\ref{fig.redshifts}c.  Here, assigned LRIS-based redshifts are compared with TKRS redshifts (black) and EGS redshifts (blue).  The red diamonds compare our $z_\mathrm{FP}$ to $z_\mathrm{DEIMOS}$ for objects for which we ultimately adopted $z_\mathrm{DEIMOS}$.  The median offset between these redshift measurements for the red diamonds is $7 \mkms$, with an RMS value of $24 \mkms$.  The median offset between $z_\mathrm{LRIS}$ and $z_\mathrm{DEIMOS}$ for the remaining spectra is $1.4 \mkms$, with an RMS value of $41 \mkms$.  Assuming that the DEIMOS measurements contribute at least $30 \mkms$ to this dispersion \citep{Willmer2006}, we find that our $z_\mathrm{LRIS}$ measurements have an RMS uncertainty of $28 \mkms$.  This is slightly larger than the RMS uncertainty suggested from our analysis of repeated observations (i.e., $22\mkms/\sqrt 2 = 15\mkms$); however, as the latter value is obtained from measurement of only eight redshifts, we conservatively adopt $28\mkms$ as our overall redshift uncertainty.

\clearpage



\section{Broad-band Photometry and SED Fitting}\label{sec.appen_photometry}

Here we describe the broad-band photometry used and results of MAGPHYS SED fitting in more detail.

\subsection{GOODS-N}

In the GOODS-N field, we use publicly-available photometry from the MOIRCS Deep Survey (MODS) \citep{Kajisawa2011}.  
The photometry catalog is $K_s$-band selected, and includes $J-$, $H-$, and $K_s-$band Subaru/MOIRCS measurements, as well as
 PSF- and aperture-matched photometry in the KPNO/MOSAIC $U$-band \citep{Capak2004}, HST/ACS $b_{435}$, $v_{606}$, $i_{775}$, 
 and $z_{850}$ bands \citep{Giavalisco2004}, and Spitzer/IRAC 3.6, 4.5, 5.8, and 8.0 $\mu$m-bands (M.\ Dickinson et al., in preparation).  
 An aperture diameter of 1.2\arcsec\ was used for each of these bands.  Prior to SED fitting, we thus apply an aperture correction calculated from the 
 ratio between the total $K_s$-band flux and the $K_s$-band aperture flux.  Also included in the MODS photometry catalog are measurements of total 24$\mu$m flux, measured from 
 Spitzer/MIPS 24 $\mu$m GOODS imaging.  We include photometry from all of the aforementioned pass bands in our SED fitting, and also 
 include total fluxes measured in the $GALEX$ NUV ($\lambda_\mathrm{eff} = 232~ \mathrm{nm}$) band and available in the source catalog 
of the public data release GR6.  We additionally assume a minimum 5\% photometric error in each passband.  

\subsection{GOODS-S}

In GOODS-S, we make use of the FIREWORKS survey \citep{Wuyts2008}, which has made publicly available a $K_s$-selected catalog that includes 
ESO/MPG 2.2m WFI photometry measured in the WFI $U_{38}BVRI$ bands, \emph{HST}/ACS $b_{435}$, $v_{606}$, $i_{775}$, and $z_{850}$ photometry, VLT/ISAAC $JHK_s$ photometry, 
and Spitzer/IRAC 3.6, 4.5, 5.8, and 8.0 $\mu$m photometry.  All of this imaging was PSF-matched prior to measurement of aperture photometry, and we again 
apply an aperture correction based on the total $K_s$-band flux, and assume a minimum photometric error of 5\%.  The FIREWORKS catalog also includes total MIPS 24 $\mu$m fluxes for each object.  All of these 
measurements are included in our SED fitting, as well as total fluxes measured in the $GALEX$ NUV band and available in the source catalog 
of the public data release GR6.

\subsection{EGS}

In the EGS, we use photometry from catalogs published in \citet{Barro2011}.  These catalogs incorporate GALEX NUV photometry available from GR3, $ugriz$ photometry from the CFHTLS\footnote{www.cfht.hawaii.edu/Science/CFHTLS-DATA/}, MMT/Megacam $u'$-band photometry, CFHT-12k $B$ and $I$-band photometry, \emph{HST}/ACS $v_{606}$ photometry, $J$, $K$, and $K_s$-band near-IR photometry from Palomar/WIRC \citep{Bundy2006} and Subaru/MOIRCS, 4 bands of IRAC imaging from \citet{Barmby2008}, and MIPS $24\mu$m imaging from the MIPS GTO and FIDEL surveys.  Photometry from additional optical and NIR pass bands are provided by \citet{Barro2011}, but we do not make use of them here.  Photometry was measured in a fixed Kron elliptical aperture in all bands.  This aperture was determined from a ground-based optical/NIR image with a PSF similar to most of the other ground-based frames.  Because the provided \citet{Barro2011} catalogs report the total magnitude of each object in each pass band, we do not apply an aperture correction prior to SED fitting, but continue to assume a minimum photometric error of 5\%.  
For just one galaxy, EGS13049649, we left out the $v_{606}$-band photometry, as it was grossly inconsistent with the other available optical photometry; all photometric measurements listed above were used for the remaining galaxies.  

\begin{figure}
\includegraphics[angle=0,width=\columnwidth]{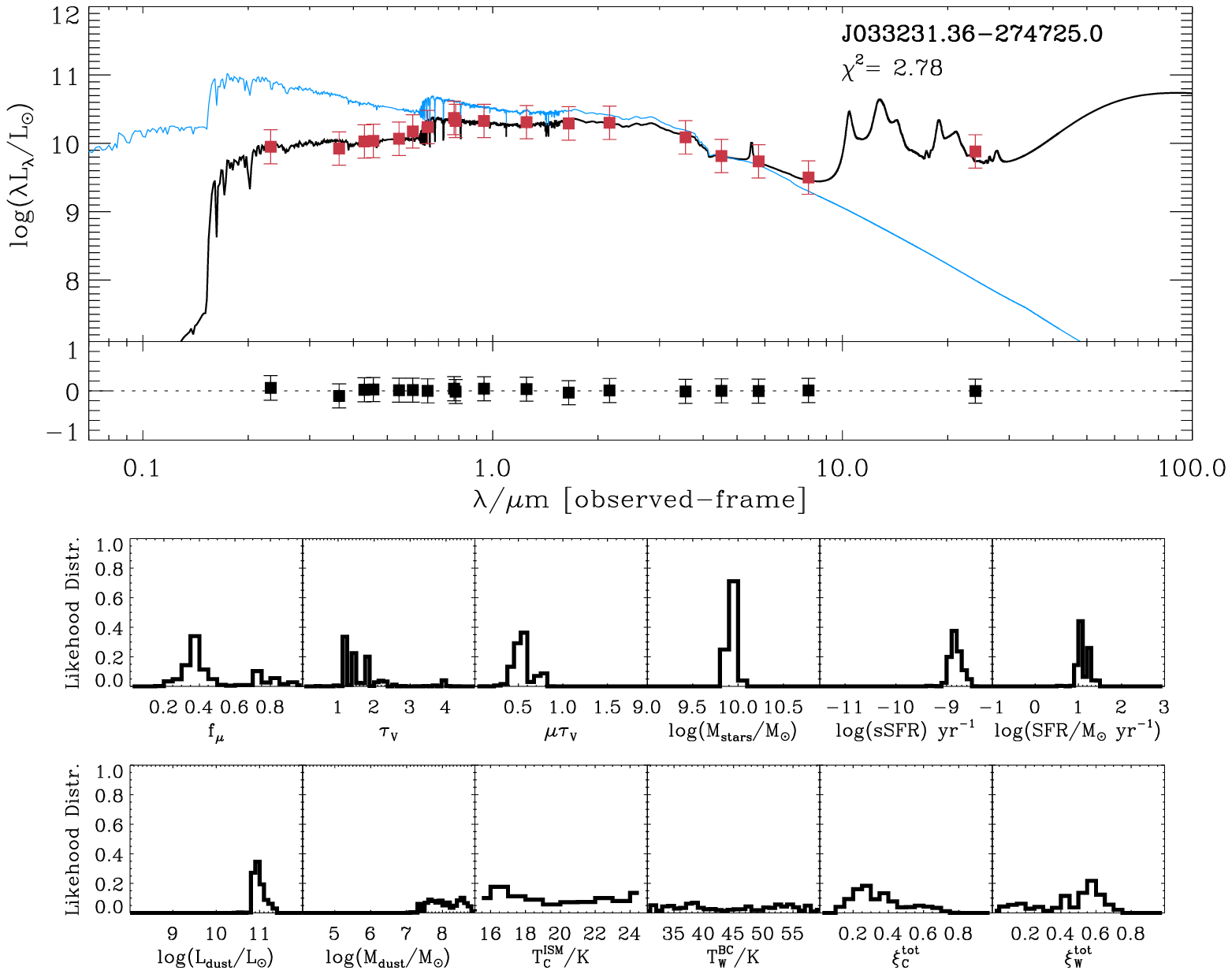}
\caption[]{Example of MAGPHYS SED fit for GOODS-S object J033231.36-274725.0.  \emph{Top:} Aperture-corrected photometry from the FIREWORKS survey \citep{Wuyts2008} is shown with magenta squares.  The best-fit SED is shown in black, and the same SED, without a dust correction, in shown in blue.  \emph{Middle:}  Residuals between the photometry and the best-fit SED.  \emph{Lower panels:}  Marginalized likelihood distributions of the following MAGPHYS model parameters (from left to right and top to bottom): fraction of dust luminosity emitted by dust in the ambient ISM ($f_{\mu}$), $V$-band optical depth of dust in front of young stars in their birth clouds ($\tau_{V}$), $V$-band optical depth of dust in front of young stars in the ambient ISM ($\mu \tau_{V}$), stellar mass ($\log M_\mathrm{stars}$), specific SFR (sSFR), SFR, total dust luminosity ($L_\mathrm{dust}$), total dust mass ($M_\mathrm{dust}$), temperature of cold dust in the ambient ISM ($\rm T_C^{ISM}$), temperature of warm dust in stellar birth clouds ($\rm T_{W}^{BC}$), 
fractional contribution of cold dust to the total dust luminosity ($\rm \xi_C^{tot}$), and fractional contribution of warm dust to the total dust luminosity ($\rm \xi_W^{tot}$).
     \label{fig.magphys}}
\end{figure}

\subsection{MAGPHYS}

The MAGPHYS SED-fitting procedure is described briefly in \S\ref{sec.mstarsfr}, and in detail in \citet{daCunha2008,daCunha2011}.  
Figure~\ref{fig.magphys} shows an example of an SED fit to the FIREWORKS broadband photometry of one of our galaxies in GOODS-S.  The black line in the upper panel shows the best-fit extinction-corrected model SED, while the blue line shows the same model before correcting for extinction.  The histograms in the lower panels show the marginalized likelihood distributions for several of the model parameters, including stellar mass ($M_\mathrm{stars}$) and total dust luminosity ($L_\mathrm{dust}$).  The histogram labeled `SFR' shows the likelihood distribution of the model SFRs averaged over the last $10^8$ years.  We adopt the median and $\pm \rm 34th-$percentiles of these histograms as the values of galaxy properties (and their uncertainties) in our preceding analysis.  

To test this method, we compare the results of MAGPHYS for galaxies in the EGS with those reported by the RAINBOW Database team in \citet[][B11]{Barro2011}.  Because we start with the photometry measured by B11, a comparison of our SFR and $M_*$ values with those of B11 exposes the systematic differences between the MAGPHYS and B11 SED modeling.  In brief, B11 fit model SEDs built using a combination of PEGASE 2.0 single stellar population emission models \citep{Fioc1997} and dust emission models from \citet{CharyElbaz2001}, \citet{DaleHelou2002}, and \citet{Rieke2009}.  The PEGASE models assume an exponential star formation history, and have variable star-formation time scale, age, metallicity, and dust attenuation.  B11 assign a best-fit model via $\chi^2$ minimization, and adopt the stellar mass value of this model.  They use several different methods to compute SFRs, but their preferred SFR measurement uses the prescription of \citet{Bell2005}: $\rm SFR = SFR_{TIR} + SFR_{UV,obs}$, where $\rm SFR_{TIR}$ is calculated from the total IR luminosity using the \citet{Kennicutt1998} calibration, and $\rm SFR_{UV,obs}$ is calculated from the rest-frame monochromatic luminosity at $0.28~\mu$m ($L(0.28)$), uncorrected for extinction.  They obtain $L(0.28)$ from their best-fit model SED, and adopt a total IR luminosity from their dust emission template fit to 8 $\mu$m, 24 $\mu$m, and 70 $\mu$m data where available.  

\begin{figure}
\centering \includegraphics[angle=90,width=5.5in]{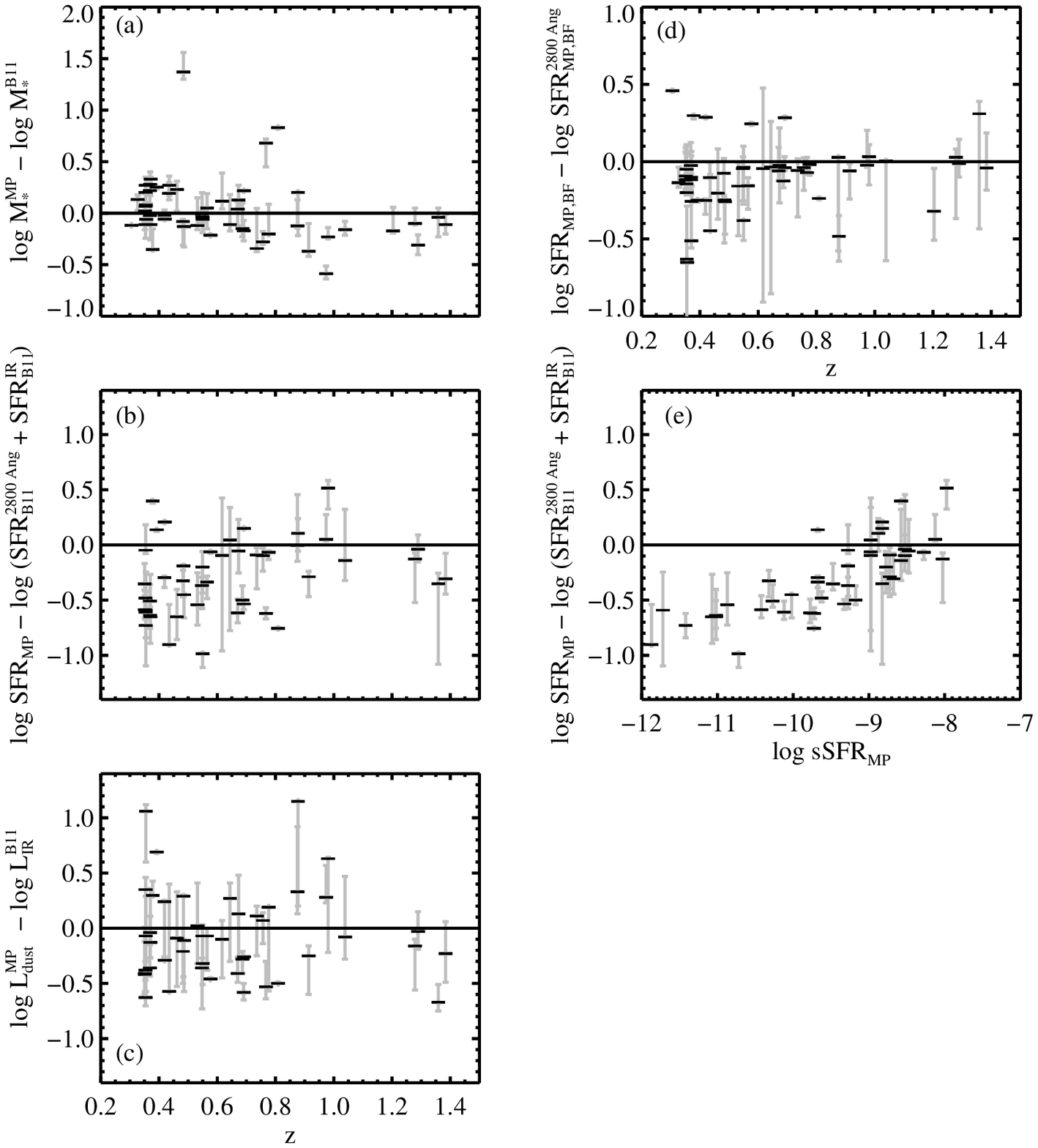}
\caption[]{Comparison of galaxy properties derived in this study to those of B11.  Gray bars in all panels show the width of the central 95\% of the likelihood distribution for each parameter computed with MAGPHYS.  \emph{(a)}  Offset between the MAGPHYS-based stellar mass ($M_*^\mathrm{MP}$) and that of B11 ($M_*^\mathrm{B11}$) as a function of redshift.  \emph{(b)}  Offset between MAGPHYS-based SFR ($\rm SFR_{MP}$) and the total, extinction-corrected SFR of B11 ($\rm SFR_{B11}^{2800\AA} + SFR_{B11}^{IR}$) as a function of redshift. \emph{(c)} Offset between the MAGPHYS-based dust (or IR) luminosity ($L_\mathrm{dust}^\mathrm{MP}$) and the total IR luminosity from B11 ($L_\mathrm{IR}^{B11}$) as a function of redshift.  \emph{(d)} Offset between the SFR of the best-fit MAGPHYS model and the SFR calculated from the 2800 \AA\ luminosity of the \emph{unextincted} best-fit MAGPHYS model ($\rm SFR_{MP,BF}^{2800\AA}$) as a function of redshift.  \emph{(e)}  Same offset shown in panel \emph{(b)}, as a function of $\rm sSFR_{MP}$.  
     \label{fig.checksfrs}}
\end{figure}

Figure~\ref{fig.checksfrs} shows the results of this comparison.  Note that the B11 results have  been corrected from a Salpeter IMF to the Chabrier IMF for consistency with MAGPHYS.  As shown in panel \emph{(a)} and discussed in \S\ref{sec.mstarsfr}, the measurements of $M_*$ are in good agreement, with an offset of $-0.038$ dex and a dispersion of 0.31 dex.  However, MAGPHYS calculates SFRs that are systematically lower than those of B11 by 0.305 dex, with a dispersion of 0.33 dex (panels \emph{b} and \emph{e}).  Furthermore, the offset increases with decreasing sSFR.  

To understand this trend, we compare the MAGPHYS value of $L_\mathrm{dust}$ to the B11 IR luminosity in panel \emph{(c)}.  Here, the two methods produce similar results, yielding an offset of -0.100 dex, and dispersion of 0.406 dex.  These small differences can be attributed to the use of the \emph{full} SED in the MAGPHYS code in constraining the amount of dust in each galaxy, rather than only the mid-IR photometry.  However, the general agreement in this quantity implies that the systematic offset between SFR values is not due to differences in the assumed IR luminosities. 

We also test the \citet{Kennicutt1998} relation between $L(0.28)$ and SFR used by B11 to calculate $\rm SFR_{UV,obs}$.  We interpolate the best-fit MAGPHYS model SED to calculate its rest-frame luminosity at 2800 \AA, and apply the relation 
$\mathrm{SFR}(M_{\odot}~\mathrm{yr}^{-1}) = 5.9 \times 10^{-10} L(0.28)/L_{\odot}$ (see eq. 4, B11).  First, we note that the offset between these values and the SFR(0.28) of B11 is only -0.048 dex, with a scatter of 0.095 dex.
We then compare this value to the SFR of the best-fit MAGPHYS model in panel \emph{(d)} of Figure~\ref{fig.checksfrs}.  This plot shows that the \citet{Kennicutt1998} relation has a substantial amount of scatter (0.236 dex, with a relatively small offset of -0.070 dex), which may be due to the time scale over which the SFR of each SED is calculated, as well as to contamination of the UV emission from old stars.  However, the small offset measured here cannot give rise to the $> 0.5$ dex offset between our SFRs and the total SFRs of B11 in systems with low sSFRs.

We instead attribute this difference to the different assumptions adopted for the source of dust heating in the two methods.  B11 assumes that all IR emission arises from reprocessed UV photons emitted by newly-formed stars.  MAGPHYS, on the other hand, accounts for dust heating from the full stellar population, and thus attributes a fraction of $L_\mathrm{dust}$ to heating by older stars.  
The contribution to dust heating from these older populations increases with increasing $M_*$, or decreasing sSFR, as is evident in panel \emph{(e)} of Figure~\ref{fig.checksfrs}.  Because MAGPHYS forces consistency between $M_*$, $L_\mathrm{dust}$, and SFR, and because even at sSFR values between $10^{-9}$ and $10^{-10}~\mathrm{yr}^{-1}$ the offset in SFRs is only $\sim 0.5$ dex, we adopt our MAGPHYS-derived SFR values without adjustment, with the caveat that they may be systematically lower than the SFRs measured in other studies at low sSFR.  








\clearpage

\section{Modeling of Synthetic Line Profiles}\label{sec.synthetic}

As discussed in \S\ref{sec.fitresults}, our sensitivity to winds must depend on spectral S/N, resolution, and on the absorption strength of gas along the line of sight which has a velocity close to that of the systemic velocity of the host galaxy.  To test these dependencies directly, here we generate a suite of synthetic \ion{Mg}{2} line profiles which include varying amounts of absorption at both systemic velocity and offset to negative velocities and fit these profiles using the same method used for the observed spectra.  Each synthetic profile includes a saturated ($N$(\ion{Mg}{2}) $= 10^{14.9}~\rm cm^{-2}$), `ISM' absorption component centered at $v = 0\mkms$ with $C_f = 1$.  
The Doppler parameter for this component, $b_{D, \rm ISM}$, varies between $20\mkms$ and $120\mkms$ in steps of $20\mkms$.  We additionally choose a maximum wind velocity, $v_\mathrm{synth}$, for each profile starting at $-50\mkms$ and increasing to $-400\mkms$ in steps of $25\mkms$.  After $v_\mathrm{synth}$ is chosen for a given profile, we include several absorption components with velocities increasing from $0\mkms$ to $v_\mathrm{synth}$ in steps of $50\mkms$.  These components are meant to crudely 
emulate a cool outflow with multiple entrained absorbing clouds having a range of velocities \citep[e.g.,][]{MartinBouche2009}.  
Each of these clouds is assigned $N$(\ion{Mg}{2}) $= 10^{14.0}~\rm cm^{-2}$, a Doppler parameter of $10\mkms$, and $C_f = 1$.  Our chosen ranges for $v_\mathrm{synth}$ and $b_{D, \rm ISM}$ result in a grid of $15 \times 6$ synthetic absorption profiles.

\begin{figure}[!h]
\subfigure{\includegraphics[angle=90,width=3.5in]{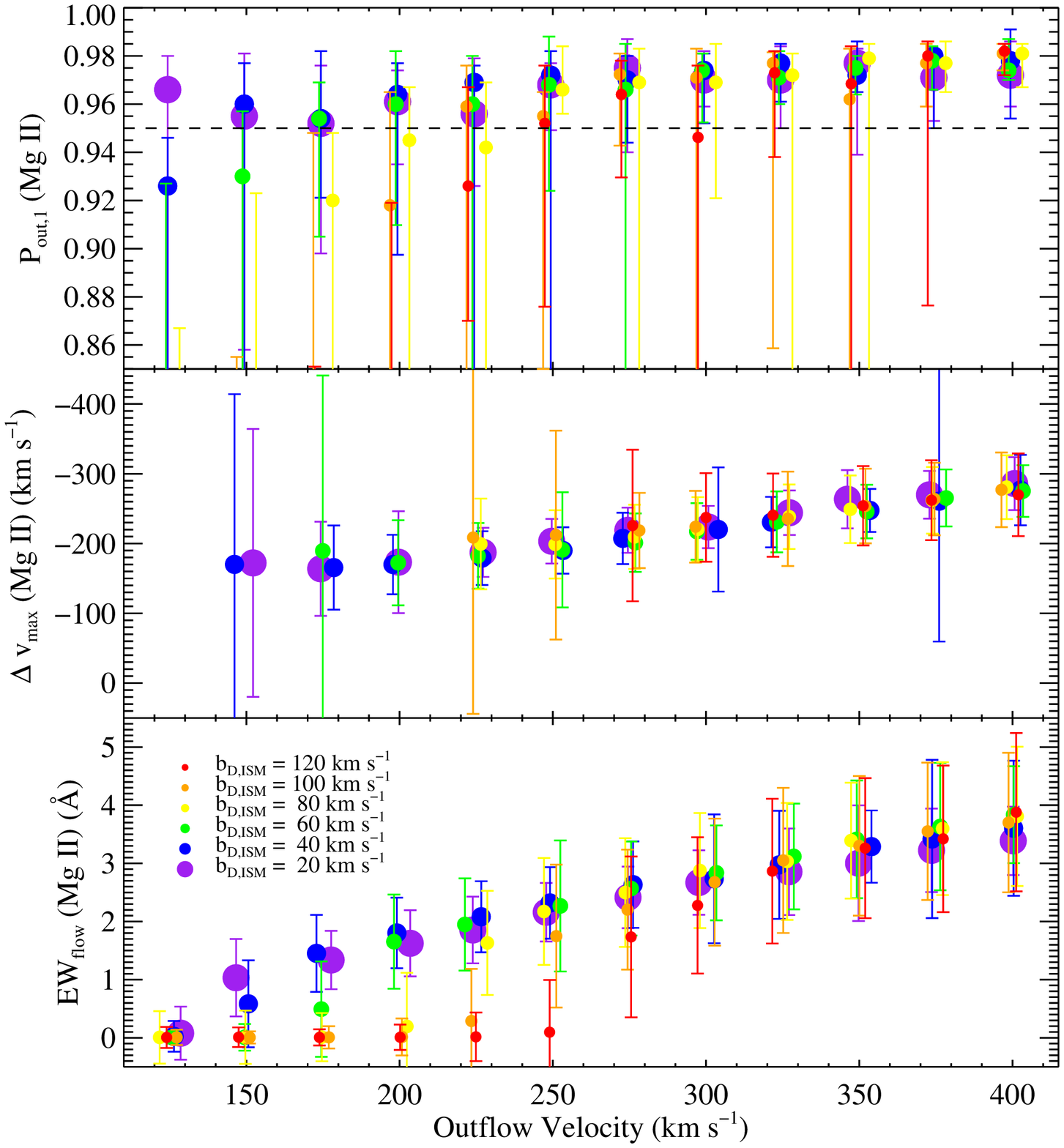}}
\subfigure{\includegraphics[angle=90,width=3.5in]{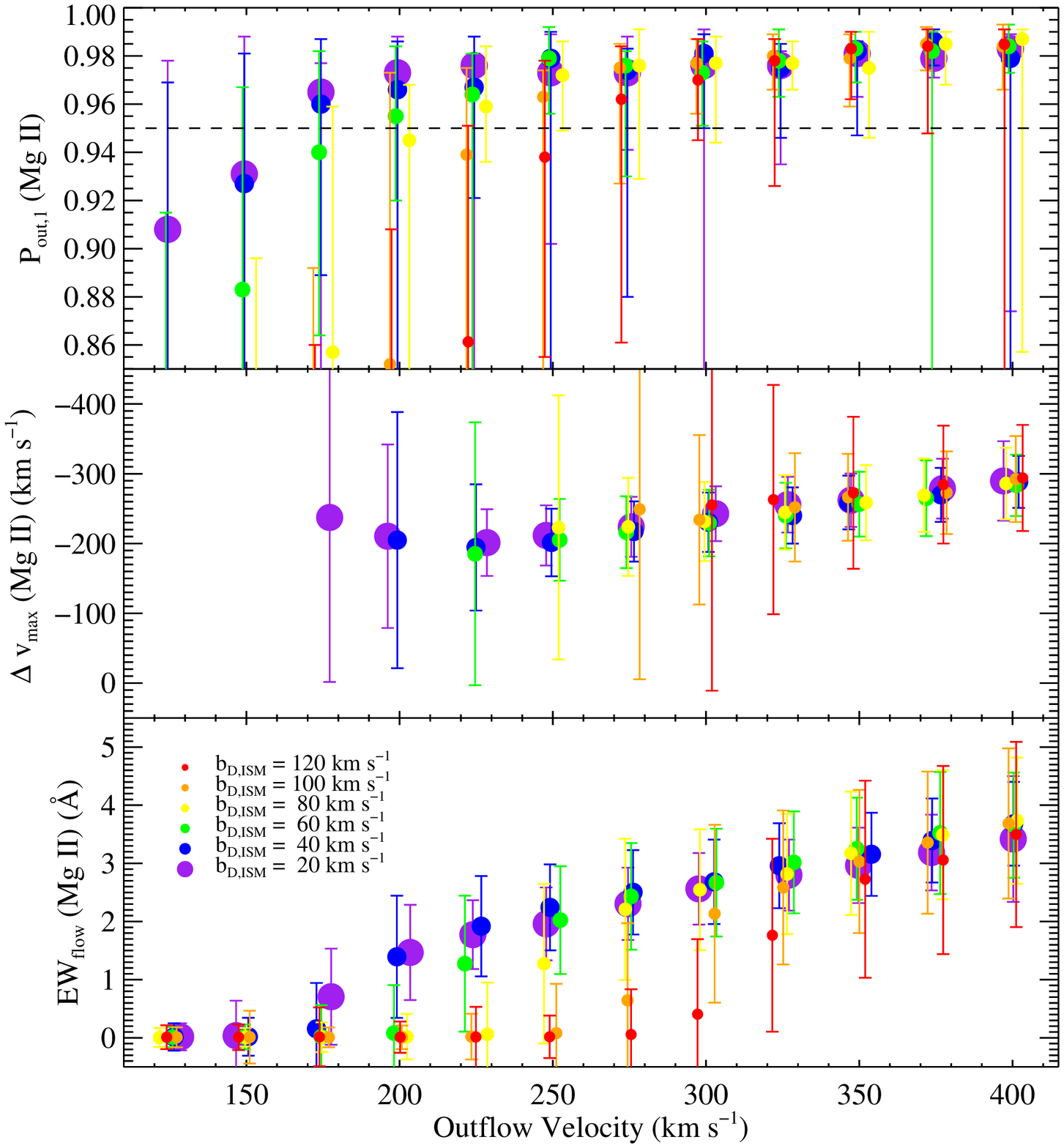}}
\caption[]{\emph{Top left:} $P_\mathrm{out, 1}$ (\ion{Mg}{2}) values for synthetic spectra described in \S\ref{sec.synthetic} as a function of input outflow velocity ($v_\mathrm{synth}$), all of which have a FWHM spectral resolution of $260\mkms$ and $\rm S/N = 9~pix^{-1}$.  Points show the median $P_\mathrm{out, 1}$ for all realizations of models with a given input $b_\mathrm{D,ISM}$, and are coded by the latter (see plot legend in bottom panel).  Error bars show the spread in $P_\mathrm{out, 1}$ obtained for all realizations.  Points are offset slightly in velocity to prevent overlap.  The dashed black line shows our threshold for wind detection.
\emph{Middle left:} Same as above, for 
$\Delta v_\mathrm{max}$ (\ion{Mg}{2}).  Error bars are calculated by adding the dispersion in $\Delta v_\mathrm{max}$ values for each model realization to the mean uncertainty in $\Delta v_\mathrm{max}$ in quadrature.   
\emph{Bottom left:} Same as above, for $\rm EW_{flow}$ (\ion{Mg}{2}).  \emph{Right:}  Same as left-hand panels, for $\rm S/N = 6~pix^{-1}$.
     \label{fig.fakewinds}}
\end{figure}

We first smooth each of these profiles to a velocity resolution close to the median velocity resolution of our sample; i.e., $260\mkms$.  
We then add noise such that the resulting spectra have $\rm S/N = 9~pixel^{-1}$,  generating 10 realizations at each grid point.   We fit these realizations with our one- and two-component models, and show the resulting values of $P_\mathrm{out, 1}$, $\Delta v_\mathrm{max}$ and $\rm EW_{flow}$ in Figure~\ref{fig.fakewinds}.  Results from a separate set of realizations at $\rm S/N = 6~pix^{-1}$ are shown in Figure~\ref{fig.fakewinds}.  

At $\rm S/N = 9~pix^{-1}$, our one-component fitting procedure consistently identifies winds ($P_\mathrm{out, 1} > 0.95$) with maximum velocities ($v_\mathrm{synth}$) greater that $250\mkms$ in most realizations, particularly at very low values of $b_{D, \rm ISM}$ ($20$ or $40\mkms$).  As $v_\mathrm{synth}$ increases beyond this threshold, all models having $b_{D, \rm ISM} \le 60\mkms$ yield detected winds.  Such models also yield increasingly stringent constraints on $\Delta v_\mathrm{max}$ and $\rm EW_{flow}$, with consistent results for these quantities over the full range of input $b_{D, \rm ISM}$ values.  We lose some of this fidelity at $\rm S/N = 6~pix^{-1}$, consistently identifying outflows at $v_\mathrm{synth} \ge 300\mkms$, and with somewhat larger error bars on $\Delta v_\mathrm{max}$ and $\rm EW_{flow}$ (see Figure~\ref{fig.fakewinds}).  However, even at this reduced S/N, in cases for which winds are detected, our two-component model continues to yield consistent values for these quantities regardless of $b_{D, \rm ISM}$.

\begin{figure}
\centering\includegraphics[angle=90,width=3.5in]{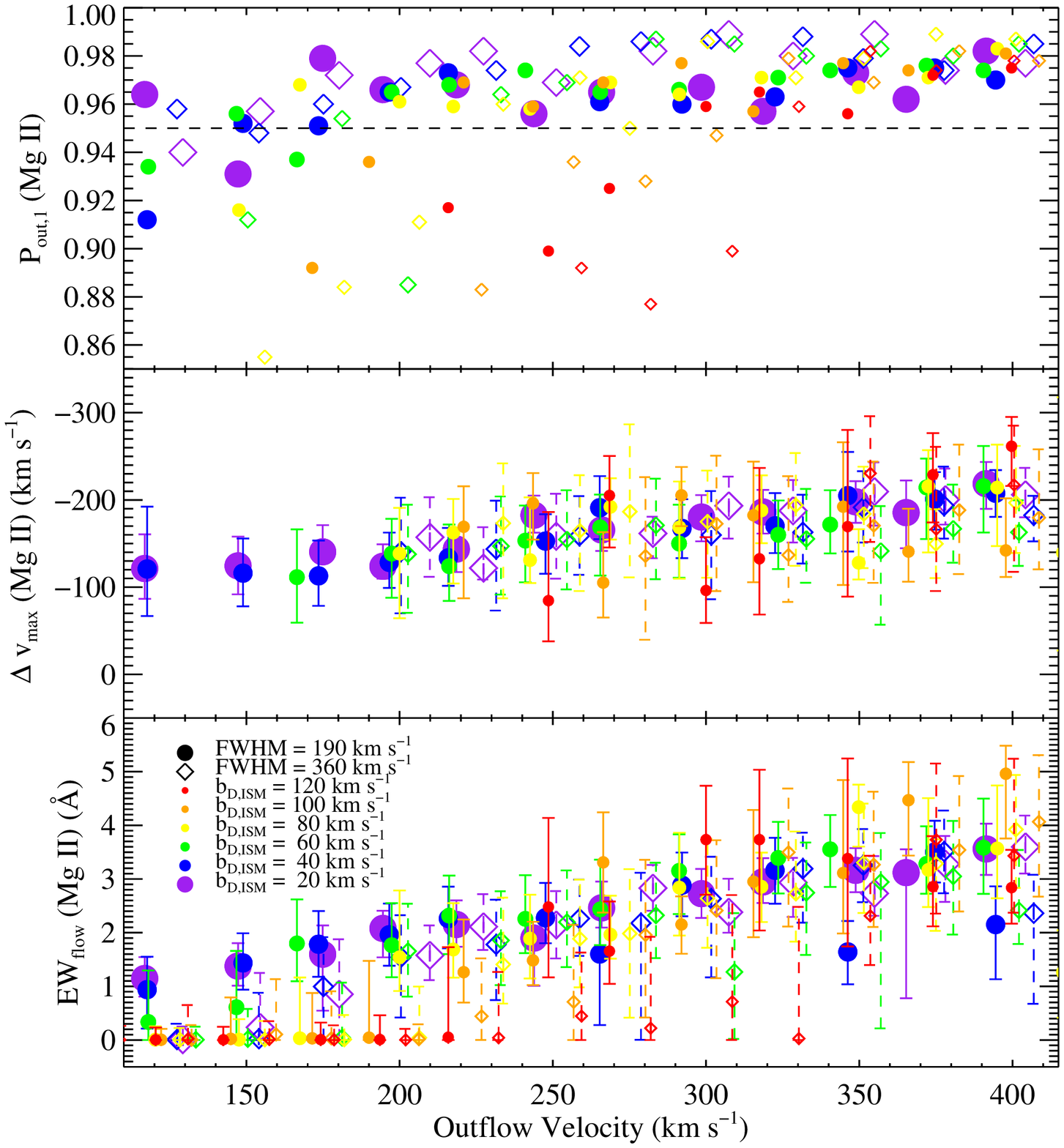}
\caption[]{\emph{Top:} $P_\mathrm{out, 1}$ values for synthetic spectra described in \S\ref{sec.synthetic} as a function of input outflow velocity ($v_\mathrm{synth}$) and spectral resolution (see plot legend in bottom panel for symbol codes).  Points are offset slightly in velocity to prevent overlap. 
\emph{Middle:} Same as above, for $\Delta v_\mathrm{max}$ (\ion{Mg}{2}). 
Points show $\Delta v_\mathrm{max}$ and corresponding errors for individual synthetic spectra with a given input $b_\mathrm{D,ISM}$ and spectral resolution.
\emph{Bottom:} Same as above, for $\rm EW_{flow}$ (\ion{Mg}{2}).
     \label{fig.fakewinds_varres}}
\end{figure}

To explore the effects of spectral resolution on our measurements, we also smooth each of the model profiles in the $15\times6$ grid described above to FWHM $= 190\mkms$ and $360\mkms$, approximately encompassing the $\pm2\sigma$ range in spectral resolutions of our data.  We generate a single realization of each of these spectra with $\rm S/N = 9~pixel^{-1}$, and apply our two-component model fitting procedure.  The results of these fits are shown in Figure~\ref{fig.fakewinds_varres}.  As shown in the top panel, spectral resolution does not affect our ability to detect winds at  $v_\mathrm{synth} > 250\mkms$ for $b_{D, \rm ISM} \le 80\mkms$.  Resolution does, however, affect our ability to recover a significant `flow' component in cases for which outflows are detected, and affects the sizes of the uncertainties on these quantities.  Please see \S\ref{sec.sensitivity} for more discussion of these effects.




\section{LRIS Spectra and Kinematic Measurement Results}

Here we show all spectra analyzed in this work, excluding those of insufficient S/N to constrain \ion{Fe}{2} or \ion{Mg}{2} absorption 
kinematics.  Objects are presented in the same order as they are listed in Tables~\ref{tab.gals}, \ref{tab.mglines}, and \ref{tab.felines}, 
starting at the top left of each panel.  \ion{Fe}{2} profiles are shown on the left of each column, and the corresponding \ion{Mg}{2} profile is shown on the right.  
The error in each pixel is shown with the gray curve.
Pixels which have been flagged prior to fitting and replaced with values at the continuum level due to the presence of either \ion{Mg}{2} emission or \ion{Mn}{2} absorption are marked in red.  If a wind has been detected in a given transition, the central velocity of the one-component fit ($\Delta v_1$) is indicated with a vertical dashed magenta line.  In cases for which a significant $\rm EW_{flow}$ was measured, $\Delta  v_\mathrm{max}$ is marked with a vertical solid magenta line.  The continuum level is shown with the horizontal dashed blue line.

\begin{figure}
\centering\includegraphics[angle=90,width=6.5in]{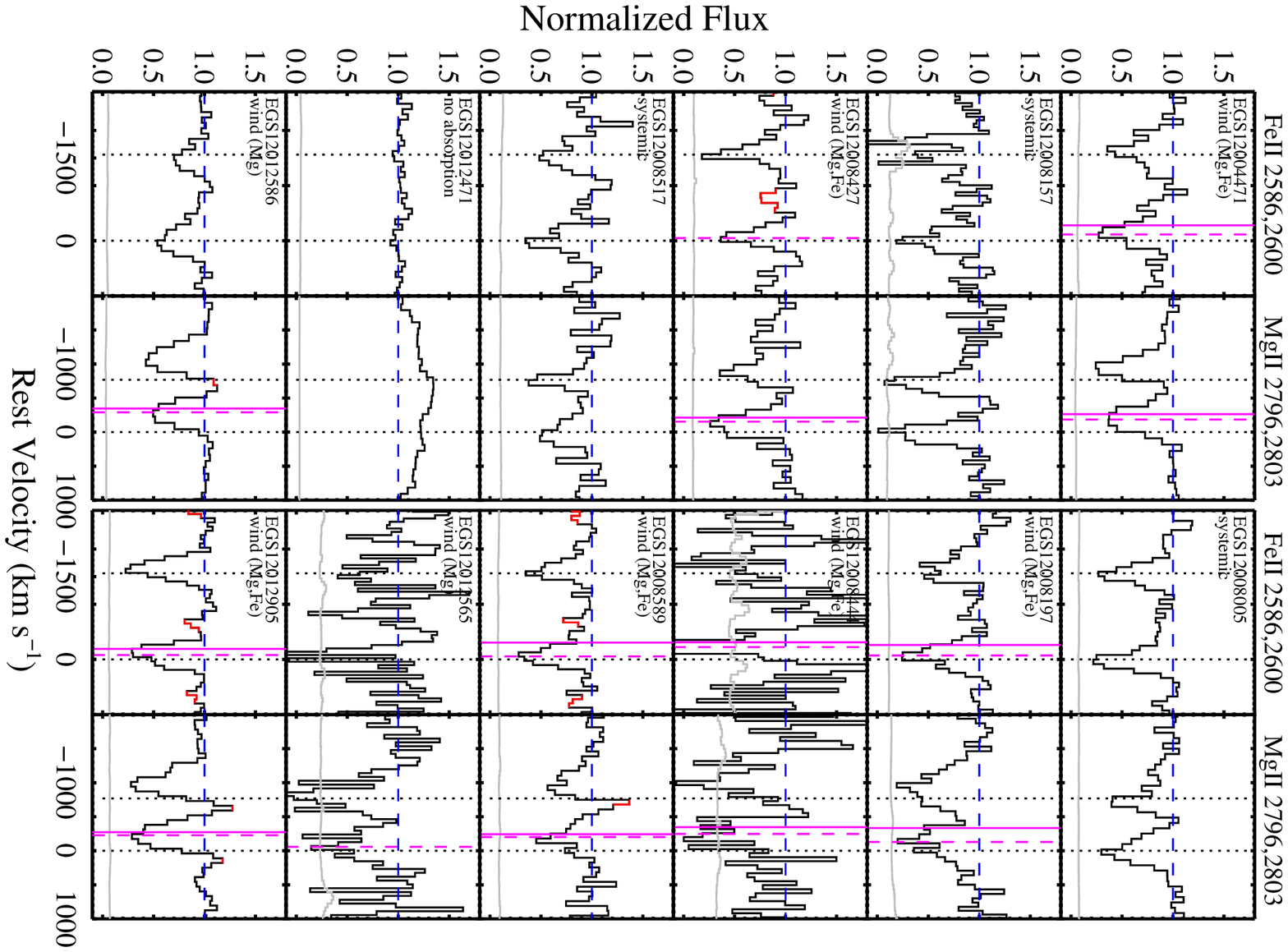}
\caption[]{\ion{Fe}{2} (left-hand panel in each column) and \ion{Mg}{2} (right-hand panel) profiles analyzed in this work.  The systemic velocity of each transition is shown with vertical dotted black lines, and the horizontal dashed blue line marks the continuum level.  The gray curve shows the error in each pixel.  Pixels which have been flagged prior to fitting and replaced with values at the continuum level due to the presence of either \ion{Mg}{2} emission or \ion{Mn}{2} absorption are marked in red.  If a wind has been detected in a given transition, the central velocity of the one-component fit ($\Delta v_1$) is indicated with a vertical dashed magenta line.  In cases for which a significant $\rm EW_{flow}$ was measured, $\Delta v_\mathrm{max}$ is marked with a vertical solid magenta line.
     \label{fig.allspecs}}
\end{figure}

\addtocounter{figure}{-1}

\begin{figure}
\centering\includegraphics[angle=90,width=6.5in]{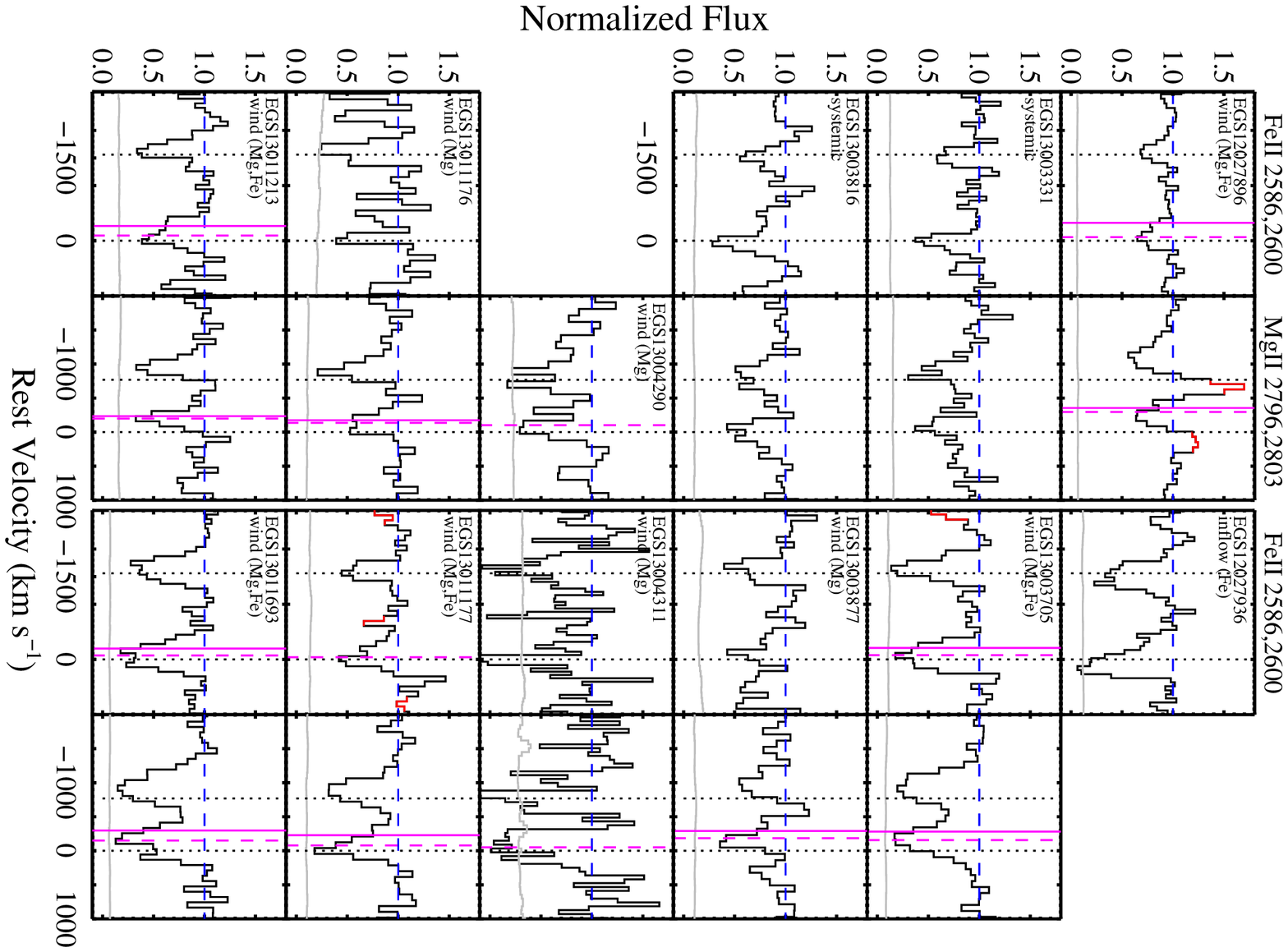}
\caption[]{ \emph{continued}
     \label{fig.allspecs1}}
\end{figure}

\addtocounter{figure}{-1}
\begin{figure}
\centering\includegraphics[angle=90,width=6.5in]{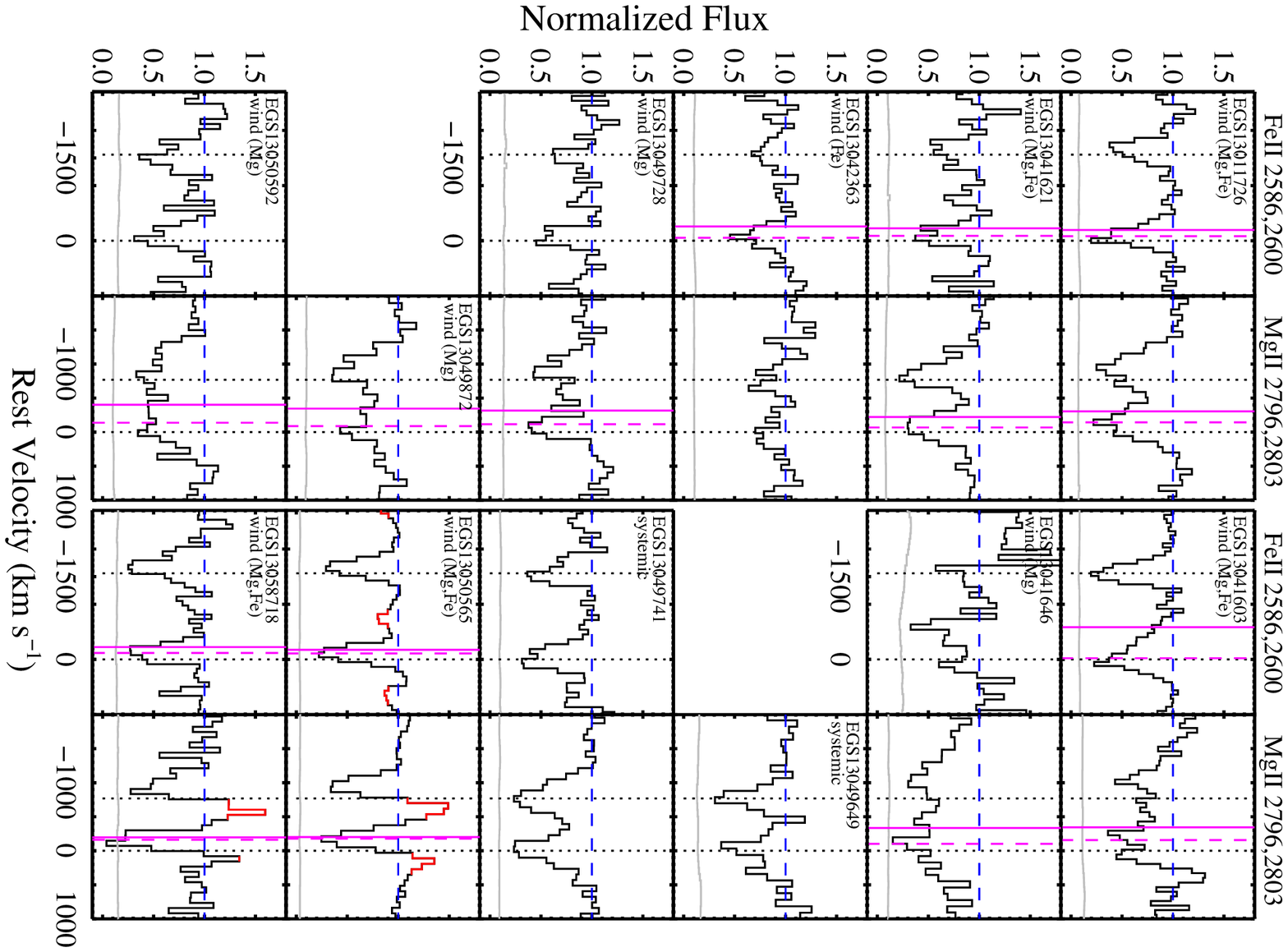}
\caption[]{ \emph{continued}
     \label{fig.allspecs2}}
\end{figure}

\addtocounter{figure}{-1}
\begin{figure}
\centering\includegraphics[angle=90,width=6.5in]{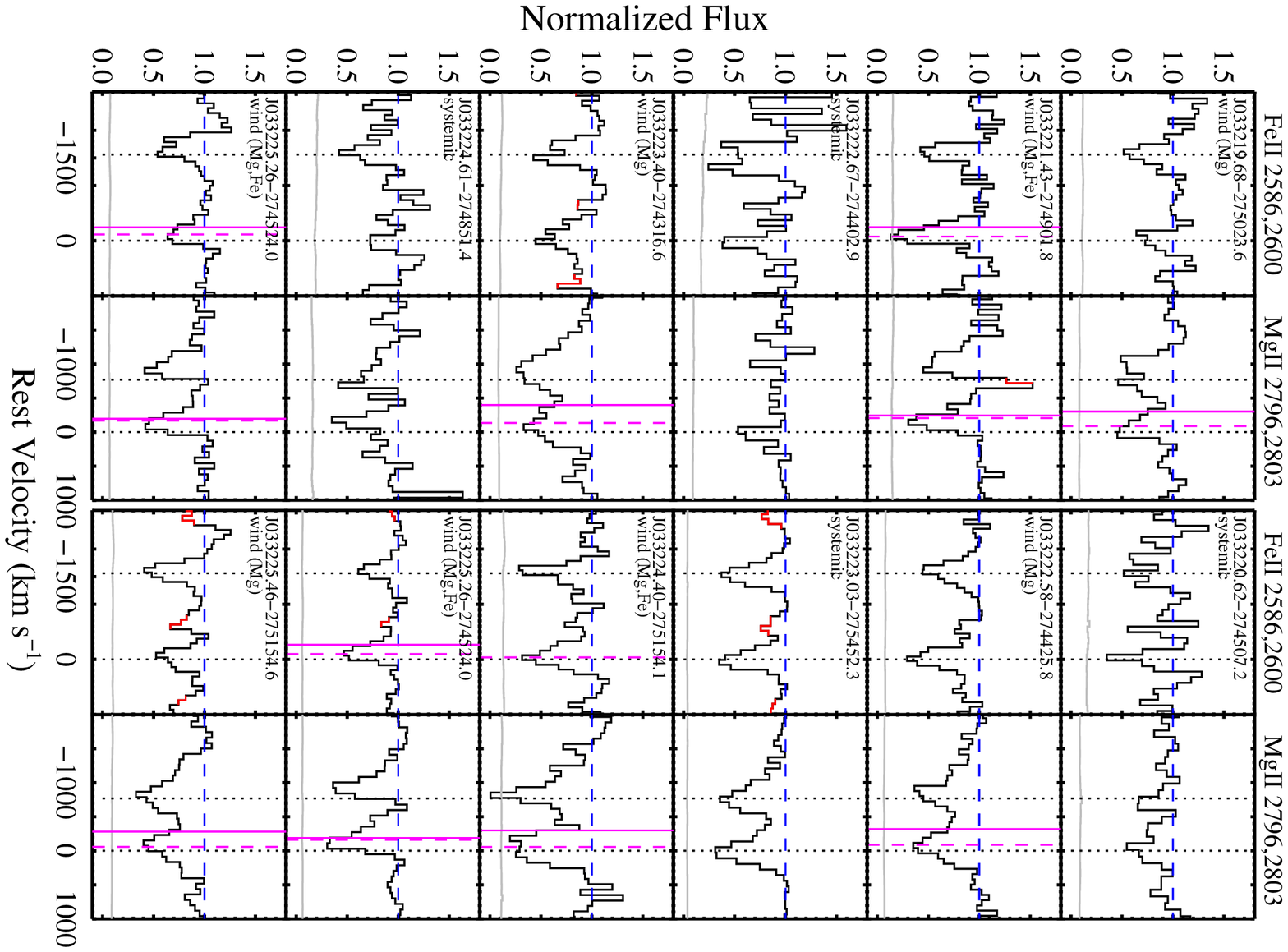}
\caption[]{ \emph{continued}
     \label{fig.allspecs3}}
\end{figure}

\addtocounter{figure}{-1}
\begin{figure}
\centering\includegraphics[angle=90,width=6.5in]{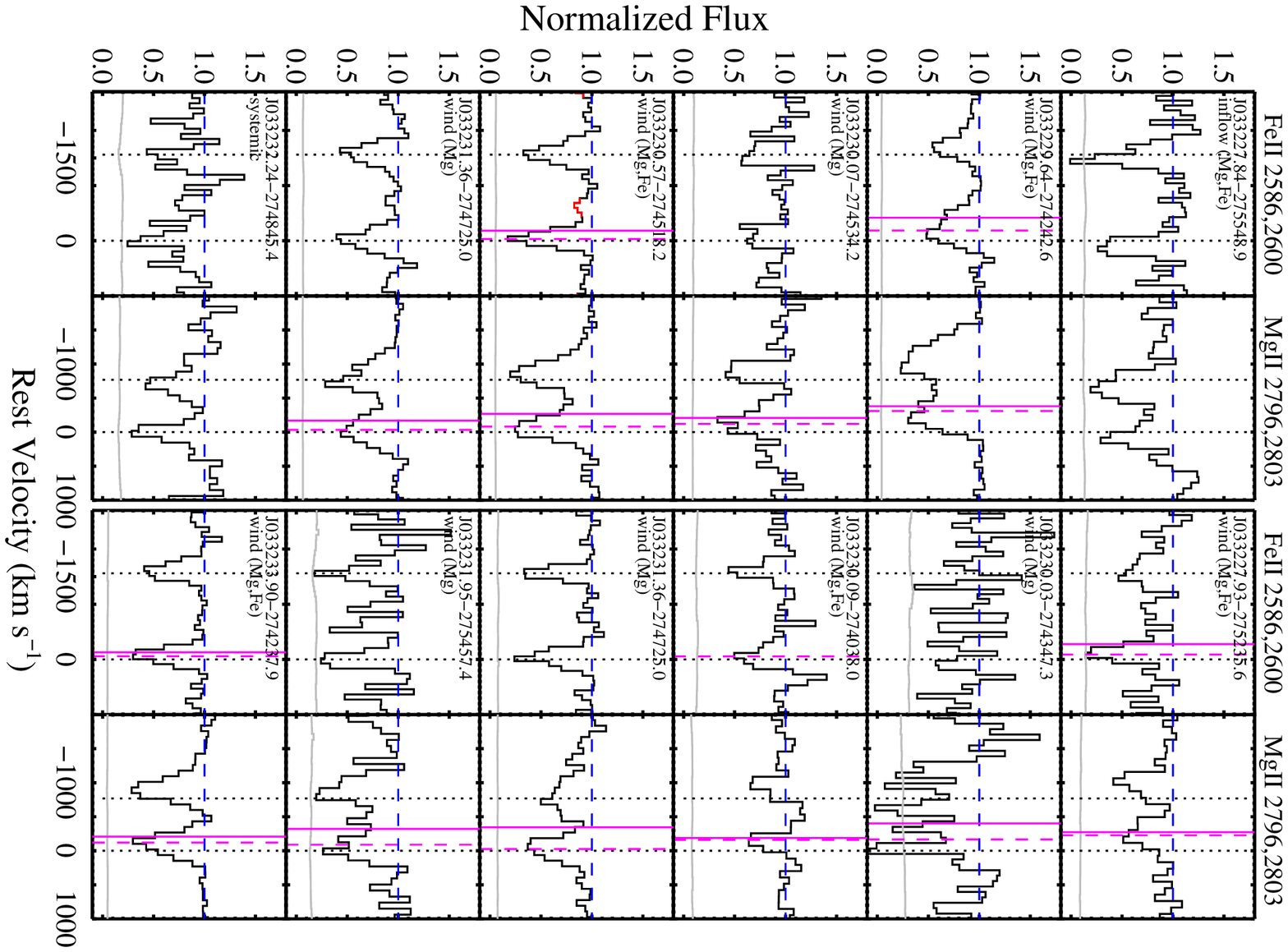}
\caption[]{ \emph{continued}
     \label{fig.allspecs4}}
\end{figure}

\addtocounter{figure}{-1}
\begin{figure}
\centering\includegraphics[angle=90,width=6.5in]{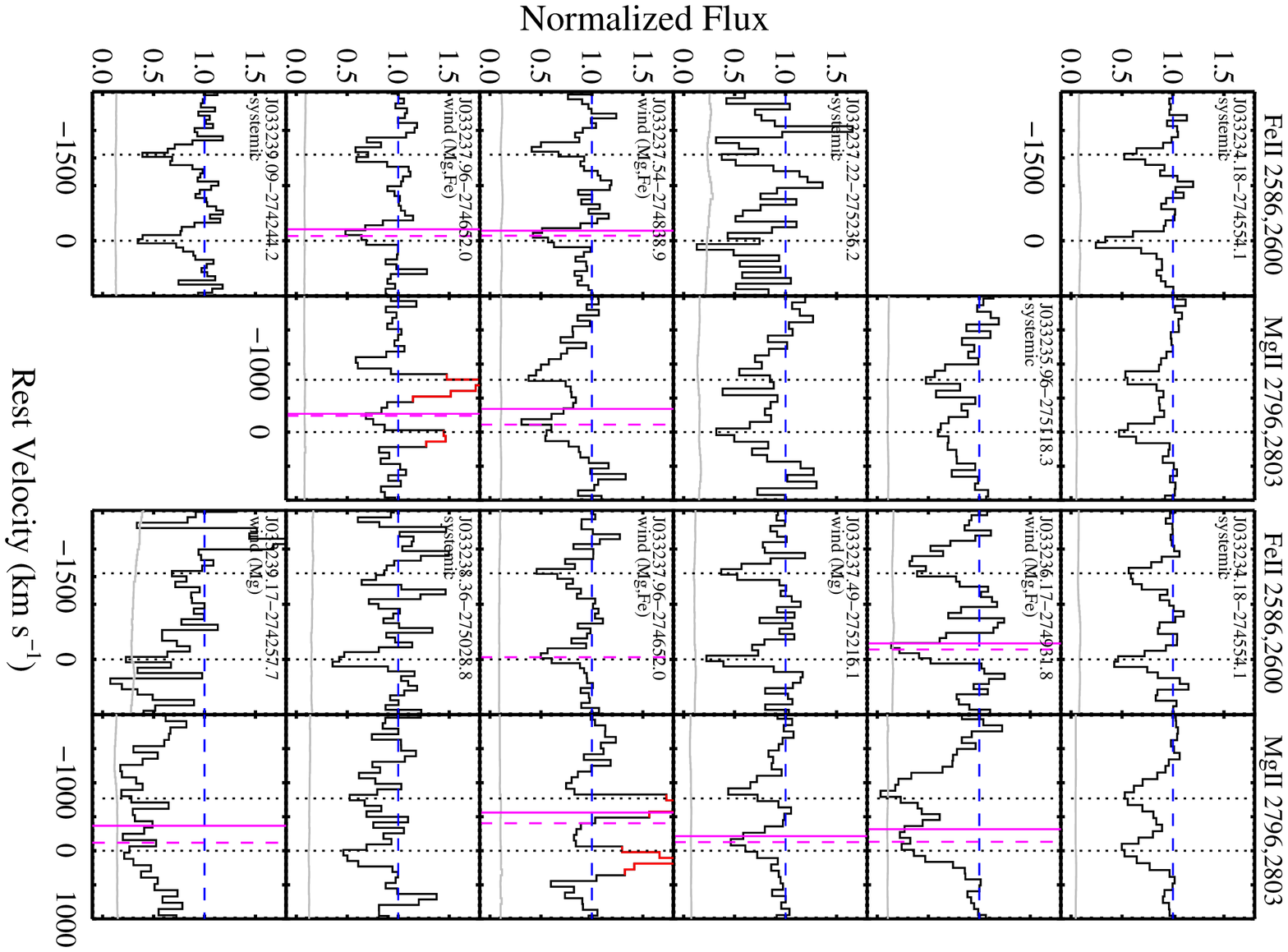}
\caption[]{ \emph{continued}
     \label{fig.allspecs5}}
\end{figure}

\addtocounter{figure}{-1}
\begin{figure}
\centering\includegraphics[angle=90,width=6.5in]{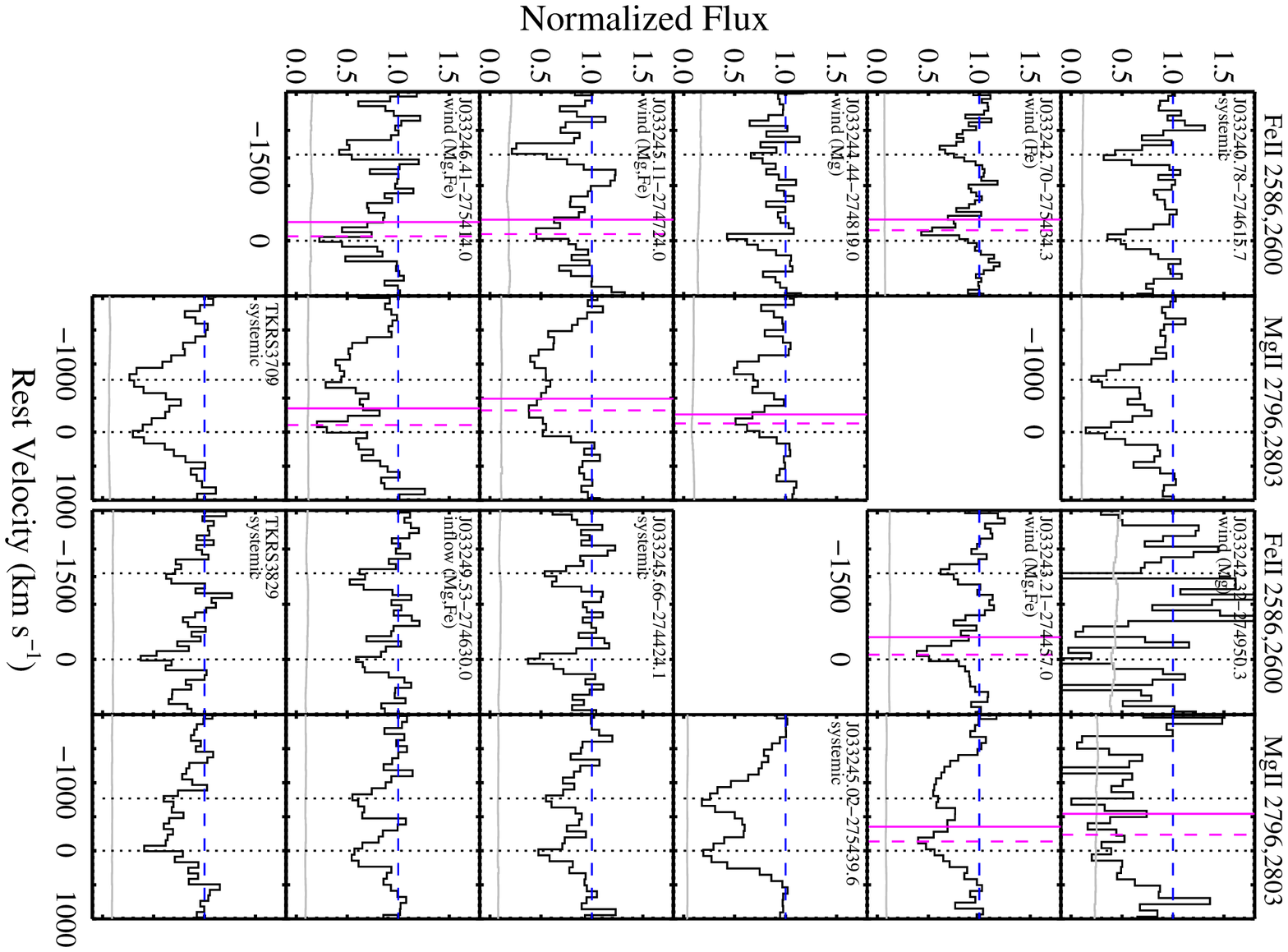}
\caption[]{ \emph{continued}
     \label{fig.allspecs6}}
\end{figure}

\addtocounter{figure}{-1}
\begin{figure}
\centering\includegraphics[angle=90,width=6.5in]{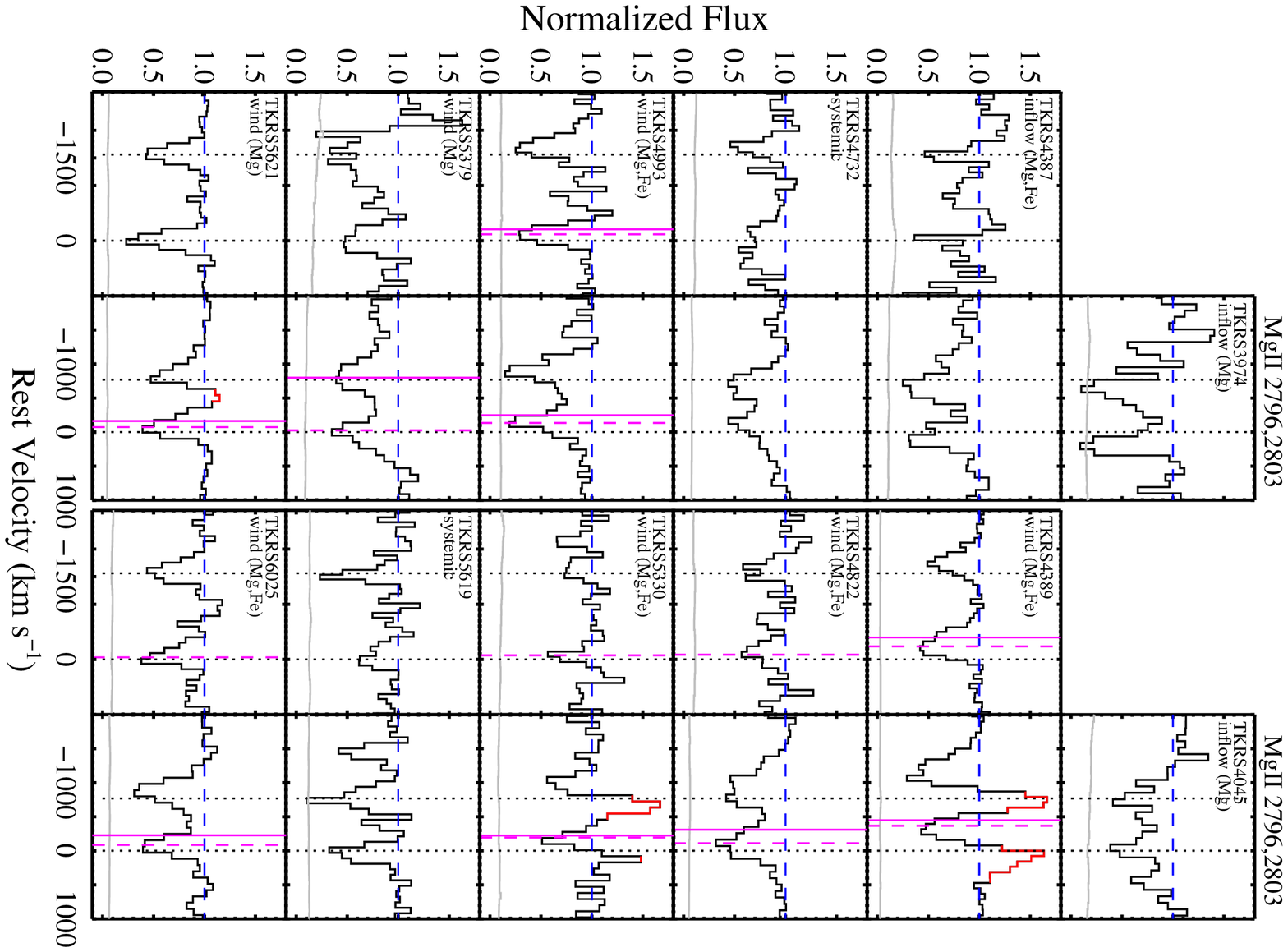}
\caption[]{ \emph{continued}
     \label{fig.allspecs7}}
\end{figure}

\addtocounter{figure}{-1}
\begin{figure}
\centering\includegraphics[angle=90,width=6.5in]{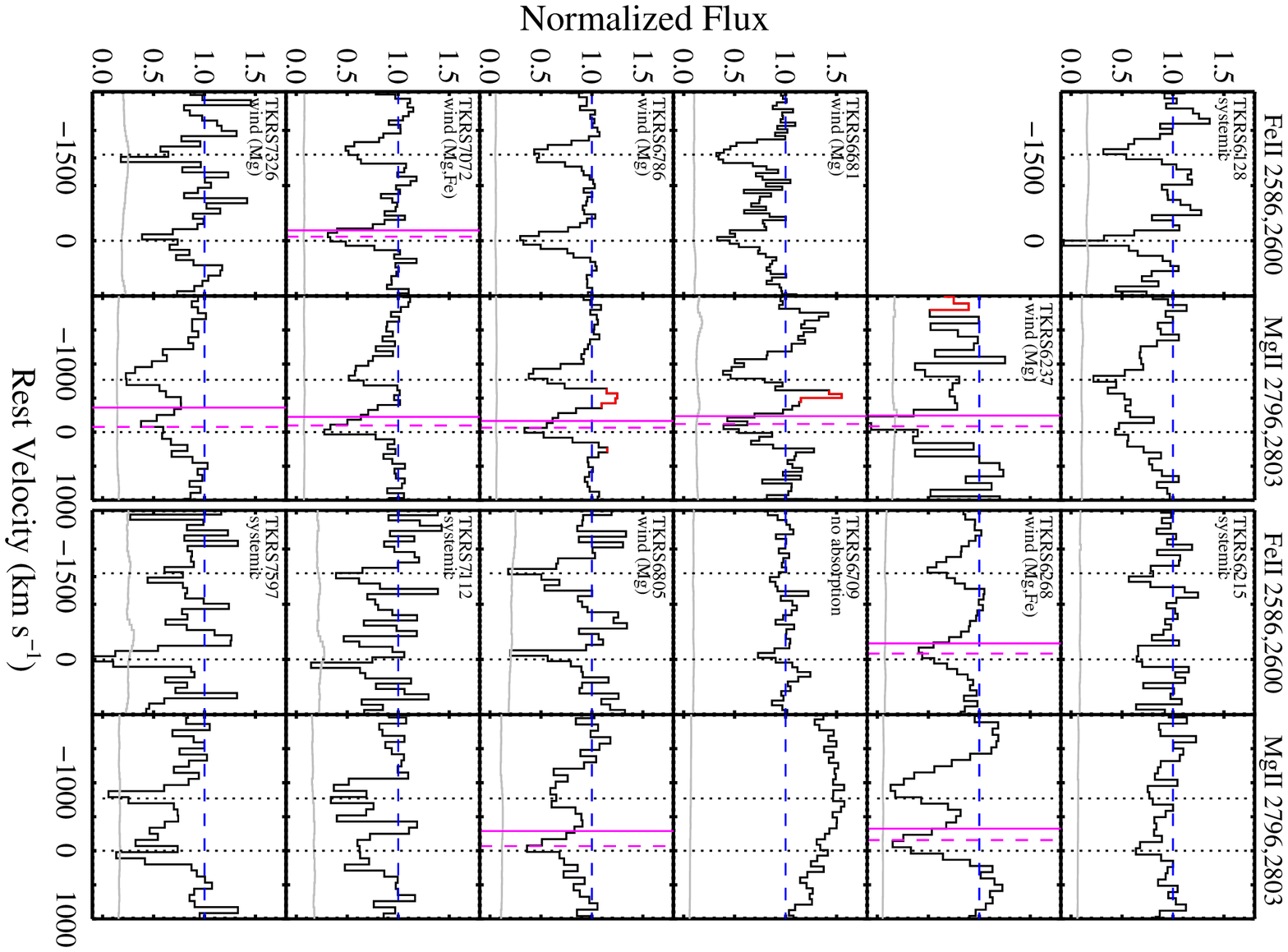}
\caption[]{ \emph{continued}
     \label{fig.allspecs8}}
\end{figure}

\addtocounter{figure}{-1}
\begin{figure}
\centering\includegraphics[angle=90,width=6.5in]{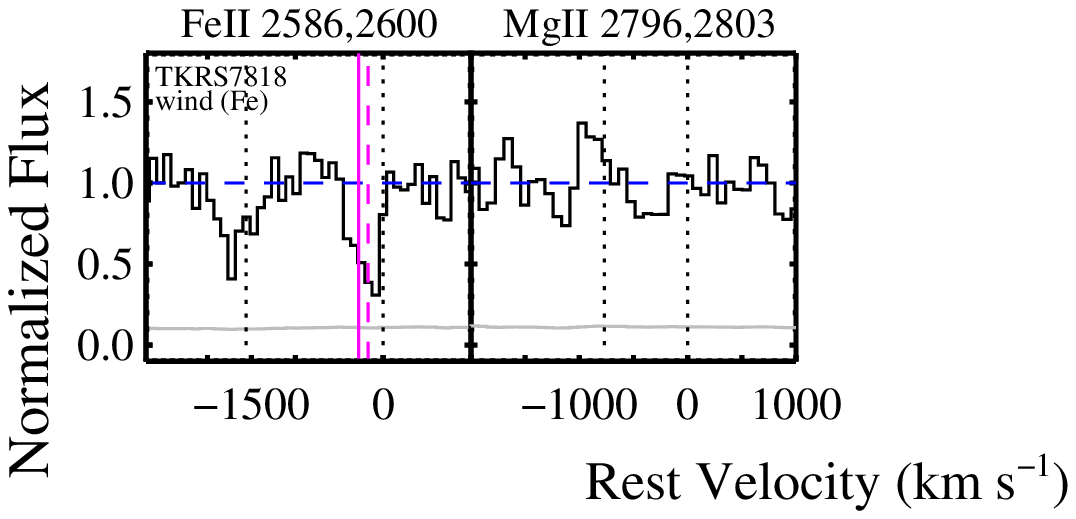}
\caption[]{ \emph{continued}
     \label{fig.allspecs9}}
\end{figure}

\end{document}